\documentclass{aa}
\usepackage{graphicx}
\usepackage{caption}
\usepackage{subcaption}
\usepackage{txfonts}
\usepackage{lscape}
\usepackage[]{natbib}

\newcommand{\codu}{CO~J=2$\rightarrow$1}
\newcommand{\cotd}{CO~J=3$\rightarrow$2}

\newcommand{\dcodu}{$^{12}$CO~J=2$\rightarrow$1}

\renewcommand{\footnote}[1]{
  \textsuperscript{ %ecriture en exposant
    \addtocounter{footnote}{1} %incrementation du compteur
    (\thefootnote) % impression au format "(compteur)"
  }
   \footnotetext{#1} % la note de bas de page
}

\begin{document}

\title{The hybrid disks: a search and study to better understand evolution of disks. }
% \title{A deep CO survey in young debris disks.}
\author{J.~P\'ericaud\inst{\ref{LAB} ,{\ref{CNRSLAB}}}\and
E.~Di~Folco\inst{\ref{LAB},\ref{CNRSLAB}}\and A.~Dutrey\inst{\ref{LAB},\ref{CNRSLAB}}\and S.~Guilloteau\inst{\ref{LAB},\ref{CNRSLAB}}\and V.~Pi\'etu\inst{\ref{IRAM}}}

\institute{Univ. Bordeaux, LAB, UMR 5804, F-33615, Pessac, France \label{LAB}
\and
CNRS, LAB, UMR 5804, F-33615, Pessac, France \label{CNRSLAB}
\and
IRAM, 300 rue de la piscine, F-38406 Saint Martin d'H\`eres, France. \label{IRAM}
}

   \date{Received July 22, 2016; accepted December 1, 2016}

% \abstract{}{}{}{}{}
% 5 {} token are mandatory

  \abstract
  % context heading (optional)
  % {} leave it empty if necessary
   {The increased sensitivity of millimeter-wave facilities now makes possible the detection of low amounts of gas in debris disks. Some of the gas-rich debris disks harbor peculiar properties, with possible pristine gas and secondary generated dust. The origin of the gas in these hybrid disks is strongly debated and the current sample is too sparse to understand this phenomenon.}
  % aims heading (mandatory)
   {More detections are necessary to increase the statistics on this population. Lying at the final stages of evolution of proto-planetary disks and at the beginning of the debris disk phase, these objects could provide new insight into the processes involved in the making of planetary systems.}
  % methods heading (mandatory)
   {We carried out a deep survey of the \codu{} and \cotd{} lines with the \textit{APEX} and \textit{IRAM} radiotelescopes in young debris disks selected according to hybrid disk properties. The survey is complemented with a bibliographic study of the ratio between the emission of the gas and the continuum (S$_{\rm CO}$/F$_{\rm cont}$) in CTTS, Herbig Ae, WTTS, hybrid, and debris disks.}
  % results heading (mandatory)
   {
   Our sub-mm survey comprises 25 stars, including 17 new targets, and we increase the sensitivity limit by a factor 2 on eight sources compared to similar published studies. We report a
   % one tentative detection,
   % to the very low number of hybrid disk candidates
   4\,$\sigma$ tentative detection of a double-peaked \codu{} line around HD~23642; an eclipsing binary located in the Pleiades. We also reveal a correlation between the emission of the CO gas and the dust continuum from CTTS, Herbig Ae and few debris disks. The observed trend of the gas to dust flux ratio suggests a concurrent dissipation of the dust and gas components. Hybrid disks systematically lie above this trend, suggesting that these systems may witness a transient phase, when the dust has evolved more rapidly than the gas, with a flux ratio S$_{\rm CO}$/F$_{\rm cont}$ enhanced by a factor of between 10 and 100 compared to standard (proto-)planetary disks.}
   {}


   \maketitle

%________________________________________________________________
%________________________________________________________________
%________________________________________________________________

\section{Introduction}

The scenarios of planet formation in disks are constrained by the observed lifetime of circumstellar material around pre-main sequence stars. This is particularly true
for the late accretion phases of giant gaseous planets.

In gas-rich disks, significant progress has been made in the past decade to find direct
evidence for disk evolution. Inner holes have been found in an increasing number of
proto-planetary Class II disks \citep[e.g. LkCa15,][]{Pietu+etal_2006}, with a frequency
as high as 20\% \citep[e.g.][]{Owen+Clarke_2012}.  In the cavity, the evolution of the
gaseous component, and how it can affect the dynamics of the small grains during the last
stages of (giant) planet formation is not yet known. Near-IR images of these transitional
disks revealed for instance smaller inner holes than those seen at millimeter wavelengths
\citep{Garufi+etal_2013}, a possible consequence of the spatial distribution of the
residual, primordial gas. % Recent observations of transitional disks obtained by ALMA
% recently provide the first images of the CO gas and dust distributions in transitional
% disks, revealing many dust assymetries  \textbf{(REF to ADD. HD142, etc... Van der Marel,
% Simon Bruderer...)}.

On the other hand, significant IR excess has been associated with many main
sequence stars \citep[e.g.][]{Eiroa+etal_2013}), without any significant reservoir of gas
except for a few younger ($< 100$ Myr) stars \citep{Dent+etal_2013, Howard+etal_2013}. In
these systems, the dust lifetime is short, and therefore the observed dust must be constantly
replenished from collisions between exo-asteroids and/or exo-comets. These so-called
debris disks represent extra-solar analogs to our Kuiper Belt.
The young $\beta$~Pictoris system is an impressive example of this sub-category. It has
an optically thin dust disk, and ALMA images of the cold CO emission reveal asymmetries
that are attributed to the destruction of cometary-like bodies, possibly trapped by an
unseen planet \citep{Dent+etal_2014}. The gas in this system is therefore not primordial.

Among young MS stars surrounded by debris disks, a handful of remarkable objects exhibit
a substantial amount of primordial gas, making them hybrid disks between proto-planetary
and debris disks: 49\,Ceti \citep[40~Myr,][]{Zuckerman+etal_1995,Moor+etal_2011}, HD~21997 \citep[30~Myr,][]{Kospal+etal_2013}, HD~141569 \citep[5~Myr,][]{Dent+etal_2005}, and HD~131835 \citep[16~Myr,][]{Moor+etal_2015}. This raises questions
over the specific physical conditions that allow CO gas disks to survive in stars as
old as 30\,Myr  since most disks dissipate in $\sim$5-10\,Myr only. %and about the
The dominant physical mechanisms at play that dissipate such old gas disks are photoevaporation
or runaway gas accretion by unseen giant planets, which may then present different
characteristics since they would accrete on a longer timescale. The paucity of such
hybrid systems may be explained by the very fast dissipation of the primordial reservoir
of gas, as soon as this process has started. They may represent the missing link between
proto-planetary and planetary systems, as suggested by the recent analysis of the \textit{ALMA}
observations of HD\,21997 by \citet{Kospal+etal_2013}, and should therefore be considered
as transient systems with ongoing dissipation of their primordial gas component and last
stages of (giant) planet formation.  However, in some cases at least, part of the
observed CO gas may be of second generation as suggested for 49~Ceti
\citep{Roberge+etal_2013}.
%Comparing gas and dust disk structures, another issue is related to the dynamical coupling between the gas and dust particles. Since it depends on the
%dust grain size, one would like to know what is the physical  link between the residual gas and the NIR dusty structures observed in these disks.
So far, all these disks have been found around Herbig Ae stars and there are no equivalent
hybrid disks observed around old T Tauri stars \citep{Lieman-Sifry+etal_2016}.

Millimeter surveys of gas in debris disks yielded low detection rates, confirming
the paucity of gas-rich disks. \citet{Dent+etal_2005} reported the first
detection of \cotd{} in 49~Ceti from \textit{JCMT} observations out of a sample of 59 HAeBe and
Vega-like stars. \citet{Pascucci+etal_2006} focused on 15 (5-100\,Myr old) stars from the
FEPS sample, but their \textit{SMT} search for \codu{} and J~=~3$\rightarrow$2 lines yielded only non
detections and upper-limits on the gas mass. \citet{Kastner+etal_2010} also failed to
detect molecular gas emission with \textit{IRAM-30m} antenna in a sample of 10 Vega-like stars.
Only recently, \citet{Moor+etal_2011} discovered a significant \cotd{} and \codu{}
emission from the 30\,Myr old star HD~21997 using the \textit{APEX}/SHFI receiver. The most sensitive survey, however, only reached upper limits on the CO gas
mass of the order of $M_{\rm CO} \lesssim  0.5-1.10^{-3}M_{\oplus}$.

We report here a more sensitive survey of 25 debris disks in \codu{} or J~=3$\rightarrow$2 using
\textit{APEX} and the \textit{IRAM-30m} radiotelescope. With these observations, we reach a sensitivity limit similar to that of the recent work of \citet{Moor+etal_2015}. We have also complemented these new observations
with a compilation \textbf{of} more recent observations of dust and gas in young disks. We report the
results in this paper. Section 2 outlines the sample of sources and the observations.
The description of the analysis by correlation is given in Section 3 while Section 4
is an open discussion of these results.

%{\bf However, detecting new hybrid disks with an adequate sensitivity using APEX is a necessary prerequisite before preparing an ALMA imaging project. }

%________________________________________________________________
%________________________________________________________________
%________________________________________________________________
%________________________________________________________________
%________________________________________________________________

\section{Samples and data reduction}
\label{sec:survey_co}
%________________________________________________________________
%________________________________________________________________
%________________________________________________________________

\subsection{APEX and IRAM observations}
\label{subsec:sample_results}

%________________________________________________________________
%________________________________________________________________
%________________________________________________________________

We performed a $^{12}$CO~J~=~3$\rightarrow$2 and $^{12}$CO~J~=~2$\rightarrow$1 survey of the sample
presented in Table \ref{table:log}. We selected 25 systems whose properties are
similar to the recently identified gas-rich debris disks, thereby focusing on A-type
stars younger than 100~Myr and with a large dust radius $R_{\rm dust}>$30~au. The sources
were collected from various \textit{IRAS}, \textit{Spitzer}, \textit{WISE,} and
\textit{Herschel} debris disks studies \citep{Rhee+etal_2007,
Wyatt+etal_2007,Zuckerman+etal_2004, Patel+etal_2014, Morales+etal_2009, Chen+etal_2006,
Riviere-Marichalar+etal_2014, Kennedy+etal_2014}.
%CO gas has already been searched for around 8 stars of our sample, but not at the sensitivity level that we present here \citep{Kastner+etal_2010, Moor+etal_2011}.
Among the 25 selected targets, 17 are new sources for the search of CO and 8 sources had
been observed in only one of these two CO transitions. We also included 2 targets with
atomic gas sprectroscopic features interpreted as the signature of Falling-Evaporating
Bodies (FEB) \citep[HD~21620 and HD~42111, ][]{Welsh+etal_2013}.

For northern targets, we used the \textit{IRAM-30m} telescope (project 172-14),
attempting to detect the \codu{} line with the EMIR receiver connected to the FTS200
backend providing 200~kHz spectral resolution. In the southern hemisphere, we
used the \textit{APEX} telescope at 345~GHz to search for the \cotd{} transition
with the SHeFI receiver (projects 094.C-0161 and 095.C-0742).

\begin{figure}
   \centering
   \includegraphics[width=0.45\textwidth,keepaspectratio]{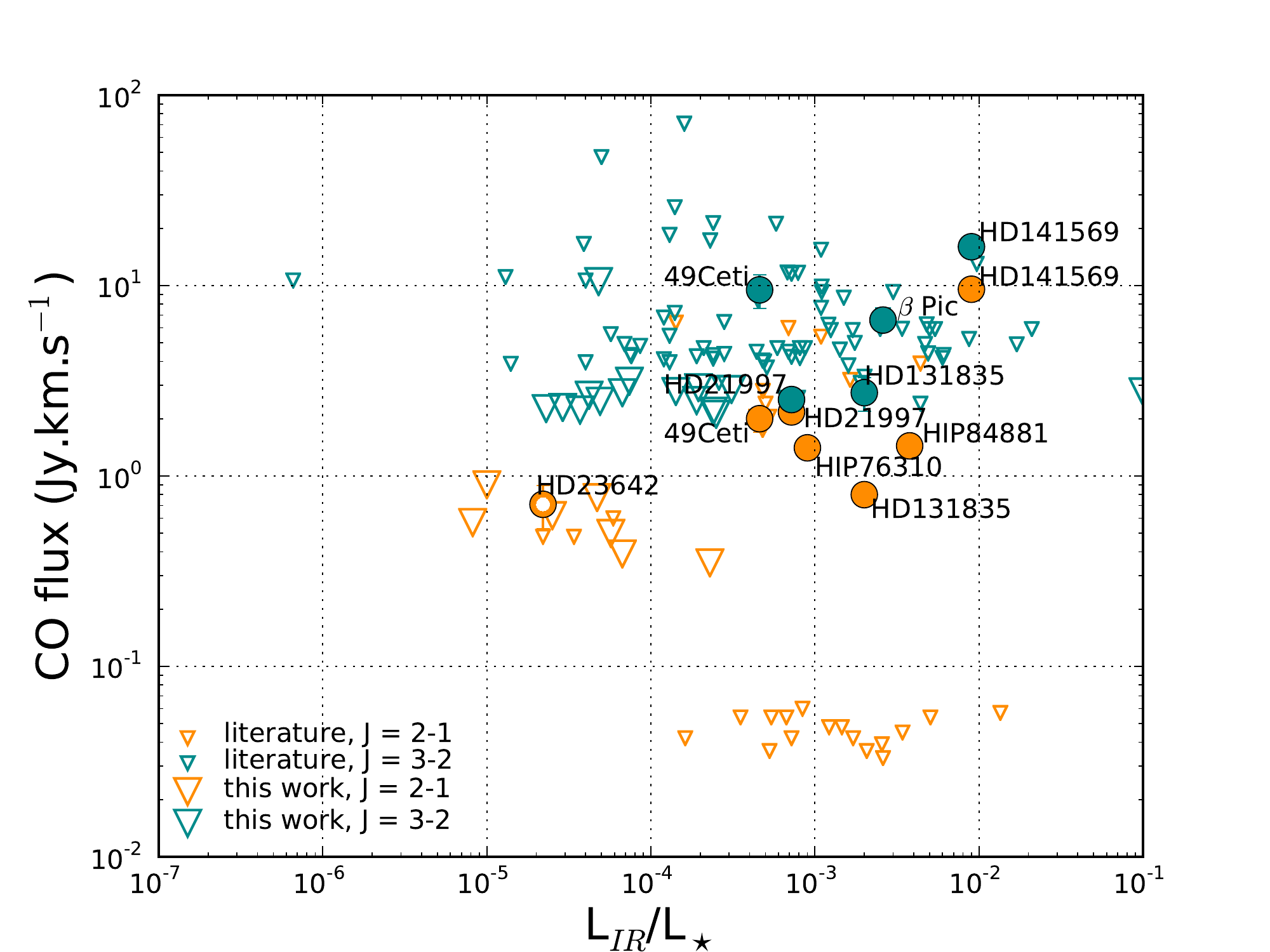}
   \caption{
   % Sum up of the search of CO gas in debris disks (Table 1). The upper limits on the integrated flux of CO is compared to
   % the infrared fractional luminosity of the debris disks (big down triangles for our survey, small down triangles for
   % litterature values:
   CO line flux as a function of infrared excess. Circles indicate CO detection, and triangles
   indicate upper limits. Large symbols are from our survey, and small symbols for literature data
   \citep{Dent+etal_2005,Kastner+etal_2010,
   Moor+etal_2011,Moor+etal_2015,Hales+etal_2014, Lieman-Sifry+etal_2016}).
   % The disks where gas has been detected (hybrids + $\beta$ Pic) are marked by circles.
   Orange symbols stand for the \codu{} measurements and green symbols for \cotd\ measurements{}. The tentative detection around HD~23642 is plotted with an empty circle.}
   \label{fig:flux_luminosity}%
\end{figure}

The data reduction was done using the GILDAS/CLASS package. Spectra have been
resampled to a $\delta v$=1 km.s$^{-1}$ channel width. The baseline subtraction and rms
calculation were performed on 800 channels, avoiding the contamination lines when
present. The conversion from the $T_A^{\ast}$ scale onto flux density (Jy) was done
with the following factors: 41\,Jy/K for
\textit{APEX}\footnote{http://www.apex-telescope.org/telescope/efficiency/} at 345~GHz and
7.8\,Jy/K for the \textit{IRAM-30m}\footnote{http://www.iram.es/IRAMES/mainWiki/Iram30mEfficiencies/} at 230~GHz. The
final spectra are plotted on Figs. \ref{fig:spectres_iram} and \ref{fig:spectres_apex}, and the upper limits on CO emission are displayed as a function of the infrared excess of disks on Fig. \ref{fig:flux_luminosity}.
The median distance of the observed stars is 69~pc, with a standard deviation of 46~pc.
Since the flux values depend on the distance of the stars, they are not directly
representative of the gas mass reservoir. We therefore present the fluxes scaled at 100~pc for all 25 stars in our sample and
other targets from the literature  in Fig.
\ref{fig:flux_luminosity_100}.

%\label{subsec:upp_limits}

% It is possible that gas produced by collisions is present in the disk we have observed, but only higher sensitive observations would detect it, as done for $\beta$ Pictoris (Dent+etal_2014).
% Whether it is the case or not, the non-detections of our survey increase the statistics on the search for CO gas in debris disks, and show that hybrid disks might represent a very peculiar or short phase in the evolution of disks around young stars.

%%%%%%%%%%%%%%%%%%%%%%%%%%%%%%%%%%%%%%%%%%%%%%%%%%%%%%%%%%%%%%%%%%%%%%%%%%%%%%%%%%%%%%%%%%%%%%%%%%%%%%%%%%%%%%%%%%%%%%%%%%
%%%%%%%%%%%%%%%%%%%%%%%%%%%%%%%%%%%%%%%%%%%%%%%%%%%%%%%%%%%%%%%%%%%%%%%%%%%%%%%%%%%%%%%%%%%%%%%%%%%%%%%%%%%%%%%%%%%%%%%%%%

    \begin{table*}
    \caption{\textbf{Observational parameters for our sample of debris disks.}}
    \label{table:log}
    \centering
    \begin{tabular}{ccccccc}     % 7 columns
    \hline\hline
      Name & LSR velocity & L$_{IR}$/L$_{\star}$ & Ref & 3$\sigma_{int}$  & M$_{H_2}$ DiskFit  & M$_{H_2}$ from S$_{CO}$  \\
           & (km.s$^{-1}$)&                      &     &(Jy.km.s$^{-1}$) & (3$\sigma$ upp. limit) M$_{\oplus}$     & (3$\sigma$ upp. limit) M$_{\oplus}$       \\
    \hline
  % \multicolumn{6}{c}{ } \\
    \multicolumn{7}{c}{\textit{IRAM-30m} - \codu{}} \\
    % \multicolumn{6}{c}{ } \\
    \hline
      HD 2772   & -5.4 & 8.2$\times$10$^{-6}$ & 1 & $<$0.569 & $<$ 0.110 & $<$ 0.454 \\
      HD 14055  & 7.5 & 6.7$\times$10$^{-5}$ & 1 & $<$0.387 & $<$ 0.008 & $<$ 0.027 \\
      \textit{HD 15115} & -0.2 & 4.8$\times$10$^{-4}$ & 2 & $<$1.89  & $<$ 0.063 & $<$ 0.229 \\
      HD 21620  & -22.8 & 2.51$\times$10$^{-5}$ & 3 & $<$0.622 & $<$ 0.383 & $<$ 0.819 \\
      HD 23642  & -4.0 & 2.2$\times$10$^{-5}$ & 1 & 0.71 $\pm$ 0.18 & 0.18 $\pm$ 0.04 & 0.57 $\pm$ 0.14 \\
      HD 31295  & -3.5 & 4.7$\times$10$^{-5}$ & 1 & $<$0.768  & $<$ 0.016 & $<$ 0.058 \\
      HD 42111  & 8.4 & 1.00$\times$10$^{-5}$ & 3 & $<$0.902  & $<$ 1.02  & $<$ 2.16  \\
      HD 159082 & 7.1 & 5.7$\times$10$^{-5}$ & 1 & $<$0.497 & $<$ 0.132 & $<$ 0.537 \\
      \textit{HD 218396}  & -12.6 & 2.29$\times$10$^{-4}$ & 4 & $<$0.349 & $<$ 0.005 & $<$ 0.032 \\

    \hline
  % \multicolumn{6}{c}{ } \\
    \multicolumn{7}{c}{\textit{APEX} - \cotd{}} \\
    % \multicolumn{6}{c}{ } \\
    \hline

      HD 225200 & 13.4 & 7.4$\times$10$^{-5}$ & 1 & $<$3.15  & $<$ 0.260 &$<$ 0.470 \\
      \textit{HD 17848} & 16.7 & 4.8$\times$10$^{-5}$ & 1 & $<$10.5  & $<$ 0.146 &$<$ 0.242 \\
      HD 24966  & 11.7 & 3.1$\times$10$^{-4}$ & 1 & $<$2.85  & $<$ 0.166 &$<$ 0.278 \\
      HD 30422  & -4.8 & 3.7$\times$10$^{-5}$ & 1 & $<$2.22  & $<$ 0.036 &$<$ 0.063 \\
      HD 35850  & 1.4 & 2.3$\times$10$^{-5}$ & 1  & $<$2.26  & $<$ 0.007 &$<$ 0.015 \\
      \textit{HD 38206} & 5.8 & 1.4$\times$10$^{-4}$ & 1  & $<$2.78  & $<$ 0.062 &$<$ 0.119 \\
      HD 54341  & 20.0$^{\ast}$ & 1.9$\times$10$^{-4}$ & 1 & $<$2.49  & $<$ 0.106 &$<$ 0.232 \\
      HD 71043  & 5.9 & 6.5$\times$10$^{-5}$ & 1 & $<$2.74  & $<$ 0.073 &$<$ 0.131 \\
      HD 71155  & -4.7 & 4.2$\times$10$^{-5}$ & 1 & $<$2.66  & $<$ 0.017 &$<$ 0.035 \\
      HD 78072  & 2.5 & 2.5$\times$10$^{-4}$ & 1 & $<$2.11  & $<$ 0.071 &$<$ 0.133 \\
      \textit{HD 136246}  & 4.2 & 4.9$\times$10$^{-5}$ & 1 & $<$2.47  & $<$ 0.281 &$<$ 0.459 \\
      \textit{HD 164249}  & 2.5 & 7.2$\times$10$^{-4}$ & 1 & $<$2.38  & $<$ 0.020 &$<$ 0.049 \\
      HD 166191 & 3.8 & 1.0$\times$10$^{-1}$ & 5 & $<$2.77  & $<$ 0.187 &$<$ 0.351 \\
      \textit{HD 181296}  & 14.3 & 2.4$\times$10$^{-4}$ & 1 & $<$2.19  & $<$ 0.014 &$<$ 0.046 \\
      \textit{HD 182681}  & 10.5 & 1.95$\times$10$^{-4}$ & 4 & $<$2.91  & $<$ 0.083 &$<$ 0.124 \\
      HD 220825 & -1.5 & 2.9$\times$10$^{-5}$ & 1 & $<$2.29  & $<$ 0.011 &$<$ 0.046 \\
    \hline
    \end{tabular}
    \tablefoot{$^{\ast}$ The radial velocity of HD~54341 was kindly provided by F.~Kiefer, see \citet{Kiefer+etal_inprep}.
    Stars in italics are systems with earlier detection limits in one CO transition, see \S \ref{subsec:sample_results}.
    References for L$_{IR}$/L$_{\star}$ : 1 - \citet{Chen+etal_2014}; 2 - \citet{Moor+etal_2011}; 3 - \citet{Roberge+etal_2008}; 4 - \citet{Rhee+etal_2007}; 5 - \citet{Schneider+etal_2013}.
    The last three colums show the 3$\sigma$ upper limits derived from our study:
    on the integrated intensity of CO, the H$_2$ mass derived from
    DiskFit model and the H$_2$ mass directly derived from the integrated flux.
    The integrated flux and the H$_2$ masses for the tentative detection around HD~23642
    are also displayed.}
    \end{table*}

%----------------------
\subsection{Compilation of existing CO and mm continuum data}
\label{subsec:compilation}

% To increase our sample,
We have compiled measurements of the \codu{} and \cotd{} emission from the
literature, as well as the continuum emission at the corresponding wavelength for a wide
variety of disks. We consider five categories: the classical T Tauri stars (CTTS), the
Herbig~AeBe, the weak-line T Tauri stars (WTTS), the hybrid disks, and finally the debris
disks.
We do not include class I objects in the sample to avoid confusion with remnant envelopes
and nearby molecular clouds.
% We have classified what is called the transitional disks in the CTTS categorie, since their gas and dust properties give evidence for a young phase of evolution.
The measurements are compiled in Table \ref{table:1}. We have retrieved the integrated emission (Jy.km.s$^{-1}$) for
the CO
lines, and the upper limits are always given at
the 3$\sigma$ level.
% The list also includes the unpublished detection of CO J~=~2--1 gas
% with ALMA around the debris star HIP~84881 (archival data, project 2012.1.00688.S, PI:
% J.~Carpenter). For this system, we have derived the CO emission from the products
% provided in the archive with CASA.
%, in order to have an estimation of the CO content of the disk; no deep analysis has been done.

We have also included binary systems. As most of them are either spectroscopic or wide binaries, the
dynamical influence on the disk should be negligible. We selected binary sources
from the \textit{ALMA} survey of \citet{Akeson+etal_2014}. The authors focussed on the analysis of
the continuum emission, and we have retrieved the \cotd{} emission of 15 sources from
the archive (project 2011.0.00150.S). We included binaries with dynamical truncation,
such as GG~Tau, since no systematic effect on the ratio $S_{\rm CO}/F_{\rm cont}$ is
seen. We have also used the data from the \textit{ALMA} project 2013.1.00163.S (PI: M. Simon),
which aims at determining the stellar mass of low-mass T Tauri stars by measuring the
Keplerian rotation of the gas. This project adds eight sources to our compilation. In
addition, for HD~21997, we have determined the upper limit on the dust emission at 1.3~mm
from the archival data (project 2011.0.00780.S).

Most of these sources are in young stellar associations. The most represented
complex is the Taurus-Auriga region, with approximatively half of the CTTS and Herbig
AeBe, and one third of WTTS. Other stars are either isolated or belong to different clouds such
as Ophiucus, upper Scorpius, TW~Hya, Orion, Chameleon.

The measurements at 1.3~mm or 0.87~mm are not available for every source. However, we
find a clear correlation between the observed \codu{}~/~\cotd{} fluxes (scaled at
100\,pc). We have thus tried to complement the missing measurements by interpolating the
flux of the missing transition on this correlation line (the procedure is described in
Appendix \ref{appendix:interpol}). This assumes that the disks have physical conditions
similar to those used to derive the correlation. The corresponding fluxes are represented
by smaller symbols in Fig.\ref{fig:co_vs_dust_21} and \ref{fig:co_vs_dust_32}.
\section{Results and analysis}

% \subsection{Upper limits on CO flux density}

%%%%%%%%%%%%%%%%%%%%%%%%%%%%%%%%%%%%%%%%%%%%%%%%%%%%%%%%%%%%%%%%%%%%%%%%%%%%%%%%%%%%%%%%%%%%%%%%%%%%%%%%%%%%%%%%%%%%%%%%%%
%%%%%%%%%%%%%%%%%%%%%%%%%%%%%%%%%%%%%%%%%%%%%%%%%%%%%%%%%%%%%%%%%%%%%%%%%%%%%%%%%%%%%%%%%%%%%%%%%%%%%%%%%%%%%%%%%%%%%%%%%%

\begin{figure*}
    \centering
    \includegraphics[width=0.95\textwidth,keepaspectratio]{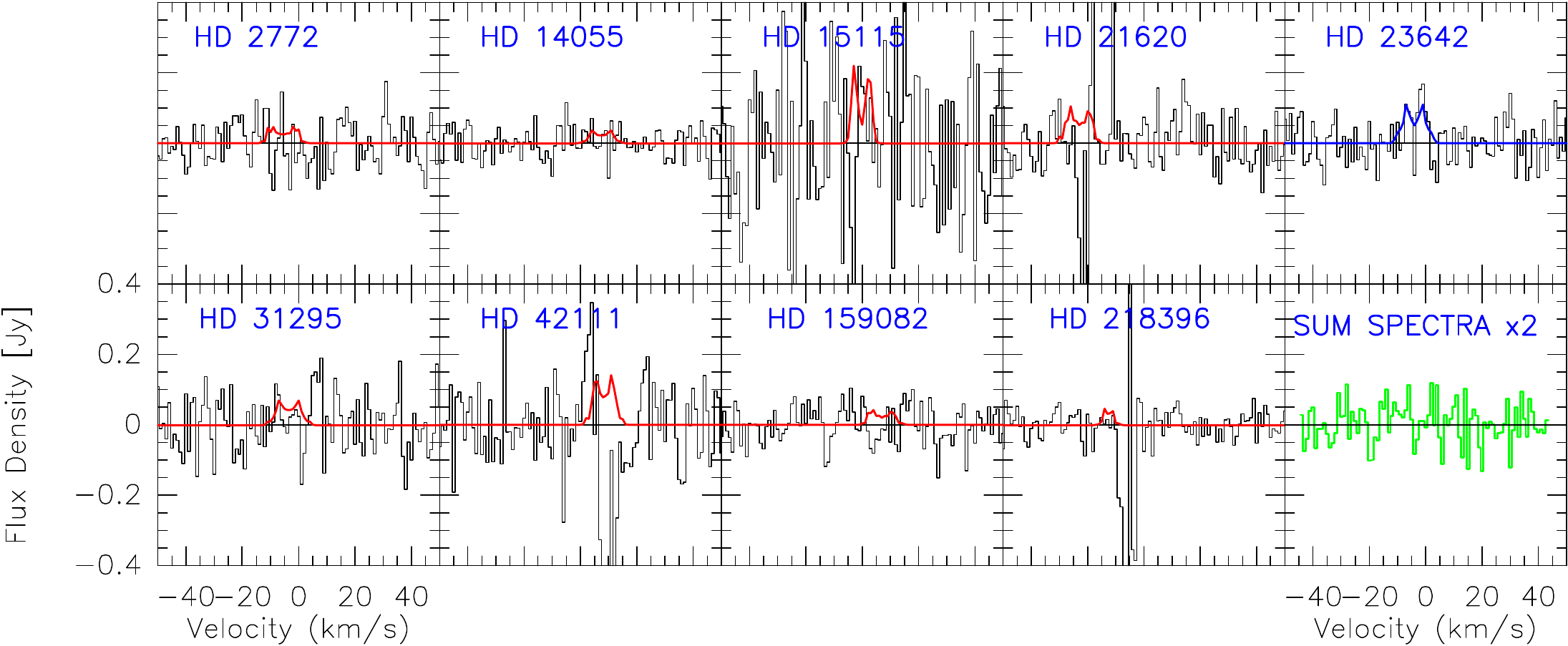}
    \caption{\codu{} spectra obtained with the \textit{IRAM-30m} telescope. Fluxes are in
    Jy with respect to local standard of rest (LSR) velocity. The last box shows the cumulated spectrum made with
    spectra shifted in velocity to set the systemic velocity to zero (note that the flux
    scale is different from the individual spectra). The sources with a high rms or
    showing strong contamination (i.e., HD~15115, HD~21620, HD~42111 and HD~218396) are
    not used to produce the cumulated spectrum. The 3$\sigma$ upper limit model
    for each spectrum is over-plotted in red (see sect. \ref{subsec:co_mass_survey})
    except for the HD~23642 spectrum, where the best-fit model for the tentative
    detection is plotted in blue.}
              \label{fig:spectres_iram}%
    \end{figure*}

% %%%%%%%%%%%%%%%%%%%%%%%%%%%%%%%%%%%%%%%%%%%%%%%%%%%%%%%%%%%%%%%%%%%%%%%%%%%%%%%%%%%%%%%%%%%%%%%%%%%%%%%%%%%%%%%%%%%%%%%%%%

% %%%%%%%%%%%%%%%%%%%%%%%%%%%%%%%%%%%%%%%%%%%%%%%%%%%%%%%%%%%%%%%%%%%%%%%%%%%%%%%%%%%%%%%%%%%%%%%%%%%%%%%%%%%%%%%%%%%%%%%%%%

   \begin{figure*}
    \centering
    \includegraphics[width=0.95\textwidth,keepaspectratio]{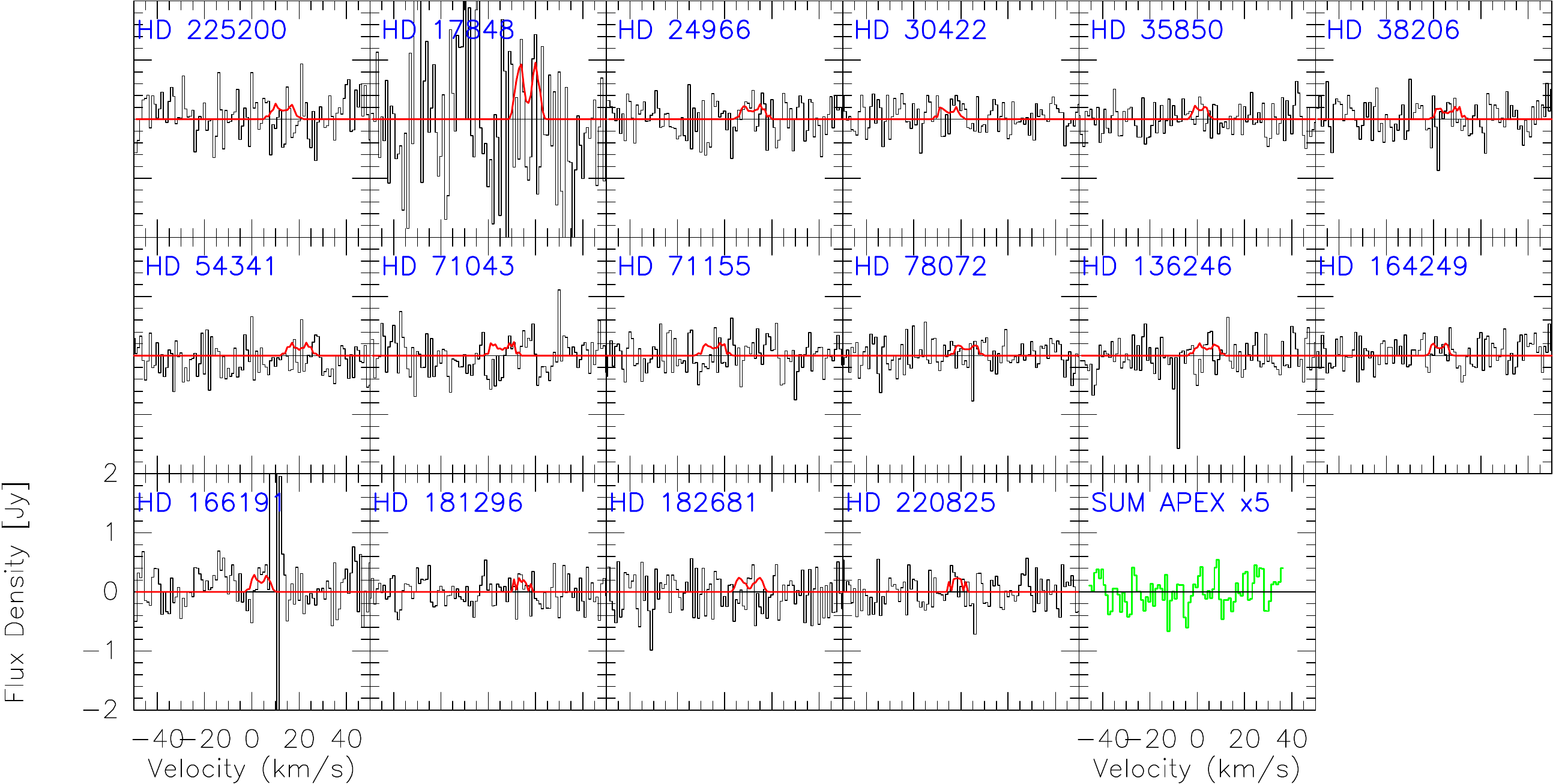}
    \caption{\cotd{} spectra obtained with the \textit{APEX} telescope. Fluxes are in Jy
    with respect to LSR velocity. The last box shows the cumulated spectrum made with
    spectra shifted in velocity to set the systemic velocity to zero (multiplied by  factor
    five). The sources with a high rms or showing strong contamination (i.e., HD~17848 and
    HD~166191) are not used to produce the cumulated spectrum. The 3$\sigma$ upper limit
    model for each spectrum is over-plotted in red (see sect.
    \ref{subsec:co_mass_survey}).}
              \label{fig:spectres_apex}%
    \end{figure*}

% %%%%%%%%%%%%%%%%%%%%%%%%%%%%%%%%%%%%%%%%%%%%%%%%%%%%%%%%%%%%%%%%%%%%%%%%%%%%%%%%%%%%%%%%%%%%%%%%%%%%%%%%%%%%%%%%%%%%%%%%%%
% %%%%%%%%%%%%%%%%%%%%%%%%%%%%%%%%%%%%%%%%%%%%%%%%%%%%%%%%%%%%%%%%%%%%%%%%%%%%%%%%%%%%%%%%%%%%%%%%%%%%%%%%%%%%%%%%%%%%%%%%%%
% %________________________________________________________________
% %________________________________________________________________
% %________________________________________________________________
\subsection{Outcome of the \textit{APEX} and \textit{IRAM} survey}
Our \textit{IRAM} and \textit{APEX} surveys were designed to provide a homogeneous sensitivity for the
whole sample, and to reach a high-enough sensitivity on the integrated intensity to
detect dim systems such as 49~Ceti at a $>5\sigma$ level. %the 8$\sigma$ level.
This corresponds, for a channel width of 1~km.s$^{-1}$, to a 6~mK rms value for T$_A^{\ast}$
with \textit{APEX} and 6.5~mK with the \textit{IRAM-30m} telescope.
%We did not firmly detect any CO emission in the present sample.
%\textbf{Nevertheless, for HD~23642, we observe what looks like a double-peaked line centered at the star velocity. Based on the integrated intensity, 0.78~$\pm$~0.24~Jy.km.s$^{-1}$, the detection level is 3.2$\sigma$. Its spectrum is presented in Fig. \ref{fig:hd23642}.}
Most targets present no CO line detection, and we report the 3$\sigma_{int}$ upper limits
on the integrated intensity, with the standard formula $\sigma_{\rm
int}=\sigma.\sqrt{\Delta v\delta v}$,  $\sigma$ being the intensity rms in a single
channel. Although typical line widths of approximately 5\,km.s$^{-1}$ are observed around A-stars
with detected CO lines, we assume here $\delta v=10$~km.s$^{-1}$ for the sake of
comparison with previous studies. Contamination from the cloud environment is seen for five
sources: HD~218396, HD~42111, HD~21620, HD~166191, and HD~136246.
%for which the rms value is calculated by masking the contaminated channels.
For each CO transition, we also stacked the spectra of all observed targets,
after correcting for the known system velocities, in order to improve the global
signal-to-noise ratio, but this did not yield any tentative detection. The
stacked spectra are displayed in green in the last insets of Fig.
\ref{fig:spectres_iram} and \ref{fig:spectres_apex}. In Table \ref{table:log}, we present
the flux density upper limits for seven new targets in the \codu{} transition, and two
stars that had already been targeted in the \cotd{} transition (namely, HD~15115 and
HD~218396). The median value is 0.8\,Jy km.s$^{-1}$ (3\,$\sigma_{\rm int}$). With
\textit{APEX}, we report upper limits for ten new targets and improve the detection limits
by a factor two at least for the other disks (median value $3\sigma_{\rm int} = 2.6$\,Jy
km.s$^{-1}$), compared to earlier studies \citet{Kastner+etal_2010},
\citet{Moor+etal_2011}, \citet{Hales+etal_2014}, except for HD~17848
where \citet{Moor+etal_2015} obtained more sensitive data.

\subsection{A possible new hybrid disk candidate in the Pleiades}
We report a tentative detection of \codu{} for the double-lined eclipsing binary
\object{HD23642} \citep[A0Vp(Si)+Am, $P= 2.46$\,d,][]{Abt+etal_1978,Torres+etal_2003,Munari+etal_2004}.
The system is classified as a debris disk with a weak infrared excess detected by
\textit{Spitzer} \citep{Su+etal_2006, Morales+etal_2009, Chen+etal_2014}. The peak
emission is found around 24\,$\mu$m, although there is only an upper limit on the flux at
70\,$\mu$m. Here we adopt the fractional excess luminosity reported by
\citep{Morales+etal_2009}, L$_{IR}$/L$_{\star}$~=~1.7$\times$10$^{-5}$, which relies on
the shape of the IRS spectrum in the 20-33\,$\mu$m range.
%This value is consistent with the luminosity  of the warm dust belt derived by \citet{Chen+etal_2014}.
The binary is composed of two A-type stars, located in the Pleiades group. Its distance
has been revised to $138\pm1.5$\,pc by \citet{Groenewegen+etal_2007}, in agreement with
the recent works on the distance of the Pleiades.

\begin{figure}
    \centering
    \includegraphics[width=0.45\textwidth,keepaspectratio]{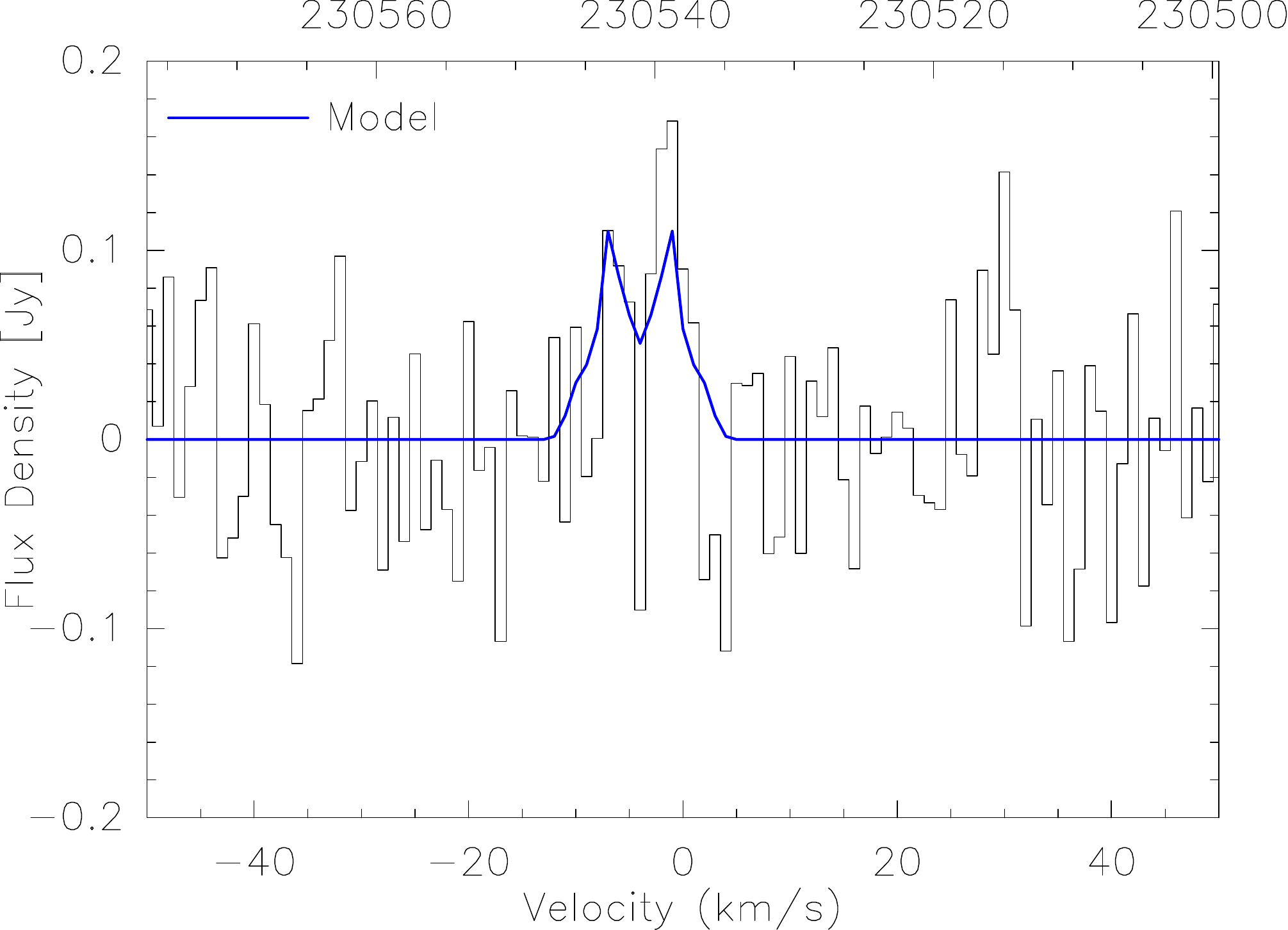}
    \caption{\codu{} spectrum of HD~23642 obtained with the \textit{IRAM-30m} telescope. Flux is in Jy with respect to LSR velocity. A line is detected at the 4$\sigma$ level, at the star velocity (-3.6~km.s$^{-1}$). The model of the line is displayed in blue.}
              \label{fig:hd23642}%
\end{figure}

The spectrum (see Fig.\ref{fig:hd23642}) presents a double-peaked line, with an intensity
integrated over the channels -8 to +2\,km.s$^{-1}$ estimated to 710 $\pm$ 180\,mJy\,km\,s$^{-1}$.
The line center is at $\sim-4$~km.s$^{-1}$, consistent with the
binary radial velocity of $V_\mathrm{LSR} = -3.6\pm0.1$\,km.s$^{-1}$ determined
by \citet{Groenewegen+etal_2007}. This signal is unlikely to be due
to contamination by a molecular cloud emission: although CO emission is present towards
the Pleiades region, it appears at LSR velocities of $\sim$7 and $10$~km.s$^{-1}$
\citep{Breger+1987,White+etal_2003}. 

%Its spectrum is presented in Fig xx.
If we interpret this signal as emission from a Keplerian disk, given the total
stellar mass derived from spectroscopic data (2.23\,M$_{\odot}$ and 1.57\,M$_{\odot}$), and the
inclination set to 78$^{\circ}$ \citep[e.g.,][]{Groenewegen+etal_2007}, the separation of
the two peaks in the CO line suggests a disk outer radius much larger than 200\,au.
Using the DiskFit tool (see Sect.\ref{subsec:co_mass_survey}), the best-fit model yields $R_{out} =450 \pm 210$\,au, and a CO gas column
density at 100\,au of $\Sigma_0 = 6.0 \times10^{14}\pm 1.3\times10^{14}$~cm$^{-2}$
($4.6\sigma$ detection). The best fit model is overplotted in Fig. \ref{fig:hd23642}.

The derived mass of gas in this system would be similar to that of the 20-40\,Myr old
systems HD~21997 and 49~Ceti, for which the gas origin is debated. Given the estimated
age of the Pleiades cluster  \citep[$\sim$125-130\,Myr, ][]{Stauffer+etal_1998,
Castellani+etal_2002}, the presence of such a large amount of remnant gas is puzzling and
makes it the oldest hybrid disk candidate. Further observations are necessary to
definitely confirm our finding, however the double-peaked line shape centered at the
expected stellar velocity suggests a robust detection.
%Indeed, if the distance of 139~pc is correct \citep{Southworth+etal_2005}, the mass of CO in the system is similar to younger hybrid disks,
%such as HD~21997 and 49~ceti where the gas origin is highly debated (primordial or produced by collisions).
%Confirmation of the detection is thus necessary before going on further considerations.
%%%%%%%%%%%%%%%%%%%%%%%%%%%%%%%%%%%%%%%%%%%%%%%%%%%%%%%%%%%%%%%%%%%%%%%%%%%%%%%%%%%%%%%%%%%%%%%%%%%%%%%%%%%%%%%%%%%%%%%%%%

% %%%%%%%%%%%%%%%%%%%%%%%%%%%%%%%%%%%%%%%%%%%%%%%%%%%%%%%%%%%%%%%%%%%%%%%%%%%%%%%%%%%%%%%%%%%%%%%%%%%%%%%%%%%%%%%%%%%%%%%%%%

\subsection{CO mass upper limits}
\label{subsec:co_mass_survey}
% %________________________________________________________________
% %________________________________________________________________
% %________________________________________________________________

The upper limits on CO integrated emission can be translated into limits on the remnant
mass of H$_2$ gas. We derive the gas mass using two methods. The first is based on
the rms obtained from the data. The second properly takes into
account the velocity gradient resulting from Keplerian shear by using a disk model.
Conversion to H$_2$ content is done using the standard value of 10$^{-4}$ for the
CO to H$_2$ abundance ratio.

\paragraph{Gas mass inferred from the rms of the flux density:}

Here we follow the approach used by many authors and derive the CO content
assuming $T_{ex}=40$\,K for the excitation temperature and  $\tau=1$ for the opacity (see
Appendix \ref{annex:mass_scoville}). These values allow a direct comparison with the
literature (although a temperature $T_{ex}= 20$~K is sometimes assumed). Results are
shown in Table \ref{table:log}.
Some sources have been observed in \codu{} and 3$\rightarrow$2. The flux ratio and H$_2$
masses derived from the two transitions are shown in Table
\ref{table:flux_mass_detections}. The discrepancy in derived masses illustrates the limits
of our assumptions. We note also that \citet{Kospal+etal_2013} have derived excitation
temperatures lower than 9~K for the $^{12}$CO in the HD~21997 disk.

Fig. \ref{fig:mass_luminosity} displays the total gas mass (H$_2$) consistently derived
using this method as a function of the dust fractional luminosity, for sources from past
surveys and from this work (larger symbols). The median value for the limits on the gas
mass (M$_{H_2}$) in our sample is  0.1\,M$_\oplus$ (3$\sigma$). The tightest constraint
is obtained for HD~14055 in \codu{} and HD~35850 in \cotd{} transition with a
limit as low as $\sim$0.03\,M$_\oplus$, that is, 2\,M$_{\rm Moon}$ of H$_2$, a value equivalent to
(even slightly less than) the amount of gas recently detected with \textit{ALMA} in the young
$\beta$~Pic system by \citet{Dent+etal_2014}.
% {\bf Warning: check in Fig5: Beta Pic mass conversion.}
% {\bf Warning 2: HD42111: large value of Mass limit due to bad influence of contaminated channels ? can we solve it?}
%Most of the reported disks have less than 1~M$\oplus$ of CO, and to the lowest upper limit around HD~35850 corresponds to the mass of $\sim$10$^8$ comets (or the mass of Pluto).

%Another way to determine
\paragraph{Gas mass inferred from a Keplerian disk model:}
The gas mass can also be estimated  by constraining the surface density in a disk
modeling approach, as done in \citet{Dutrey+etal_2011} and \citet{Chapillon+etal_2012}.
For this purpose, we used the DiskFit code \citep{Pietu+etal_2007}, dedicated to the
simulation of Keplerian disks. We describe the radial physical properties of disks as
power-laws, in a similar analysis to resolved disks. The disks are assumed to be in Keplerian
rotation around their central stars, the mass of stars being estimated from their stellar
types.\footnote{\url{http://www.uni.edu/morgans/astro/course/Notes/section2/spectralmasses.html}}
%We have modeled a disk with all the parameters fixed except the surface density, which is determined by minimisation with the uv tables created with the single-dish spectra.
We assume for the disk model an outer radius $R_{out}= 200$~au, a typical value for the
resolved hybrid disks, and an inner radius $R_{in}= 5$~au. For six disks, the inclination
is known from resolved scattered light observations: 80$^{\circ}$ for HD~182681,
73$^{\circ}$ for HD~17848 \citep{Moor+etal_2015}, 56.7$^{\circ}$ for HD~71155
\citep{Booth+etal_2013}, 86$^{\circ}$ for HD~15115 \citep{Mazoyer+etal_2014},
26$^{\circ}$ for HD~218396 \citep{Matthews+etal_2014}. For HD~181296, we have set the
inclination to the 20$^{\circ}$ upper limit determined by \citet{Smith+etal_2009}. The
most probable value of 60$^{\circ}$ is assumed for the other disks. The temperature law
is set to 30~K at 100~au with the exponent of the radial variation $q= 0.4$. With an assumed exponent of $p=1.5$ for the surface density \citep[see][for case
studies]{Pietu+etal_2007}, the surface density at 100~au $\Sigma_0$ is the only free
parameter.
%For contaminated spectra, we considere two different cases: 1 -
% In the fit procedure, we remove the contaminated channels when the contamination line is
% farther than 3\,km/s from the systemic velocity (i.e., for HD~136246, HD~166191 and
% HD~218396). In the other cases, our fit includes the contribution from the cloud.

%  and
% the true upper limit on the gas mass might be much smaller than the reported value
% (in particular for HD42111 whose mass limit is not taken into account in the following).

The model spectra corresponding to the $3 \sigma$ upper limits on $\Sigma_0$ are
superimposed in red to the spectra in Fig.\ref{fig:spectres_iram} and \ref{fig:spectres_apex}.
In addition, the best fit model for the tentative detection in HD~23643 is displayed in
blue.
%Since the surface density is the only free parameter in the model,
% When the noise within the line width around the systemic velocity is dominated by negative channels,
% the model may derive a negative surface density, eventually resulting in an artificial absorption line.
%With only non-detections, the value of the surface density doens't make sense.
%Only its error bar, determined from the rms level, can give an upper limit on the surface density.
% In the case of non detections,
The corresponding 3 $\sigma$ upper limits on the total disk masses are reported
in Table \ref{table:log} (see Appendix~\ref{annex:mass_diskfit} for details).
%The calculations to retrieve the mass of $H_2$ are presented in  and the resulting upper
The derived gas masses are a factor of two or three lower than from the simple uniform $T_\mathrm{ex}$
approach (see Table \ref{table:log}), as a result of different assumptions. As it accounts
for density and temperature gradients, the method using DiskFit should be considered
as more reliable.

% %________________________________________________________________
% %________________________________________________________________
% %________________________________________________________________

\subsection{Detection statistics}
\label{subsec:statistics}
%\label{subsec:survey_other_studies}

\begin{figure}
   \centering
   \includegraphics[width=0.45\textwidth,keepaspectratio]{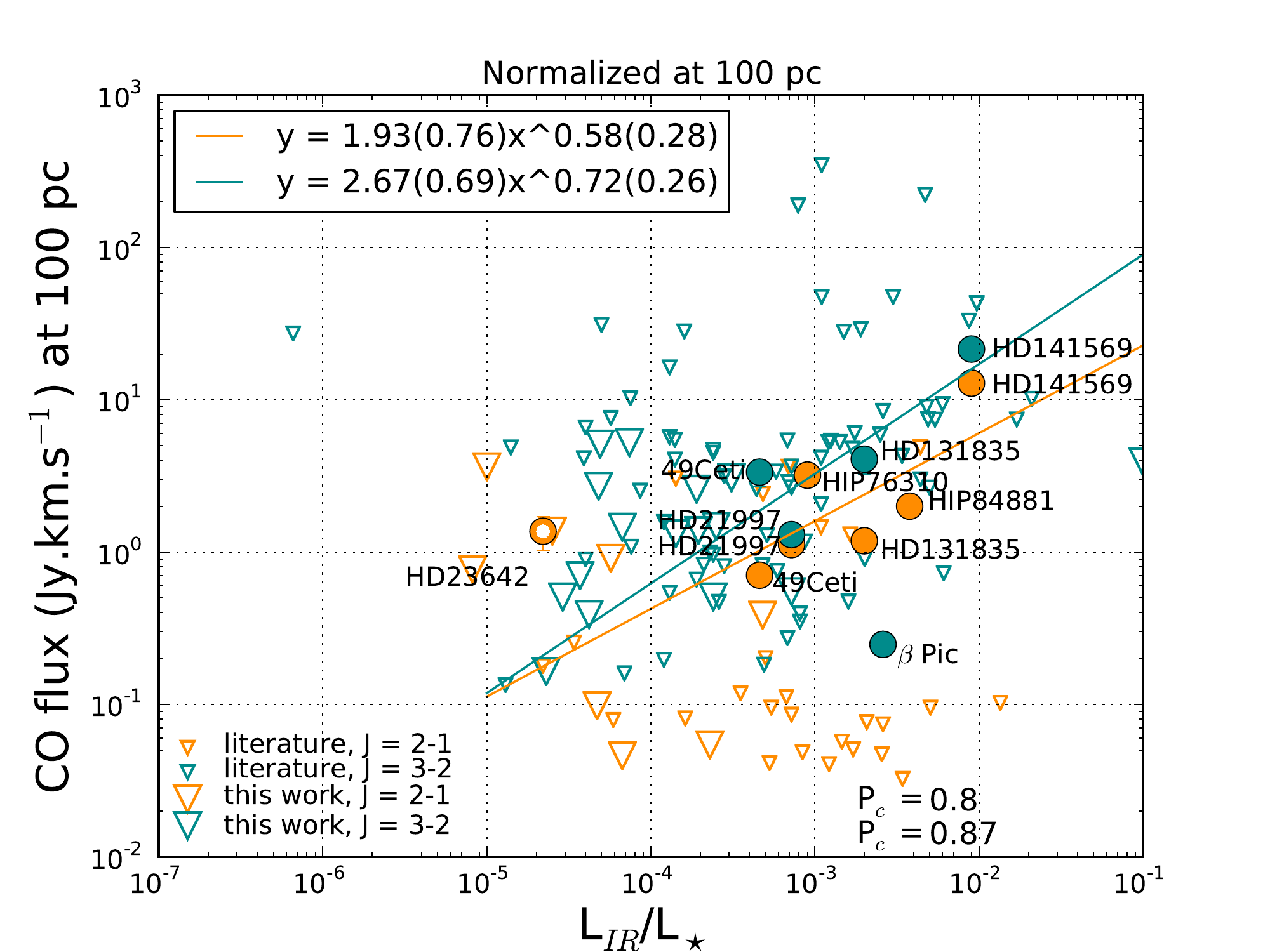}
   \caption{Integrated flux of CO normalized at 100~pc compared to the infrared fractional luminosity of the debris disks. The same disks and symbol/color code are used as in Fig. \ref{fig:flux_luminosity}.}
              \label{fig:flux_luminosity_100}%
\end{figure}

Our survey adds 17 new systems to the sample of young debris disks which have been
searched for molecular gas. Taking into account these new constraints in addition to the
surveys of \citet{Kastner+etal_2010}, \citet{Moor+etal_2011}, \citet{Hales+etal_2014},
the recent survey of \citet{Moor+etal_2015} with the discovery of HD131835, and the \textit{ALMA}
detections of gas around HIP~84881 and HIP~76310 in the survey of \citet{Lieman-Sifry+etal_2016}, we can derive an approximate estimate of
the occurrence of CO gas in young debris disks: $7.1 \pm 2.6$\,\% (1$\sigma$ uncertainty, 7 detections among 98 observed systems).
This value can be considered as a lower limit since only less than half of the targets have been observed with enough sensitivity to detect a CO luminosity as large as that of HD~21997/49~Ceti (see Fig. \ref{fig:flux_luminosity_100}). The occurrence of a CO gas disk as bright as HD~21997/49~Ceti (after scaling at 100\,pc) is 12.2~$\pm$~4.7\% (1$\sigma$) (6/49 detections).

\begin{figure}
   \centering
   \includegraphics[width=0.45\textwidth,keepaspectratio]{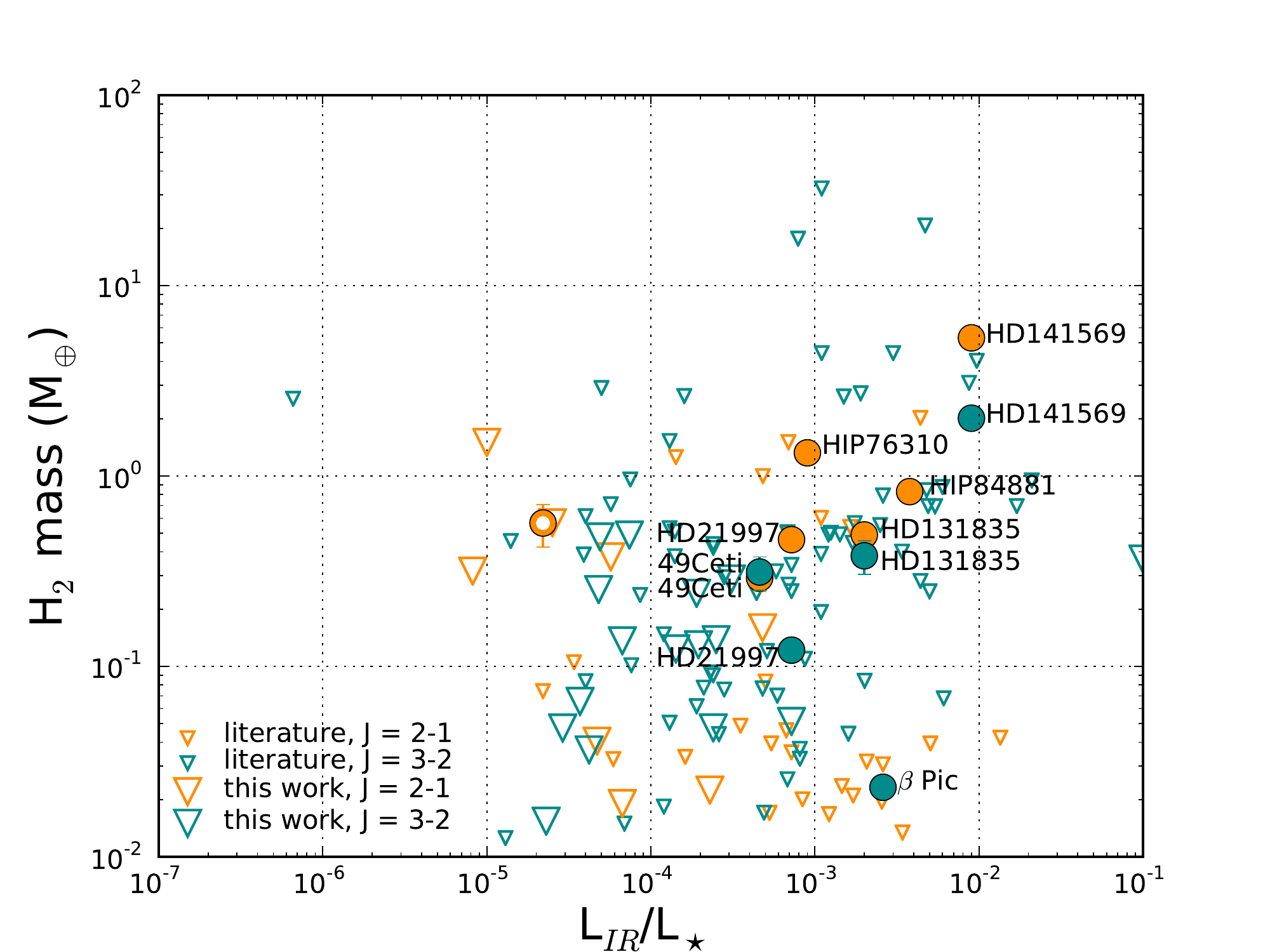}
   \caption{Mass of H$_2$ (from \cite{Scoville+etal_1986}) compared to the infrared fractional luminosity of the debris disks. The same disks and symbol/color code are used as in Fig. \ref{fig:flux_luminosity}.}
              \label{fig:mass_luminosity}%
\end{figure}

This value suggests that such systems are very rare, likely because
they represent a very short transient phase in the evolution of circumstellar disks. It
should also be recognised that this detection frequency does not rely on a homogeneous
sample: the observed targets are located in different regions of the sky, may span a
broad range of ages, and the various surveys have reached different sensitivity limits.
The occurrence of hybrid disks (with secondary dust and possible primordial CO gas) is
even smaller, since the origin of the detected CO gas is still debated for most gas-rich
debris disks (HD~141569, HD~21997, 49~Ceti and HD~131835). Several authors favor a
secondary origin \citep{Moor+etal_2011, Zuckerman+etal_2012} for some of these systems,
as is the case for the famous $\beta$\,Pictoris disk, where the CO gas detected by \textit{ALMA} may have been
produced by a massive collision of planetesimals \citep{Dent+etal_2014}. Besides
$\beta$ Pic, the detected systems with the weaker CO emission (or gas mass) are HD~21997
and 49~Ceti (see Fig.~\ref{fig:flux_luminosity_100} and \ref{fig:mass_luminosity}).
These two are therefore the most probable candidates for a secondary origin of the gas,
although the gas mass remains one order of magnitude larger than in $\beta$\,Pic.

Among the stars in our study, we note that 17 disks (45 when including the surveys from
the literature), that is, 68\% of our sample (resp. 48\%), are constrained to CO gas masses
smaller than that of 49~Ceti, and one of them, HD~35850, shows even less CO gas than
$\beta$~Pictoris.

% %________________________________________________________________
% %________________________________________________________________
% %________________________________________________________________

% %%%%%%%%%%%%%%%%%%%%%%%%%%%%%%%%%%%%%%%%%%%%%%%%%%%%%%%%%%%%%%%%%%%%%%%%%%%%%%%%%%%%%%%%%%%%%%%%%%%%%%%%%%%%%%%%%%%%%%%%%%
% %%%%%%%%%%%%%%%%%%%%%%%%%%%%%%%%%%%%%%%%%%%%%%%%%%%%%%%%%%%%%%%%%%%%%%%%%%%%%%%%%%%%%%%%%%%%%%%%%%%%%%%%%%%%%%%%%%%%%%%%%%

% %%%%%%%%%%%%%%%%%%%%%%%%%%%%%%%%%%%%%%%%%%%%%%%%%%%%%%%%%%%%%%%%%%%%%%%%%%%%%%%%%%%%%%%%%%%%%%%%%%%%%%%%%%%%%%%%%%%%%%%%%%
% %%%%%%%%%%%%%%%%%%%%%%%%%%%%%%%%%%%%%%%%%%%%%%%%%%%%%%%%%%%%%%%%%%%%%%%%%%%%%%%%%%%%%%%%%%%%%%%%%%%%%%%%%%%%%%%%%%%%%%%%%%

\subsection{CO to dust flux ratio correlations}

To better understand the peculiarity of hybrid disks, we compare their properties with
those of other (proto-planetary) disks. For this purpose, we have chosen to study the
ratio of the emission of CO over the emission of the dust, $\frac{S_{\rm CO}}{F_{\rm
cont}}$. In the literature, we find many studies considering the ratio of masses, but the
mass is always a calculated value, depending on models and on assumptions about
excitation conditions and  dust-to-mass ratio, etc. On the contrary, the $\frac{S_{\rm
CO}}{F_{\rm cont}}$, which is an equivalent width, is a distance-independent product of
observations that is not affected by hidden assumptions or biased by interpretation
models. It may be impacted by the diversity of optical depths in the disks, but
we show in the following that this ratio presents some characteristic trends that
can be explained easily when considering the possible opacity conditions.
%is that the we study a ratio of two measurements, i.e. no assumptions.

\begin{figure*}
   \centering
   \begin{subfigure}{0.49\textwidth}
   \centering
   \includegraphics[width=0.99\textwidth,keepaspectratio]{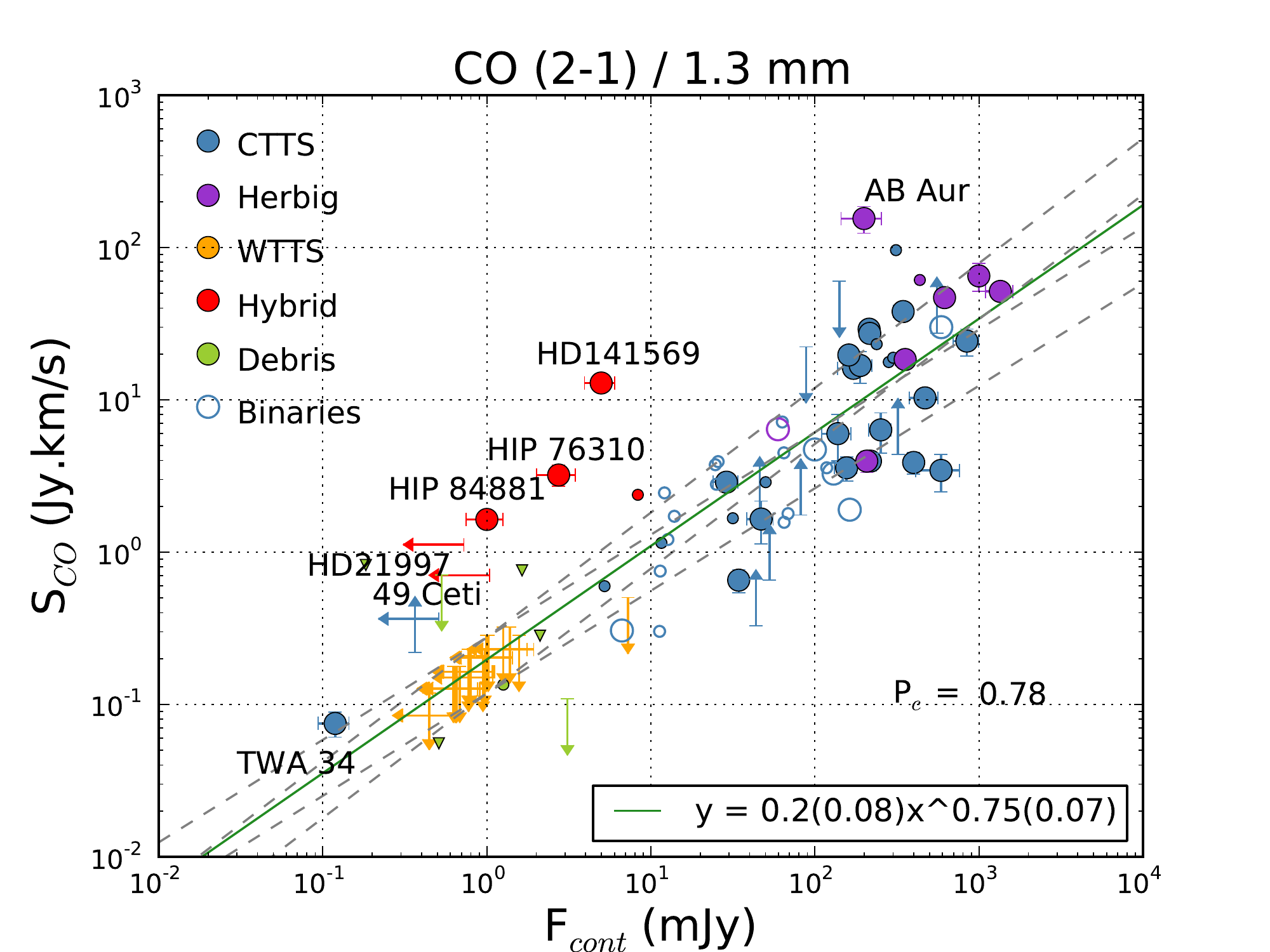}
  \caption{}
              \label{fig:co_vs_dust_21}%
  \end{subfigure}
  \begin{subfigure}{0.49\textwidth}
   \centering
   \includegraphics[width=0.99\textwidth,keepaspectratio]{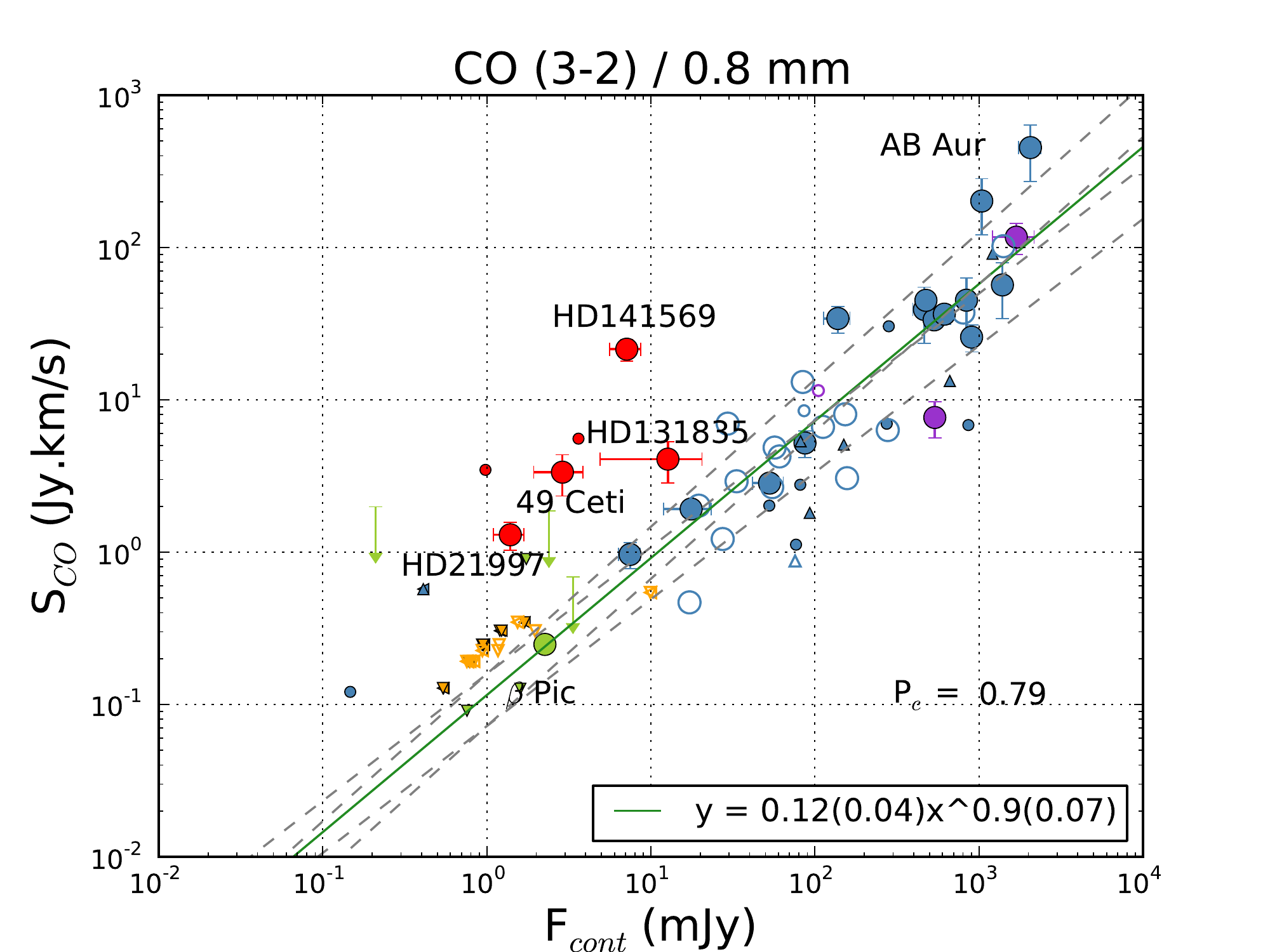}
  \caption{}
               \label{fig:co_vs_dust_32}%
  \end{subfigure}
     \caption{(a) \codu{} emission plotted with the corresponding continuum measurement,
     the fluxes being normalized at 100~pc. Blue corresponds to CTTS, violet to Herbig,
     orange to WTTS, red to hybrid and green to debris disks. The open circles mark the
     binary systems. The Pearson coefficient of correlation of the data, P$_c$, is
     indicated in the lower right-hand corner.\\ (b) As for Fig.~\ref{fig:co_vs_dust_21} but
     for the \cotd{}~/~0.8~mm emission.}
\end{figure*}

The diagrams on Fig. \ref{fig:co_vs_dust_21} and \ref{fig:co_vs_dust_32} summarize the results
displayed in Table
\ref{table:1} by plotting the CO integrated emission as a function of the dust emission
at the corresponding wavelength, and scaled at 100~pc. As some data have no
given error bars, or very small ones, we have added a 10 \% error bar
on the whole sample in order to homogenize the sample and better account for calibration uncertainties.
Different colors are used to represent the five categories of disks: blue indicates the CTTS, violet the
Herbig~AeBe disks, orange the WTTS, red the hybrid disks and green the debris disks.
Only 13 sources are common to both diagrams. As for the transitional disks, we do not
distinguish them from the CTTS category as their position on the
diagrams does not depart from the CTTS group.
For the \codu{}~/~1.3\,mm emission, we have considered 33 sources where both gas and
dust are detected. Among them, six sources are displayed with a lower limit arrow, meaning
that the CO emission might be contaminated by the environment. 6 sources have only an
upper limit on the gas emission, 3 on the dust emission (one having a lower limit on the
gas), and 20 sources have upper limits both on the gas and the dust emission (mostly WTTS
from \citet{Hardy+etal_2015}). For the \cotd{}~/~0.8\,mm emission, 42 sources have
the continuum and gas emission detected, 3 have only upper limits on the gas emission and
1 has upper limits on the gas and dust emissions.
\\

% %________________________________________________________________
% %________________________________________________________________
% %________________________________________________________________

\label{sec:ratio}

There is a clear correlation between the emission of the gas and that of dust at mm wavelengths,
with naturally decreasing flux densities for more evolved systems, such as in WTTS or debris disks.
The correlation is seen both at 1.3\,mm and 0.8\,mm.

The sample spans at least three orders of magnitude in flux density and we observe an
almost linear correlation between dust and gas emissions on Fig.~\ref{fig:co_vs_dust_21}
and \ref{fig:co_vs_dust_32}, with no clear evidence for a slope break (although there are
many more sources in the upper right-hand corner of the diagram than in the faint dust/CO emission
region where the optically thin regime is expected to be found). There is no obvious difference between
CTTS, HAeBe, single, and wide binary stars. A few sources lie well above the
global trend: most of them are suspected hybrid disks (HIP~84881, 49~Ceti, HD~21997
and the most prominent
outlier HD~141569) with the exception of the Herbig~Ae (envelope
embedded) star AB~Aur.
% The trend suggests that the CO flux density is about ten times smaller than the dust emission in the 0.8-1.3\,mm domain.
The dispersion around the best-fit line is less than one order of magnitude.
The absence of a break in the correlation suggests that the majority of the detected gas
disks might still be in the optically thick regime, while optically thin gaseous disks
(expected at the last stages of the proto-planetary disk dispersal or in young debris
disks) remain mostly undetected so far.

Finally, we notice on Fig.\ref{fig:flux_luminosity_100} that there could also be a
correlation between the CO integrated flux (scaled at 100\,pc) and the dust fractional
luminosity for hybrid disks. However, this trend is based on very few detected gas-rich
debris disks so far, and their fractional luminosity spans over approximately one decade only. If
it were confirmed, this trend combined with the sensitivity limits of current
surveys (at the level of 49\,Ceti emission) could partly explain the very low occurrence
of CO detections in disks with weak IR excess ($L_{IR}/ L_{\star} < 5.10^{-4}$). We note
that the new candidate HD~23642 does not fit well on that possible trend, although its IR
excess is still poorly constrained at far-IR wavelengths. % and could be somewhat larger.
The binary nature of the source (with a larger stellar mass and a suggested larger disk
than other hybrid disks) may also be responsible for the discrepancies.
% \textbf{J'ai rajoute que pas de bar d'erreur sur LIR ds la table 1.}

\subsection{Search for correlations between the CO/dust flux ratio and the stellar parameters}
\label{subsec:correlation_star}

\begin{figure}
   \centering
   \includegraphics[width=0.48\textwidth,keepaspectratio]{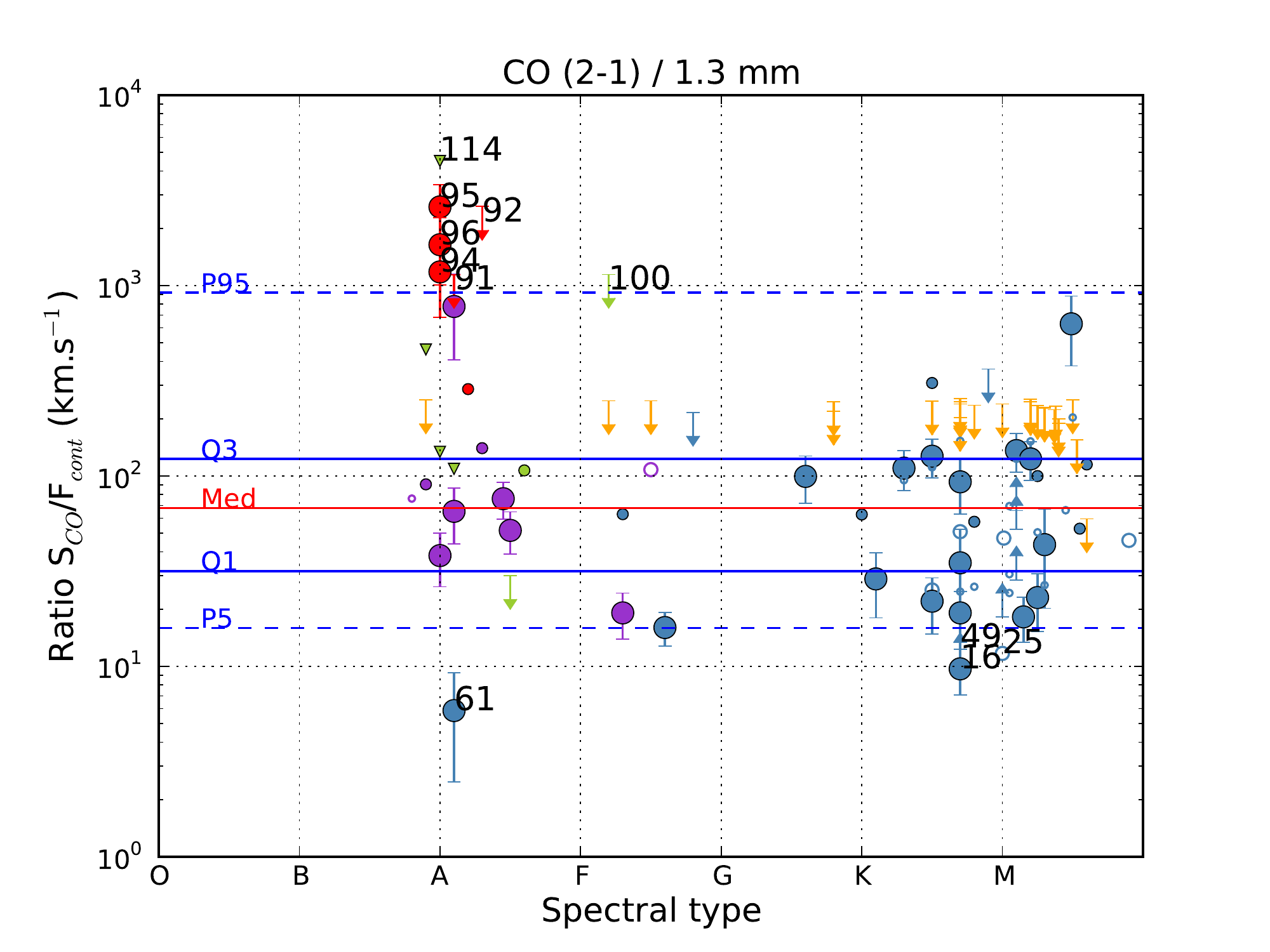}
   \caption{Ratio S$_{\rm CO}$/F$_{\rm cont}$ plotted against the spectral type of the
   stars, at 1.3 mm. The color code is the same as in Fig. \ref{fig:co_vs_dust_21} and
   \ref{fig:co_vs_dust_32}: blue for CTTS, violet for Herbig, orange for WTTS, red
   for hybrid and green for debris disks. Some statistics are represented by the horizontal lines:
   the red full line shows the median value of the ratio for the distribution of points,
   the blue full lines represent the first and third quartiles (50\% of the points are
   between these two lines) and the blue dashed lines show the 5th and 95th percentiles
   (90\% of the points between the lines).}
              \label{fig:ratio_spt}%
\end{figure}

\begin{figure}
   \centering
      \includegraphics[width=0.48\textwidth,keepaspectratio]{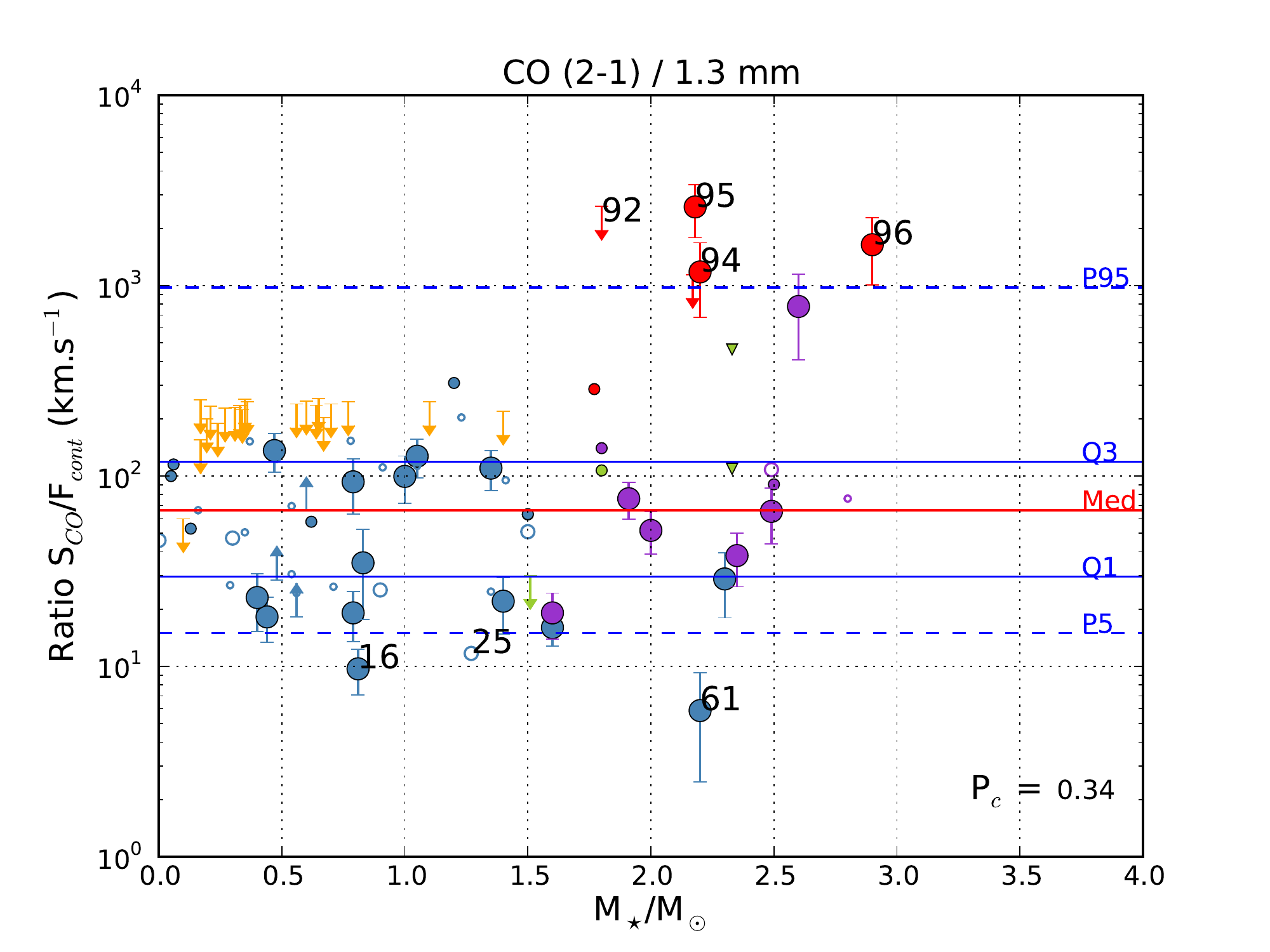}
    \caption{Ratio S$_{\rm CO}$/F$_{\rm cont}$ at 1.3\,mm plotted against the stellar mass.
    The color code and lines are the same as in Fig. \ref{fig:ratio_spt}. 
    The Pearson coefficient of correlation of the data, P$_c$, is indicated in the lower right-hand corner.}
              \label{fig:ratio_mstar}%
\end{figure}

\begin{figure}
   \centering
      \includegraphics[width=0.48\textwidth,keepaspectratio]{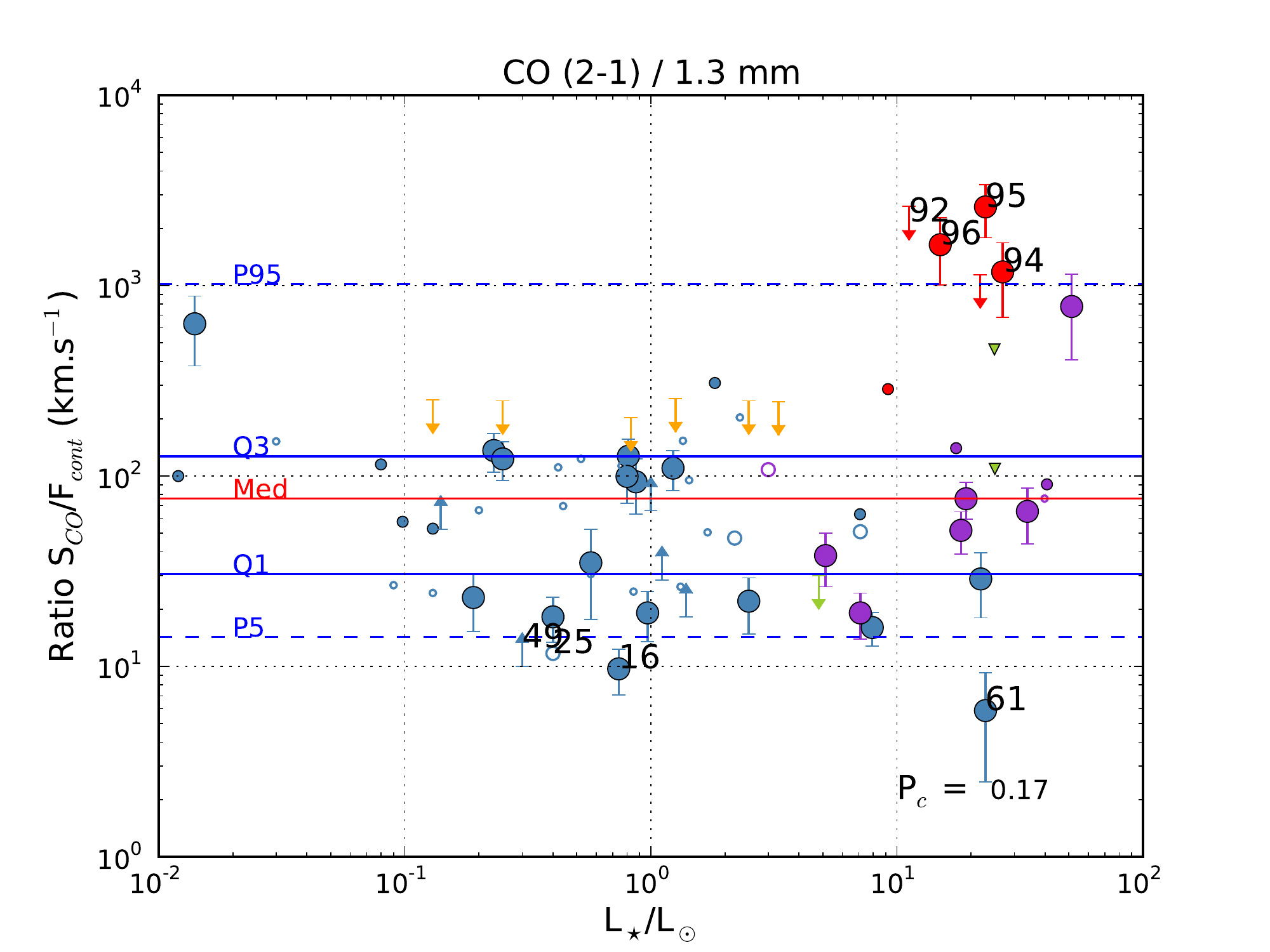}
    \caption{As Fig. \ref{fig:ratio_mstar} but as a function of stellar luminosity.}
              \label{fig:ratio_lstar}%
\end{figure}

\begin{figure}
   \centering
      \includegraphics[width=0.48\textwidth,keepaspectratio]{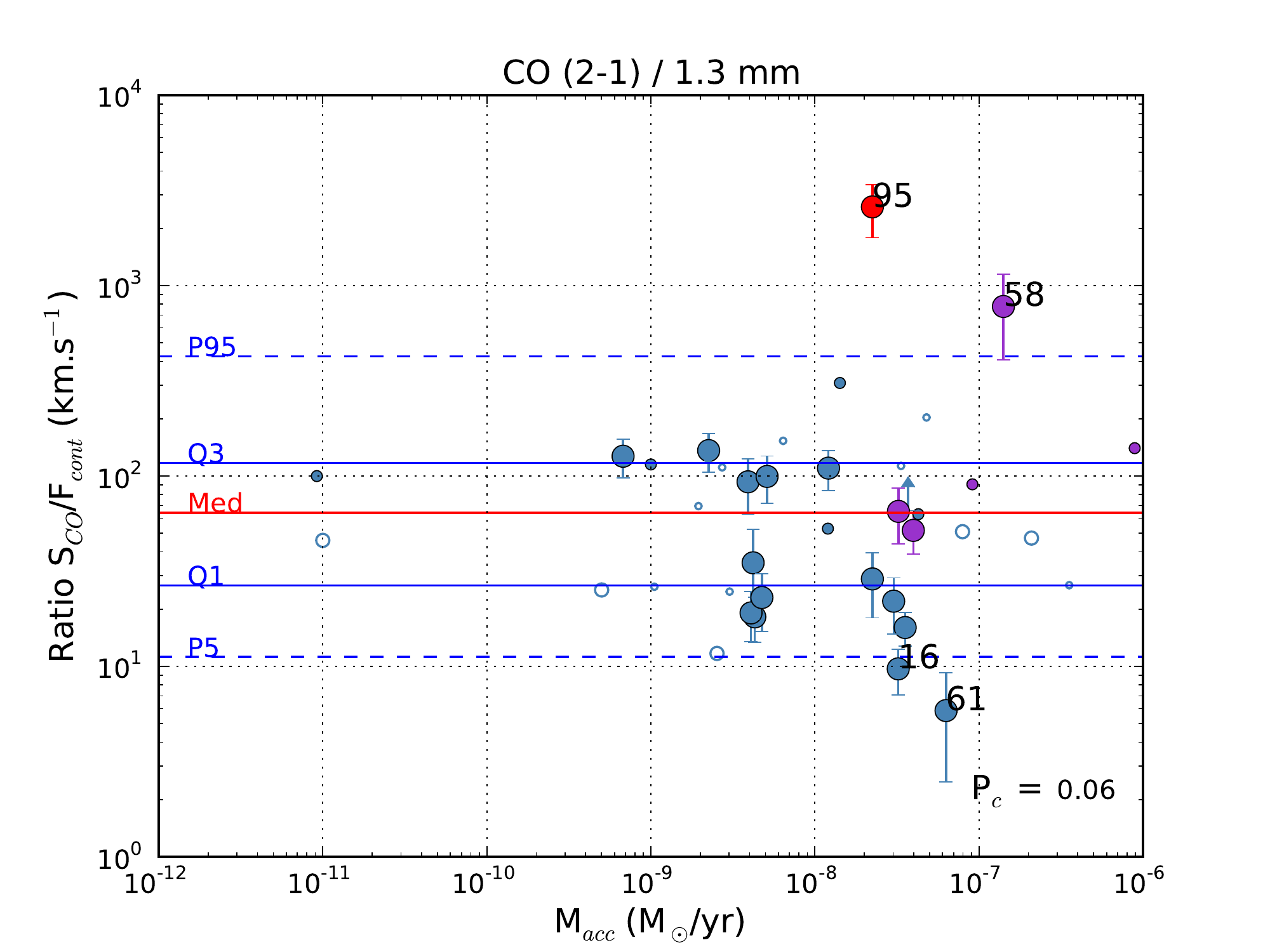}
    \caption{As Fig. \ref{fig:ratio_mstar} but as a function of accretion rate.}
              \label{fig:ratio_macc}%
\end{figure}

\begin{figure}
   \centering
      \includegraphics[width=0.48\textwidth,keepaspectratio]{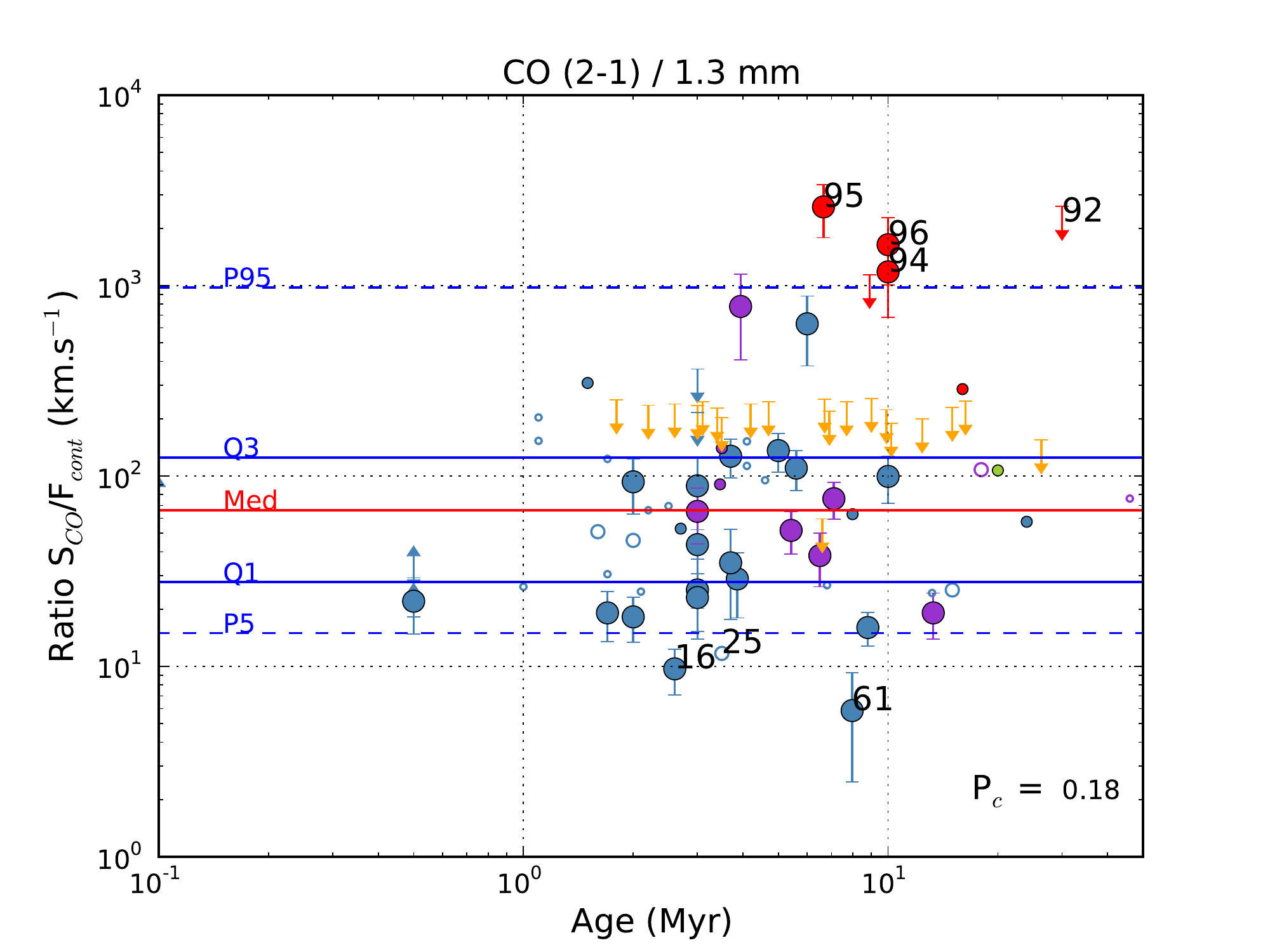}
    \caption{As Fig. \ref{fig:ratio_mstar} but as a function of system age.}
              \label{fig:ratio_age}%
\end{figure}

The equivalent width, that is, the ratio $S_{\rm CO}/F_{\rm cont}$, has been
calculated for the whole sample of 103 sources. We obtain an estimate of $S_{\rm
CO}/F_{\rm cont}$ for 73 different sources; 46 at 1.3~mm and 44 at 0.8~mm. The
ratio was not calculated when only upper limits exist for both the gas and the continuum.
For some sources, the flux was interpolated at 1.3 or 0.8\,mm based on the observed
correlations (see Appendix B). The ratio $S_{\rm CO}/F_{\rm cont}$ was then plotted
against the spectral type (Fig. \ref{fig:ratio_spt}), the stellar mass (Fig.
\ref{fig:ratio_mstar}), the stellar luminosity (Fig. \ref{fig:ratio_lstar}), the
accretion rate (Fig. \ref{fig:ratio_macc}), and the age (Fig. \ref{fig:ratio_age}) of the
stars when they are known. We have retrieved the spectral type for 98 sources, the
stellar mass for 85 sources and the stellar luminosity for 71 sources.

In the figures, the median value, the first and third quartiles, Q1 and Q3 (respectively
delimiting values below which 25\% and 75\% of the sources lie), and the 5th and 95th
percentiles, P5 and P95 (respectively delimiting values below which 5\% and 95\% of the
sources lie) are drawn for a better understanding. The name of the sources is specified
only for values out of P5 and P95.

\paragraph{Spectral type, M$_{\star}$ and L$_{\star}$:}
\label{subsub:spt}
% Muzerolle 2010
% Sicilia-Aguilar 2008
% mm dust : not affected by stellar properties because in the midplane (Andrews+etal_2005, 631)
% CO dependance on stellar prop ?
There is no correlation between the ratio $S_{\rm CO}/F_{\rm cont}$ and either
the spectral type (Fig. \ref{fig:ratio_spt}), stellar mass (Fig.\ref{fig:ratio_mstar}),
or stellar luminosity (Fig.\ref{fig:ratio_lstar}). However, A stars (heavier
than 2~M$_{\odot}$) tend to have a higher $S_{\rm CO}/F_{\rm cont}$ ratio. This trend is mainly
dominated by the hybrid disks, where the ratio is approximately 1000 km.s$^{-1}$, compared to the median
value $\sim$100 km.s$^{-1}$. Among the Herbig AeBe, only AB Aur shows a high ratio,
most likely because of the presence of its envelope (see \S \ref{subsub:high}).

\paragraph{Accretion rate:} We have found in the literature the accretion rates for 40
sources. We do not observe any correlation in the corresponding diagram (Fig. \ref{fig:ratio_macc}). We do not observe any correlations in the diagrams.  This may partly reflect the
lack of data (only 39\% of the sources have measured accretion rates), in particular
for the older population (hybrids and debris).  At mm wavelengths, the emission is mostly
dominated by the outer parts of the disks, and the sensitivity on inner parts is limited.
Since the accretion tracers reflect phenomena occuring close to the star, it is not
surprising that the global emission of the disk is not correlated to the the local inner
accretion signatures. Measurements or at least upper limits on the accretion rates on
more evolved stars might help, but it seems that accretion signatures are not a good
tracer for evaluating the global evolution of the material around disks.

\paragraph{Age:}
\label{subsub:age}
We have retrieved the age of 82 sources and present the relationship between the ratio $S_{\rm
CO}/F_{\rm cont}$ and age in Fig. \ref{fig:ratio_age}. A trend is observed for older
disks to have higher  $S_{\rm CO}/F_{\rm cont}$ ratios. Nevertheless, as already noted
in paragraph \ref{subsub:spt}, the trend is mainly driven by the hybrid disks,
and possibly affected by the difficult determination of ages of young stars.
% See Jeffries 2011

\subsection{Outliers}

Here we list the disks that seem to have an equivalent width $S_{\rm CO}/F_{\rm cont}$ departing
from the observed median value 100 km.s$^{-1}$, that is, lying below the 5$^\mathrm{th}$ quartile and
above the 95$^\mathrm{th}$ quartile in the different diagrams presented in section
\ref{subsec:correlation_star}.
With the exception of AB~Aur, all sources having a high flux ratio belong to the category
of hybrid disks.

\subsubsection{Objects with high equivalent width}
\label{subsub:high}

\paragraph{HD 141569}
The star of spectral type B9.5V/A0Ve is located 116~pc and has two M-star companions
that may be unbound \citep{Reche+etal_2009}. The age of the triple system is approximately 5
$\pm$ 3\,Myr \citep{Weinberger+etal_2000, Merin+etal_2004}. With a stellar mass of
2M$_{\odot}$, the HD~141569A star+disk system appears in an intermediate evolutionary
stage compared to the younger gas-rich system AB Aurigae \citep{Pietu+etal_2005,
Tang+etal_2012} and the more evolved $\beta$ Pictoris disk. The debris disk of HD~141569 \citep[$L_{IR}/L_{\star} = 9\times10^{-3}$]{Meeus+etal_2012}, presents a complex
multiple-ring architecture with spirals and gaps imaged in scattered light by the HST
\citep{Augereau+etal_1999, Weinberger+etal_1999, Mouillet+etal_2001,
Boccaletti+etal_2003, Clampin+etal_2003}. A large quantity of molecular and atomic gas is
still present \citep{Zuckerman+etal_1995, Brittain+etal_2003, Dent+etal_2005,
Goto+etal_2006, Brittain+etal_2007, Salyk+etal_2011, Thi+etal_2014}. The CO gas extends
out to $\sim$250~au \citep{Flaherty+etal_2016,Pericaud+etal_inprep}.

\paragraph{HD 21997}
 A gaseous disk around HD~21997 discovered by \citet{Moor+etal_2011} in a survey
of gas around young debris disks ($L_{IR}/L_{\star} = 7.2\times10^{-4}$,
\citet{Chen+etal_2014}). Located at 72~pc, the A3 star is a member of the Columba
association, which is estimated to be $\sim$~30~Myr old
\citep{Moor+etal_2006,Torres+etal_2008}. Given the age of the star and the amount of CO,
\citet{Zuckerman+etal_2012} have proposed that the gas could be produced by collisions of
comets. Nevertheless, the detailed study of the system with \textit{ALMA} resolved
observations \citep{Moor+etal_2013,Kospal+etal_2013} reveals that the gas and the dust
are not co-located, which could imply that they have a different origin: the gas might
still be primordial while the dust is of second generation.

\paragraph{49 Ceti}
The system is very similar to HD~21997: the spectral type of this 59~pc distant star is A1, and
it is associated with the 40~Myr Argus group
\citep{Zuckerman+etal_2012,Torres+etal_2008}. It is one of the first debris disks
($L_{IR}/L_{\star} = 4.6\times10^{-4}$, \citet{Chen+etal_2014}) where CO gas was
detected \citep{Zuckerman+etal_1995}. The origin of the gas is still debated, and the
comparison of the resolved \textit{ALMA} observations of the dust with the gas
\citep{Hughes+etal_2008} will help to decipher its origin, currently proposed to be either primordial or secondary.

\paragraph{HD 131835}
The gas disk around HD~131835 was discovered using the same approach as for HD~21997,
during a survey to detect gas in debris disks \citep{Moor+etal_2015}. They have detected
the J~=~3$\rightarrow$2 transition at a 5$\sigma$ level, but no significant emission is observed for
the \codu{} and CO J~=~4$\rightarrow$3. In addition to the gas detection, they have deeply
studied the disk thanks to the \textit{Herschel}, \textit{Spitzer,} and \textit{FEROS}
instruments. The dust is resolved at 70 and 100~$\mu m$, extending, on average, to 170~au,
with a fractional dust luminosity $L_{IR}/L_{\star}$ of 2.0$\times$10$^{-3}$
\citep{Chen+etal_2014}. This A2-type star (122~pc) is young (16~Myr,
\citet{Pecaut+etal_2012}), and thus represents another example of a gas-rich debris disk,
possibly of the hybrid kind: the analysis of the gas by \citet{Moor+etal_2015} does not
rule out a primordial origin.

%\paragraph{TWA~34}
%The star has been discovered thanks to its infrared excess, detected with \textit{WISE} \citep{Schneider+etal_2012}, and has been observed with \textit{Herschel} \citep{Liu+etal_2015}. In the recent survey of 15 low-mass stars with ALMA conducted by \citet{Rodriguez+etal_2015}, the 47 pc distant TW~34 star is the only one where gas has been detected, in a small CO disk (20-40~au or less) in keplerian rotation.

\paragraph{HIP~84881}
\label{sec:hip84881}
The disk of HIP~84881 has been revealed by the analysis of \textit{WISE} photometric data \citep{Rizzuto+etal_2012}. Its infrared fractional luminosity is of the same order of magnitude as HD~141569, HD~131835, and $\beta$~Pic ($f=3.78\time10^{-3}$, \citet{McDonald+etal_2012}). This has motivated the search for gas in this young debris disk with \textit{ALMA}. Both the CO and continuum are detected at 1.3~mm \citep{Lieman-Sifry+etal_2016}. The dust disk has a marginally resolved inner radius, $<$20~au and extends out to 150~$\pm$30~au.

\paragraph{HIP~76310}
In addition to HIP~84881, HIP~76310 is the other new hybrid disk discovered by \citet{Lieman-Sifry+etal_2016}. The infrared excess has been detected by \textit{Spitzer} \citep{Carpenter+etal_2006}, and is the strongest excess among A-B stars in Upper Sco (L$_{IR}$/L$_{\star}$~=~9$\times$10$^{-4}$ after \citet{Chen+etal_2014}). The millimeter emission at 1.2~mm was detected first by \citet{Mathews+etal_2012} with the IRAM 30~m, and recently by \textit{ALMA}, as well as \dcodu{} emission \citep{Lieman-Sifry+etal_2016}. The dusty disk has a resolved inner cavity of $\sim$70~au.

\paragraph{AB~Aurigae} This is a younger Herbig Ae system of the Taurus region, $\sim$4~Myr old. The
A1-type star is surrounded by a large flared disk of gas and dust. A 100~au inner
depletion of material is observed \citep{Pietu+etal_2005}, as well as spirals both in the
gas and dust warped distributions
\citep{Fukagawa+etal_2004,Hashimoto+etal_2011,Tang+etal_2012}. A large envelope remains
present around the system (r$\sim$1300~au, \citet{Grady+etal_1999}), from which material
could be accreted despite its evolved age \citep{Tang+etal_2012}. Consequently, the high
ratio of CO emission compared to the dust might be affected by this remnant envelope.

%\paragraph{GK Tau a}
%\paragraph{SZ 91}
%\paragraph{RY Tau}
%\paragraph{DR Tau}
%\paragraph{DS Tau}
%\paragraph{$\rho$ Oph 102}
%\paragraph{DM Tau}
%\paragraph{LkCa 15}
%\paragraph{GM Aur}

\subsubsection{Objects with low equivalent width}

\paragraph{HD 245185}
This A type star HD~245185 is part or the $\lambda$ Orionis cloud, located 450~pc away,
even if the membership of the Taurus region is not excluded \citep{Ansdell+etal_2015}.
Several studies have demonstrated that the star is accreting material from a primordial disk
\citep{Finkenzeller+etal_1984, Herbst+etal_1999, de_Winter+etal_2001,
Hernandez+etal_2004, Acke+etal_2005, Wade+etal_2007, Hernandez+etal_2010,
Donehew+etal_2011}. \citet{Ansdell+etal_2015} have observed the detected \codu{}
emission and the dust at 850~$\mu m$. They observe a mid-IR dip in its spectral energy
distribution (SED), which could be the sign of a gap in the disk.

\paragraph{DL Tau}

A dust disk around DL~Tau was first detected by \citet{Beckwith+etal_1990}, followed by
the observation of CO gas \citep{Koerner+etal_1995, Simon+etal_2000}, CN
\citep{Guilloteau+etal_2014}, and H$_2$CO \citep{Guilloteau+etal_2013}, despite strong
confusion. While the CO and CN extends out to $\sim$460~au, the dust disk is smaller,
with an outer radius of 150-180~au \citep{Guilloteau+etal_2011,Kwon+etal_2015}.
Contamination by the molecular cloud may have affected the integrated line flux.

\paragraph{DQ Tau} This is a spectroscopic binary, part of the Taurus-Auriga complex, implying a distance
of $\sim$140~pc and a young age (1-3~Myr). With an orbital period of 15.8 days, the
stellar separation is $\sim$0.1~au \citep{Mathieu+etal_1997}. The authors deduce that the
disk truncation effect by the binary is limited to the inner 0.4~au region, and that no
effect is visible on the SED compared to a single classical T Tauri disk. Warm CO and
dust have been detected in this region (R$_{CO} \sim$0.1~au, \citet{Carr+etal_2001},
R$_{dust} \sim$0.2~au, \citet{Boden+etal_2009}). The system is strongly accreting material
from the circumbinary disk, as attested by the variability of the infrared and radio
emissions \citep{Salter+etal_2008,Bary+etal_2014}. The gas emission is faint, and only
$^{12}$CO has been detected (CN and H$_2$CO search led to non-detections, strong
$^{13}$CO confusion, \citet{Guilloteau+etal_2013}). \citet{Williams+etal_2014} estimate
the gas disk mass to be lower than 10$^{-3}$~M$_{\odot}$ and the dust mass lower than
9$\times$10$^{-5}$~M$_{\odot}$.

\paragraph{BP Tau}

Like DQ~Tau, BP~Tau is a member of the Taurus cloud, but is a single star harboring a CO
disk extending out to 120~au, and a dust disk larger than 100~au
\citep{Dutrey+etal_2003}. This makes BP~Tau one of the most compact disks detected.
\citet{Dutrey+etal_2003} argue that the $^{12}$CO is optically thin, which could explain
its low position on the diagram.

\begin{longtab}
    \begin{longtable}{lcccccc} 
        % 6 columns
    \caption{\label{table:1} $^{12}$CO and continuum emission used to plot Tables 7a and 7b. Objects are sorted first by category and then by increasing R.A. Values are presented here as given in the literature. For some rare cases, there are no errorbars. Values in italics correspond to extrapolation from the other frequency. Values given with the symbol ``$<$'' followed by an errorbar (e.g. \textit{$<$1.5 $\pm$ 0.5}) correspond to sources observed with ALMA where confusion with surrounding molecular clouds partly affects the emission from the source. Upper limits are 3$\sigma$.} \\
    \hline\hline
      ID & Name & \codu{} & Continuum & \cotd{} & Continuum & References  \\
         &  & (Jy.km.s$^{-1}$) & 1.3 mm (mJy) & (Jy.km.s$^{-1}$) & 0.8 mm (mJy)  & \\
    \hline
    \endfirsthead
    \caption{continued.}\\
    \hline\hline
      ID & Name & \codu{} & Continuum & \cotd{} & Continuum & References  \\
         &  & (Jy.km.s$^{-1}$) & 1.3 mm (mJy) & (Jy.km.s$^{-1}$) & 0.8 mm (mJy)  & \\
    \hline
    \endhead
    \hline
    \endfoot
    % \multicolumn{6}{c}{ } \\
    \multicolumn{6}{c}{CTTS disks} \\
    % \multicolumn{6}{c}{ } \\
    \hline

1 & CIDA 1            & \textit{0.9 $\pm$ 0.8} & \textit{16 $\pm$ 8} & 1.45 $\pm$ 0.05  & 27 $\pm$ 3     & 1 \\
2 & $^\dagger$IRAS 04113+2758   & \textit{10 $\pm$ 12} & \textit{196 $\pm$ 96} & 19.1             & 410 $\pm$ 5    & 2 \\
3 & CY Tau            & 2.02 $\pm$ 0.08  & 111 $\pm$ 3       & \textit{3.6 $\pm$ 1.0} & 140 $\pm$ 5    & 3, 4, 5 \\
4 & $^\dagger$FQ Tau            & \textit{0.2 $\pm$ 0.2} & \textit{6 $\pm$ 3} & 0.239            & 8.8 $\pm$ 1.2  & 2  \\
5 & BP Tau            & 1.11 $\pm$ 0.08  & 58.2 $\pm$ 1.3    & \textit{1.90 $\pm$ 0.59} & 130 $\pm$ 7    & 3, 4, 5 \\
6 & RY Tau            & 6.6              & 229 $\pm$ 17      & 123 $\pm$ 37     & 560 $\pm$ 30   & 6, 7, 8, 5 \\
7 & $^\dagger$FV Tau A          & \textit{1.3 $\pm$ 1.5} & 6.17 $\pm$ 0.16   & 2.17             & 31 $\pm$ 1.4   & 9, 2 \\
8 & $^\dagger$FW Tau c          & 0.156            & 3.4 $\pm$ 0.2     & \textit{0.2 $\pm$ 0.1} & \textit{4.9 $\pm$ 2.4} & 51 \\
9 & $^\dagger$FX Tau            & \textit{0.9 $\pm$ 1.0} & 7.1 $\pm$ 0.15    & 1.49             & 17 $\pm$ 0.9   & 9, 2   \\
10 & $^\dagger$DK Tau            & \textit{0.9 $\pm$ 1.1} & 35 $\pm$ 7        & 1.56             & 66 $\pm$ 1.0   & 9, 2  \\
11 & $^\dagger$HK Tau A          & \textit{2.3 $\pm$ 2.7} & 33 $\pm$ 0.15     & 4.1              & 78 $\pm$ 1.1   & 9, 2 \\
12 & $^\dagger$HK Tau B          & \textit{1.9 $\pm$ 2.2} & 12.6 $\pm$ 0.15   & 3.39             & 57 $\pm$ 1.3   & 9, 2 \\
13 & $^\dagger$V710 Tau          & \textit{1.8 $\pm$ 2.1} & 59.18 $\pm$ 0.33  & 3.23             & 142 $\pm$ 1.5  & 9, 2  \\
14 & $^\dagger$GG Tau            & \textit{13 $\pm$ 16} & \textit{546 $\pm$ 298} & 53 $\pm$ 15      & 1255 $\pm$ 57  & 8, 5   \\
15 & GK Tau A          & \textit{2.00 $\pm$ 2.33}  & 5.33 $\pm$ 0.56   & 3.56             & 15 $\pm$ 1     & 9, 2    \\
16 & DL Tau            & 1.98 $\pm$ 0.12  & 204 $\pm$ 2       & \textit{3.5 $\pm$ 1.1} & 440 $\pm$ 40   & 3, 4, 10 \\
17 & $^\dagger$HN Tau            & \textit{1.4 $\pm$ 1.7} & 12.8 $\pm$ 0.2    & 2.48             & 37 $\pm$ 1.1   & 9, 2    \\
18 & DM Tau            & 14.87 $\pm$ 0.12 & 109 $\pm$ 2.4     & 20 $\pm$ 6       & 237 $\pm$ 12   & 11, 4, 8, 5  \\
19 & $^\dagger$IT Tau B          & \textit{0.6 $\pm$ 0.7} & \textit{6.5 $\pm$ 3.1} & 1.03             & 10 $\pm$ 0.9   & 2      \\
20 & AA Tau            & 8.2 $\pm$ 0.16   & 88 $\pm$ 9        & \textit{15.5 $\pm$ 4.7} & 144 $\pm$ 5    & 11, 5  \\
21 & $^\dagger$HO Tau            & \textit{0.8 $\pm$ 0.9} & 17.06 $\pm$ 0.27  & 1.36             & 38 $\pm$ 1     & 9, 2 \\
22 & $^\dagger$HBC 411 B          & \textit{0.4 $\pm$ 0.4} & 5.8 $\pm$ 0.27    & 0.624            & 14 $\pm$ 1     & 9, 2 \\
23 & LkCa 15           & 13.94 $\pm$ 0.15 & 110 $\pm$ 2       & 23 $\pm$ 7       & 428 $\pm$ 11   & 11, 4, 8, 5 \\
24 & J04442713+2512164 & \textit{0.6 $\pm$ 0.6} & \textit{5.9 $\pm$ 3.6} & 0.98 $\pm$ 0.03  & 9 $\pm$ 2      & 1  \\
25 &$^\dagger$DQ Tau            & 0.97 $\pm$ 0.11  & 83.1 $\pm$ 2.8    & \textit{1.7 $\pm$ 0.6} & \textit{161 $\pm$ 86} & 3, 4 \\
26 & DR Tau            & \textit{49 $\pm$ 65} & 159 $\pm$ 11      & 103 $\pm$ 31     & 533 $\pm$ 7    & 5, 8 \\
27 & $^\dagger$DS Tau            & \textit{3.7 $\pm$ 4.3} & 19.94 $\pm$ 0.25  & 6.69             & 43 $\pm$ 1.4   & 9, 2  \\
28 & GM Aur            & 19.41 $\pm$ 0.11 & 176 $\pm$ 5.3     & 29.0 $\pm$ 8.7   & 707 $\pm$ 4    & 11, 4, 8, 12 \\
29 & V836 Tau          & 0.84 $\pm$ 0.18  & 24 $\pm$ 2        & \textit{1.42 $\pm$ 0.65} & \textit{42 $\pm$ 23} & 13  \\
30 & $^\dagger$IRAS 05022+2527   & \textit{2.9 $\pm$ 3.4} & 35.22 $\pm$ 0.26  & 5.21             & 87 $\pm$ 1.3   & 9, 2 \\
31 & HD 34282          & 4.2 $\pm$ 0.1    & 110 $\pm$ 10      & \textit{7.7 $\pm$ 2.2} & \textit{218 $\pm$ 131} & 2\\
32 & HD 294268         & $<$ 0.9          & 5 $\pm$ 0.8       & \textit{$<$1.5 $\pm$ 1.8} & 19.4 $\pm$ 5.4   & 52 \\
33 & V2731 Ori         & 0.95 $\pm$ 0.13  & 10.7 $\pm$ 0.8    & \textit{1.6 $\pm$ 0.6} & 32.6 $\pm$ 4.8 & 52 \\
34 & J05391573-0230568 & $<$ 2.43         & 8 $\pm$ 1         & \textit{$<$4.3 $\pm$ 1.9} & 13 $\pm$ 2.7 & 52 \\
35 & V602 Ori          & 0.34 $\pm$ 0.08  & 7.8 $\pm$ 0.8     & \textit{0.5 $\pm$ 0.3} & 17.4 $\pm$ 4.2 & 52 \\
36 & V606 Ori          & 0.36 $\pm$ 0.07  & 14.3 $\pm$ 0.8    & \textit{0.6 $\pm$ 0.2} & 31.4 $\pm$ 4.8 & 52 \\
37 & TWA 34            & 0.34 $\pm$ 0.03  & 0.54 $\pm$ 0.06   & \textit{0.55 $\pm$ 0.17}  & \textit{0.66 $\pm$ 0.32} & 15 \\
38 & TW Hya            & 12.4 $\pm$ 1     & 540 $\pm$ 30      & 45 $\pm$ 4       & 1620 $\pm$ 50  & \\ %Qi+etal_2004, Qi+etal_2006\\
39 & T Cha             & \textit{7 $\pm$ 7} & 81.8              & 12.46 $\pm$ 0.11 & 198 $\pm$ 4    & \\ % xx, Andrews+etal_2009  \\
40 & HD 135344         & \textit{10 $\pm$ 10} & \textit{153 $\pm$ 75} & 18.6 $\pm$ 0.8   & 314 $\pm$ 4.5  & 16, 17 \\
41 & HD 142527         & 17.4 $\pm$ 0.01  & 1090 $\pm$ 0.3    & \textit{34 $\pm$ 10} & \textit{2700 $\pm$ 1500} & 18      \\
42 & HD 143006         & 4.28 $\pm$ 0.04  & 43 $\pm$ 3        & 3 $\pm$ 1        & \textit{78 $\pm$ 43} & 19, 20, 16 \\
43 & J16042165-2130284 & \textit{11 $\pm$ 11} & \textit{113 $\pm$ 53} & 21.4 $\pm$ 0.2   & 226 $\pm$ 1    & 21     \\
44 & Sz 91             & 4.93 $\pm$ 0.02  & 40 $\pm$ 1        & 8.55             & 34.5 $\pm$ 2.9 & 53, 22, 23  \\
45 & $^\dagger$AS 205 A (N)      & 19.23 $\pm$ 0.27 & 377 $\pm$ 2       & 65               & 905 $\pm$ 7    & 24, 25 \\
46 & $^\dagger$AS 205 B (S)      & 3.02 $\pm$ 0.22  & 64 $\pm$ 1        & \textit{5.4 $\pm$ 1.8} & 55 $\pm$ 7     & 24, 25 \\
47 & GSS 26            & $>$4             & 176               & \textit{$>$7.3 $\pm$ 19.9} & \textit{364 $\pm$ 267} & 2     \\
48 & GSS 39            & $>$25            & 304               & \textit{$>$49.9 $\pm$ 2.1} & 663 $\pm$ 3    & 2, 25  \\
49 & WL 18             & $>$0.3           & 24                & \textit{$>$0.5 $\pm$ 1.7} & \textit{42 $\pm$ 28} & 2      \\
50 & WL 14             & $>$0.2           & $<$0.2            & \textit{$>$0.3 $\pm$ 1.7} & \textit{$<$0.2 $\pm$ 1.2} & 2     \\
51 & ISO-Oph 102       & \textit{0.3 $\pm$ 0.3} & 2.9 $\pm$ 1.2 & 0.53 $\pm$ 0.05  & 4.1 $\pm$ 0.2  & 26     \\
52 & YLW 16c           & $>$1.6           & 45                & \textit{$>$2.8 $\pm$ 1.8} & \textit{82 $\pm$ 58} & 2   \\
53 & ROX 25            & $>$0.6           & 29                & \textit{$>$1.0 $\pm$ 1.7} & \textit{51 $\pm$ 35} & 2   \\
54 & Flying Saucer     & $>$2.1           & 32                & \textit{$>$3.7 $\pm$ 1.9} & \textit{57 $\pm$ 39} & 2   \\
55 & AS 209            & 7.4              & 300               & 16.5             & 577 $\pm$ 3    & \\ %Andre\&Montmerle94, Andrews09}  \\
56 & $^\dagger$V4046 Sgr         & 11.38 $\pm$ 0.16 & 451 $\pm$ 20      & \textit{22 $\pm$ 7} & \textit{1020 $\pm$ 600} & 27, 28    \\

    \hline
    % \multicolumn{6}{c}{ } \\
    \multicolumn{6}{c}{Herbig AeBe disks} \\
    % \multicolumn{6}{c}{ } \\
    \hline

57 & $^\dagger$V892 Tau          & \textit{22 $\pm$ 29} & \textit{293 $\pm$ 167} & 85.4 $\pm$ 26    & 638 $\pm$ 54   & 8, 5  \\
58 & AB Aur            & 80$^{1}$         & 103 $\pm$ 18      & 153 $\pm$ 4      & 359 $\pm$ 67   & 2, 5, 16 \\
59 & MWC 480           & 22 $\pm$ 0.2     & 289.3 $\pm$ 2.5   & 52 $\pm$ 1.1     & \textit{626 $\pm$ 336} & 11, 4, 16 \\
60 & MWC 758           & \textit{7.9 $\pm$ 8.9} & 56 $\pm$ 1        & 15 $\pm$ 2       & 217 $\pm$ 40   & 29, 16, 30 \\
61 & HD 245185         & 0.17 $\pm$ 0.03  & 29 $\pm$ 6        & \textit{0.3 $\pm$ 0.1} & 74.1 $\pm$ 4.2 & 14 \\
62 & CQ Tau            & 3.1 $\pm$ 0.18   & 162 $\pm$ 2       & 5.95 $\pm$ 0.96  & 421 $\pm$ 9    & 11, 4, 16,  31 \\
63 & HD 100546         & \textit{49 $\pm$ 62} & \textit{540 $\pm$ 272} & 158 $\pm$ 24     & 1240           & 32, 33 \\
64 & $^\dagger$AK Sco            & 2.21 $\pm$ 0.01  & 32.65 $\pm$ 0.07  & \textit{5.7 $\pm$ 1.9} & \textit{52 $\pm$ 38} & 34 \\
65 & HD 163296         & 46 $\pm$ 5       & 705 $\pm$ 12      & 109 $\pm$ 11     & 1910 $\pm$ 20  & 35, 36 \\
66 & HD 169142         & 2.72 $\pm$ 0.7   & 169 $\pm$ 5       & 32.6 $\pm$ 2.5   & \textit{349 $\pm$ 190} & 37, 16 \\

    \hline
    % \multicolumn{6}{c}{ } \\
    \multicolumn{6}{c}{WTTS disks} \\
    % \multicolumn{6}{c}{ } \\
    \hline

67 & J04182147+1658470 & $<$0.09          & $<$0.436             & \textit{$<$0.14} & \textit{$<$0.53} & 38  \\
68 & $^\dagger$J04192625+2826142 & $<$0.09          & 0.533                & \textit{$<$0.14} & \textit{0.66 $\pm$ 0.38} & 38  \\
69 & J04242321+2650084 & $<$0.09          & $<$0.426             & \textit{$<$0.14} & \textit{$<$0.51} & 38  \\
70 & J04314503+2859081 & $<$0.09          & $<$0.435             & \textit{$<$0.14} & \textit{$<$0.53} & 38  \\
71 & J04325323+1735337 & $<$0.09          & $<$0.438             & \textit{$<$0.14} & \textit{$<$0.53} & 38  \\
72 & J04330422+2921499 & $<$0.09          & $<$0.431             & \textit{$<$0.14} & \textit{$<$0.52} & 38  \\
73 & J04364912+2412588 & $<$0.09          & $<$0.435             & \textit{$<$0.14} & \textit{$<$0.53} & 38  \\
74 & J04403979+2519061 & $<$0.09          & $<$0.431             & \textit{$<$0.14} & \textit{$<$0.52} & 38  \\
75 & $^\dagger$J04420548+2522562 & $<$0.09          & $<$0.423             & \textit{$<$0.14} & \textit{$<$0.51} & 38  \\
76 & J08413703-7903304 & $<$0.09          & $<$0.476             & \textit{$<$0.14} & \textit{$<$0.58} & 38 \\
77 & J08422372-7904030 & $<$0.09          & $<$0.472             & \textit{$<$0.14} & \textit{$<$0.58} & 38 \\
78 & J11073519-7734493 & $<$0.09          & $<$0.541             & \textit{$<$0.14} & \textit{$<$0.67} & 38 \\
79 & $^\dagger$J11124268-7722230 & $<$0.09          & $<$0.494             & \textit{$<$0.14} & \textit{$<$0.60} & 38 \\
80 & $^\dagger$J16002612-4153553 & $<$0.09          & 0.696                & \textit{$<$0.14} & \textit{0.88 $\pm$ 0.5} & 38 \\
81 & J16010896-3320141 & $<$0.09          & $<$0.442             & \textit{$<$0.14} & \textit{$<$0.53} & 38 \\
82 & J16031181-3239202 & $<$0.09          & $<$0.453             & \textit{$<$0.14} & \textit{$<$0.55} & 38 \\
83 & $^\dagger$J16085553-3902339 & $<$0.09          & 1.813                & \textit{$<$0.14} & \textit{2.5 $\pm$ 1.5} & 38 \\
84 & $^\dagger$J16124119-1924182 & $<$0.09          & $<$0.459             & \textit{$<$0.14} & \textit{$<$0.56} & 38 \\
85 & $^\dagger$J16220961-1953005 & $<$0.09          & $<$0.483             & \textit{$<$0.14} & \textit{$<$0.59} & 38 \\
86 & J16223757-2345508 & $<$0.09          & $<$0.461             & \textit{$<$0.14} & \textit{$<$0.56} & 38 \\
87 & J16251469-2456069 & $<$0.09          & $<$0.453             & \textit{$<$0.14} & \textit{$<$0.55} & 38 \\
88 & $^\dagger$J16275209-2440503 & $<$0.09          & $<$0.442             & \textit{$<$0.14} & \textit{$<$0.53} & 38 \\
89 & $^\dagger$J19002906-3656036 & $<$0.09          & 0.569                & \textit{$<$0.14} & \textit{0.7 $\pm$ 0.3} & 38 \\
90 & $^\dagger$J19012901-3701484 & $<$0.09          & $<$0.465             & \textit{$<$0.14} & \textit{$<$0.56} & 38 \\

    \hline
    % \multicolumn{6}{c}{ } \\
    \multicolumn{6}{c}{Hybrid disks} \\
    % \multicolumn{6}{c}{ } \\
    \hline

91 & 49 Ceti           & 2 $\pm$ 0.3      & $<$2.1            & 9.5 $\pm$ 1.9    & 8.2 $\pm$ 1.9  & 39, 40  \\
92 & HD 21997          & 2.17 $\pm$ 0.23  & $<$ 1             & 2.52 $\pm$ 0.27  & 2.69 $\pm$ 0.3 & 41, 2, 42 \\
93 & HD 131835         & 0.798 $\pm$ 0.035   & \textit{5.6 $\pm$ 4.9} & 2.74 $\pm$ 0.55  & 8.5 $\pm$ 4.4  & 54, 43, 44  \\
94 & HIP 76310         & 1,41 $\pm$ 0.08  & 1.2 $\pm$ 0.2     & \textit{2.4 $\pm$ 0.5} & 1.6 $\pm$ 1.1 & 54 \\
95 & HD 141569         & 9.57 $\pm$ 0.03  & 3.7 $\pm$ 0.4     & 14.6 $\pm$ 1     & 5.3 $\pm$ 0.6  & 45, 16, 46 \\
96 & HIP 84881         & 1.18 $\pm$ 0.04  & 0.72 $\pm$ 0.11   & \textit{2.5 $\pm$ 1} & \textit{0.7 $\pm$ 0.4} & 54 \\

    \hline
    % \multicolumn{6}{c}{ } \\
    \multicolumn{6}{c}{Debris disks} \\
    % \multicolumn{6}{c}{ } \\
    \hline

97 & HD 225200         & \textit{$<$1.7 $\pm$ 0.5} & --                & $<$3.03          & --             & 2 \\
98 & $^\dagger$HD 2772           & $<$0.84          & --                & \textit{$<$1.4 $\pm$ 1.8} & --             & 2 \\
99 & HD 14055          & $<$0.47          & \textit{4.3 $\pm$ 2.4} & \textit{$<$0.8 $\pm$ 1.7} & 6.4 $\pm$ 1.1  & 2, 47 \\
100 & HD 15115          & $<$2.48          & 2.6 $\pm$ 0.6     & \textit{$<$4.4 $\pm$ 1.9} & 8.5 $\pm$ 1.2  & 2, 48, 47 \\
101 & HD 17848          & \textit{$<$5.4 $\pm$ 0.5} & --                & $<$10.06         & --             & 2 \\
102 & $^\dagger$HD 21620          & $<$0.91          & --                & \textit{$<$1.5 $\pm$ 1.8} & --             & 2 \\
103 & $^\dagger$HD 23642          & 0.710 $\pm$ 0.18 & --                & \textit{$<$1.4 $\pm$ 1.8} & --             & 2 \\
104 & HD 24966          & \textit{$<$1.6 $\pm$ 0.5} & --                & $<$2.75          & --             & 2 \\
105 & HD 30422          & \textit{$<$1.2 $\pm$ 0.5} & --                & $<$2.13          & --             & 2 \\
106 & HD 31295          & $<$1.1           & --                & \textit{$<$1.9 $\pm$ 1.8} & --             & 2 \\
107 & HD 35850          & \textit{$<$1.3 $\pm$ 0.5} & --                & $<$2.17          & $<$5.4         & 2 \\
108 & HD 38206          & \textit{$<$1.5 $\pm$ 0.5} & --                & $<$2.67          & --             & 2 \\
109 & $\beta$ Pic       & \textit{3.6 $\pm$ 3.9} & \textit{34 $\pm$ 18} & 6.59 $\pm$ 0.69  & 60 $\pm$ 6     & 49 \\
110 & HD 42111          & $<$1.21          & --                & \textit{$<$2.1 $\pm$ 1.8} & --             & 2 \\
111 & HD 54341          & \textit{$<$1.4 $\pm$ 0.5} & --                & $<$2.39          & --             & 2 \\
112 & HD 71043          & \textit{$<$1.5 $\pm$ 0.5} & --                & $<$2.64          & --             & 2 \\
113 & HD 71155          & \textit{$<$1.5 $\pm$ 0.5} & --                & $<$2.56          & --             & 2 \\
114 & HD 78702          & \textit{$<$1.2 $\pm$ 0.5} & \textit{0.3 $\pm$ 1.9} & $<$2.03          & 0.3 $\pm$ 2.3  & 2 \\
115 & HD 136246         & \textit{$<$1.4 $\pm$ 0.5} & --                & $<$2.37          & --             & 2   \\
116 & $^\dagger$HD 159082         & $<$0.66          & --                & \textit{$<$1.1 $\pm$ 1.8}& --             & 2 \\
117 & HD 164249         & \textit{$<$1.3 $\pm$ 0.5} & --                & $<$2.29          & --             & 2 \\
118 & HD 166191         & \textit{$<$1.5 $\pm$ 0.5} & --                & $<$2.66          & --             & 2 \\
119 & HD 181296         & \textit{$<$1.2 $\pm$ 0.5} & \textit{9.1 $\pm$ 3.7} & $<$2.11          & 14.4           & 2  \\
120 & HD 182681         & \textit{$<$1.6 $\pm$ 0.5} & \textit{3.4 $\pm$ 2.1} & $<$2.79          & 5 $\pm$ 1.3    & 2, 47 \\
121 & HD 218396         & $<$0.5           & 20                & \textit{$<$0.8 $\pm$ 1.7} & 10.3 $\pm$ 1.8 & 2, 50 \\
122 & HD 220825         & \textit{$<$1.3 $\pm$ 0.5} & --                & $<$2.2           & --             & 2 \\

    \hline
    \end{longtable}    
    \tablefoot{$^{1}$ The flux of AB Aur is contaminated by the envelope. We have
    estimated the emission of the disk to be $\approx$~80 Jy.km.s$^{-1}$ from
    interferometric observations with PdBI.
    \\
    $^\dagger$: binary stars.
    \\
    References: 1 - \citet{Ricci+etal_2014}; 2 - this work; 3 -
    \citet{Williams+etal_2014}; 4 - \citet{Guilloteau+etal_2011}; 5 -
    \citet{Andrews+etal_2005}; 6 - \citet{Koerner+etal_1995}; 7 -
    \citet{Beckwith+etal_1990}; 8 - \citet{Thi+etal_2001}; 9 - \citet{Akeson+etal_2014};
    10 - \citet{Mannings+etal_1994}; 11 - \citet{Oberg+etal_2010}; 12 -
    \citet{Andrews+etal_2007}; 13 - \citet{Duvert+etal_2000}; 14 -
    \citet{Ansdell+etal_2015}; 15 - \citet{Rodriguez+etal_2015}; 16 -
    \citet{Dent+etal_2005}; 17 - \citet{Brown+etal_2009}; 18 - \citet{Perez+etal_2015};
    19 - \citet{Zuckerman+etal_1995}; 20 - \citet{Natta+etal_2004}; 21 -
    \citet{Zhang+etal_2014}; 22 - \citet{Tsukagoshi+etal_2014}; 23 -
    \citet{Romero+etal_2012}; 24 - \citet{Salyk+etal_2014}; 25 -
    \citet{Andrews+etal_2009}; 26 - \citet{Ricci+etal_2012}; 27 -
    \citet{Kastner+etal_2008}; 28 - \citet{Jensen+etal_1996}; 29 -
    \citet{Chapillon+etal_2008}; 30 - \citet{Isella+etal_2010}; 31 -
    \citet{Banzatti+etal_2011}; 32 - \citet{Panic+etal_2010}; 33 -
    \citet{Walsh+etal_2014}; 34 - \citet{Czekala+etal_2015}; 35 -
    \citet{Rosenfeld+etal_2013}; 36 - \citet{Isella+etal_2007}; 37 -
    \citet{Raman+etal_2006}; 38 - \citet{Hardy+etal_2015}; 39 - \citet{Hughes+etal_2008};
    40 - \citet{Sheret+etal_2004}; 41 - \citet{Kospal+etal_2013}; 42 -
    \citet{Moor+etal_2013}; 43 - \citet{Moor+etal_2015}; 44 - \citet{Nilsson+etal_2010};
    45 -\citet{Pericaud+etal_inprep}; 46 - \citet{Sylvester+etal_2001}; 47 -
    \citet{Panic+etal_2013}; 48 - \citet{MacGregor+etal_2015}; 49 -
    \citet{Dent+etal_2014}; 50 - \citet{Williams+etal_2006}; 51 -
    \citet{Caceres+etal_2015}; 52 - \citet{Williams+etal_2013}; 53 -
    \citet{Canovas+etal_2016}; 54 - \citet{Lieman-Sifry+etal_2016}.  }

    \end{longtab}

\begin{longtab}
    \begin{longtable}{lcccccccc}     % 6 columns
    \caption{\label{table:star_param} Stellar parameters. The classification is the same as in Table 2.} \\
    \hline\hline
      ID & Name & Distance & Spectral & M$_{\star}$ & L$_{\star}$ & $\dot{M}_{acc}$ & Age & References  \\
         &  & (pc)     & type     & (M$_{\odot}$) & (L$_{\odot}$) & (M$_{\odot}$/yr) & Myr & \\
    \hline
    \endfirsthead
    \caption{continued.}\\
    \hline\hline
      ID & Name & Distance & Spectral & M$_{\star}$ & L$_{\star}$ & $\dot{M}_{acc}$ & Age & References  \\
         &  & (pc)     & type     & (M$_{\odot}$) & (L$_{\odot}$) & (M$_{\odot}$/yr) & Myr & \\
    \hline
    \endhead
    \hline
    \endfoot
    % \multicolumn{6}{c}{ } \\
    \multicolumn{8}{c}{CTTS disks} \\
    % \multicolumn{6}{c}{ } \\
    \hline

1 & CIDA 1            & 140  & M5.5 & 0.13 & 0.13  & --                           & 2.7      & 1, 2 \\
2 & IRAS 04113+2758   & 140  & --   & --   & --    & --                           & --       & \\
3 & CY Tau            & 140  & M1.5 & 0.44 & 0.40  & 4.08 $\times$ 10$^{-9}$      & 5        & 1, 2, 3 \\
4 & FQ Tau            & 140  & M3   & 0.29 & 0.09  & 3.55 $\times$ 10$^{-7}$      & 6.8      & 4, 2, 5\\
5 & BP Tau            & 76.8 & K7   & 0.79 & 0.97  & 4.08 $\times$ 10$^{-9}$      & 1.7      & 6, 2, 3 \\
6 & RY Tau            & 192  & K1   & 2.3  & 21.9  & 2.24 $\times$ 10$^{-8}$      & 3.86     & 6, 7, 8 \\
7 & FV Tau A          & 140  & M5   & 1.23 & 2.3   & 4.79 $\times$ 10$^{-8}$      & 3.7      & 4, 2, 9 \\
8 & FW Tau c          & 140  & M9   & 0.001 & --   & 1.00 $\times$ 10$^{-11}$     & 2        & 48, 49 \\
9 & FX Tau            & 140  & M1   & 0.48 & 0.52  & 2.23 $\times$ 10$^{-9}$      & 1.7      & 10, 2, 5 \\
10 & DK Tau            & 140  & K8   & 0.71 & 1.32  & 1.05 $\times$ 10$^{-9}$      & 1        & 10, 2, 3 \\
11 & HK Tau A          & 140  & M0.5 & 0.54 & 0.44  & 1.95 $\times$ 10$^{-9}$      & 2.5      & 1, 2, 3 \\
12 & HK Tau B          & 140  & M2   & 0.37 & 0.03  & --                           & 4.1      & 1, 2 \\
13 & V710 Tau          & 140  & M0.5 & 0.54 & 0.57  & --                           & 1.7      & 1, 2 \\
14 & GG Tau            & 140  & K7   & 1.35 & 0.85  & 3.02 $\times$ 10$^{-9}$      & 2.1      & 11, 2, 9 \\
15 & GK Tau A          & 140  & K7   & 0.78 & 1.35  & 6.45 $\times$ 10$^{ -9}$      & 1.1      & 5, 2 \\
16 & DL Tau            & 140  & K7   & 0.81 & 0.74  & 3.22 $\times$ 10$^{-8}$      & 2.6      & 1, 2, 3 \\
17 & HN Tau            & 140  & K5   & 0.91 & 0.42  & 2.72 $\times$ 10$^{-9}$      & 5.4      & 12, 2, 3 \\
18 & DM Tau            & 140  & M1   & 0.47 & 0.23  & 2.25 $\times$ 10$^{-9}$      & 5        & 5, 2, 3  \\
19 & IT Tau B          & 140  & K3   & 1.41 & 1.43  & --                           & 4.6      & 10, 2 \\
20 & AA Tau            & 140  & K7   & 0.79 & 0.87  & 3.91 $\times$ 10$^{-9}$      & 2        & 5, 2, 3 \\
21 & HO Tau            & 140  & M0.5 & 0.56 & 0.13  & --                           & 13.2     & 1, 2 \\
22 & hbc 411           & 140  & --   & --   & --    & --                           & --       &  \\
23 & LkCa 15           & 140  & K5   & 1.05 & 0.81  & 6.77 $\times$ 10$^{-10}$     & 3.7      & 5, 2, 3 \\
24 & J04442713+2512164 & 140  & M7.3 & --   & --    & --                           & --       & \\
25 & DQ Tau            & 140  & M0   & 1.27 & 0.4   & 2.53 $\times$ 10$^{-9}$      & 3.5      & 13, 14, 2, 3 \\
26 & DR Tau            & 140  & K5   & 1.2  & 1.82  & 1.42 $\times$ 10$^{-8}$      & 1.5      & 1, 2, 3 \\
27 & DS Tau            & 140  & K5   & 1.05 & 0.76  & 3.35 $\times$ 10$^{-8}$      & 4.1      & 1, 2, 3 \\
28 & GM Aur            & 140  & K3   & 1.35 & 1.23  & 1.21 $\times$ 10$^{-8}$      & 5.6      & 15, 2, 3 \\
29 & V836 Tau          & 140  & K7   & 0.83 & 0.57  & 4.20 $\times$ 10$^{-9}$      & 3.7      & 1, 2, 3 \\30 & IRAS 05022+2527   & 140  & M0   & --   & --    & --                           & --       & \\
31 & HD 34282          & 350  & A0   & 2.35 & 5.13  & $<$1.95 $\times$ 10$^{-8}$   & 6.5      & 50, 19, 40 \\
32 & HD 294268         & 420  & F8   & --   & --    & --                           & --       & 53  \\
33 & V2731 Ori         & 420  & --   & --   & --    & --                           & --       &    \\
34 & J05391573-0230568 & 420  & K9   & --   & --    & --                           & --       & 53   \\
35 & V602 Ori          & 420  & M3   & --   & --    & --                           & --       & 53   \\
36 & V606 Ori          & 420  & --   & --   & --    & --                           & --       &    \\
37 & TWA 34            & 47   & M4.9 & --   & 0.014 & --                           & 6        & 16, 17 \\
38 & TW Hya            & 54   & M2.5 & 0.4  & 0.19  & 4.75 $\times$ 10$^{-9}$      & 3        & 18, 3 \\
39 & T Cha             & 164  & K0   & 1.5  & --    & --                           & --       &  \\
40 & HD 135344         & 140  & F3   & 1.5  & 7.08  & 4.27 $\times$ 10$^{-8}$      & 8        & 19, 20 \\
41 & HD 142527         & 140  & F6   & 1.6  & 7.94  & 3.55 $\times$ 10$^{-8}$      & 8.8      & 19, 20 \\
42 & HD 143006         & 82   & G6   & 1    & 0.8   & 5.11 $\times$ 10$^{-9}$      & $>$10 00 & 21, 3 \\
43 & J16042165-2130284 & 145  & --   & --   & --    & --                           & --       &  \\
44 & Sz 91             & 200  & M2   & --   & 0.25  & --                           & --       & 17 \\
45 & AS 205 A (N)      & 125  & K7   & 1.5  & 7.1   & 7.94 $\times$ 10$^{-8}$      & 1.6      & 22, 23, 24  \\
46 & AS 205 B (S)      & 125  & M0.1 & 0.3  & 2.19  & 2.09 $\times$ 10$^{-7}$      & --       & 17, 23 \\
47 & GSS 26            & 135  & M0   & 0.56 & 1.39  & --                           & 0.5      & 25, 26 \\
48 & GSS 39            & 135  & M1   & 0.6  & 1     & 3.70 $\times$ 10$^{-8}$      & 1        & 27, 24, 28  \\
49 & WL 18             & 135  & K7   & --   & 0.3   & --                           & --       & 29, 30 \\
50 & WL 14             & 135  & --   & --   & --    & --                           & --       &  \\
51 & ISO-Oph 102       & 135  & M6   & 0.06 & 0.08  & 1.00 $\times$ 10$^{-9}$      & --       & 31 \\
52 & YLW 16c           & 135  & M1   & 0.48 & 1.11  & --                           & 0.5      & 25, 26 \\
53 & ROX 25            & 135  & --   & --   & --    & --                           & --       &  \\
54 & Flying Saucer     & 120  & M1   & --   & 0.14  & --                           & --       & 32  \\
55 & AS 209            & 125  & K5   & 1.4  & 2.5   & 3.02  $\times$ 10$^{-8}$     & 0.5      & 33, 23, 24 \\
56 & V4046 Sgr         & 54   & --   & --   & --    & --                           & --       &   \\

    \hline
    % \multicolumn{6}{c}{ } \\
    \multicolumn{8}{c}{Herbig AeBe disks} \\
    % \multicolumn{6}{c}{ } \\
    \hline

57 & V892 Tau          & 140  & B8   & 2.8  & 38.8  & --                           & 46       & 7, 2 \\
58 & AB Aur            & 139  & A1   & 2.6  & 51.3  & 1.41  $\times$ 10$^{-7}$     & 3.94     & 7, 19 \\
59 & MWC 480           & 146  & A4.5 & 1.91 & 19.1  & $<$5.89  $\times$ 10$^{-8}$  & 7.1      & 34, 2, 8 \\
60 & MWC 758           & 279  & A3   & 1.8  & 17.4  & 8.90  $\times$ 10$^{-7}$     & 3.5      & 35, 36, 37, 38 \\
61 & HD 245185         & 450  & A1   & 2.2  & 22.9  & --                           & 7.97     & 7  \\
62 & CQ Tau            & 113  & F3   & 1.6  & 7.1   & $<$5.01  $\times$ 10$^{-9}$  & 13.27    & 7, 8 \\
63 & HD 100546         & 97   & B9   & 2.5  & 40.7  & 9.12  $\times$ 10$^{-8}$     & 3.46     & 39, 40, 20 \\
64 & AK Sco            & 103  & F5   & 2.49 & --    & --                           & --       & 41  \\
65 & HD 163296         & 119  & A1   & 2.49 & 33.9  & 3.23  $\times$ 10$^{-8}$     & 3        & 19, 40, 20 \\
66 & HD 169142         & 145  & A5   & 2    & 18.2  & 3.98  $\times$ 10$^{-8}$     & 5.42     & 19, 40 \\

    \hline
    % \multicolumn{6}{c}{ } \\
    \multicolumn{8}{c}{WTTS disks} \\
    % \multicolumn{6}{c}{ } \\
    \hline

67 & J04182147+1658470 & 135 & K5    & --   & --    & --                           &          &  \\
68 & J04192625+2826142 & 135 & K7    & 0.67 & --    & $<$ 1.00 $\times$ 10$^{-11}$ & 3.5      & 42, 43 \\
69 & J04242321+2650084 & 135 & M2    & 0.35 & --    & $<$ 1.00 $\times$ 10$^{-11}$ & 6.7      & 42, 43 \\
70 & J04314503+2859081 & 135 & F5    & --   & --    & $<$ 1.00 $\times$ 10$^{-11}$ & --       & 42, 43 \\
71 & J04325323+1735337 & 135 & M2    & 0.36 & --    & --                           & 3.1      & 42 \\
72 & J04330422+2921499 & 135 & B9    & --   & --    & $<$ 1.00 $\times$ 10$^{-11}$ & --       & 42, 43 \\
73 & J04364912+2412588 & 135 & F2    & --   & --    & $<$ 1.00 $\times$ 10$^{-11}$ & --       & 42, 43 \\
74 & J04403979+2519061 & 135 & M5    & 0.17 & --    & $<$ 1.00 $\times$ 10$^{-11}$ & --       & 42 \\
75 & J04420548+2522562 & 135 & K7    & 0.65 & --    & --                           & --       & 42 \\
76 & J08413703-7903304 & 97  & M3    & 0.27 & --    & --                           & 3.4      & 42 \\
77 & J08422372-7904030 & 97  & M3    & 0.31 & --    & --                           & 15       & 42 \\
78 & J11073519-7734493 & 160 & M4    & 0.20 & --    & --                           & 12.4     & 42 \\
79 & J11124268-7722230 & 160 & G8    & 1.4  & --    & --                           & 6.9      & 42 \\
80 & J16002612-4153553 & 150 & M5.3  & 0.17 & --    & --                           & 26.3     & 42 \\
81 & J16010896-3320141 & 150 & G8    & 1.1  & --    & --                           & 7.7      & 42 \\
82 & J16031181-3239202 & 150 & K7    & 0.7  & --    & --                           & 2.6      & 42 \\
83 & J16085553-3902339 & 200 & M6    & 0.1  & --    & --                           & 6.6      & 42 \\
84 & J16124119-1924182 & 119 & K8    & 0.64 & --    & --                           & 2.2      & 42 \\
85 & J16220961-1953005 & 119 & M3.7  & 0.34 & --    & --                           & 9.9      & 42 \\
86 & J16223757-2345508 & 119 & M2.5  & 0.33 & --    & --                           & 3        & 42 \\
87 & J16251469-2456069 & 119 & M0    & 0.56 & --    & --                           & 4.2      & 42 \\
88 & J16275209-2440503 & 129 & K7    & 0.77 & --    & --                           & 4.7      & 42 \\
89 & J19002906-3656036 & 129 & M4    & 0.24 & --    & --                           & 10.2     & 42 \\
90 & J19012901-3701484 & 116 & M3.8  & 0.21 & --    & --                           & --       & 42 \\

    \hline
    % \multicolumn{6}{c}{ } \\
    \multicolumn{8}{c}{Hybrid disks} \\
    % \multicolumn{6}{c}{ } \\
    \hline

91 & 49 Ceti           & 59  & A1    & 2.17 & 21.8  & --                           & 8.9      & 44 \\
92 & HD 21997          & 72  & A3    & 1.8  & --    & --                           & 30       & 45 \\
93 & HD 131835         & 122 & A2    & --   & --    & --                           & 16       & 46 \\
94 & HIP 76310         & 151 & A0    & 2.2  & 26.9  & --                           & 10       & 52, 51\\
95 & HD 141569         & 116 & A0    & 2.18 & 22.9  & 2.24 $\times$ 10$^{-8}$      & 6.65     & 34, 40, 20 \\
96 & HIP 84881         & 118 & A0    & 2.9  & 15    & --                           & 10       & 51 \\
    \hline
    % \multicolumn{6}{c}{ } \\
    \multicolumn{8}{c}{Debris disks} \\
    % \multicolumn{6}{c}{ } \\
    \hline

97 & HD 225200         & 129 & A0    & --   & --    & --                           & --       & \\
98 & HD 2772           & 116 & B8    & --   & --    & --                           & --       & \\
99 & HD 14055          & 34  & A1    & --   & --    & --                           & --       & \\
100 & HD 15115          & 45  & F2    & --   & --    & --                           & --       & \\
101 & HD 17848          & 51  & A2    & --   & --    & --                           & --       & \\
102 & HD 21620          & 149 & A0    & --   & --    & --                           & --       & \\
103 & HD 23642          & 139 & A0    & --   & --    & --                           & --       & \\
104 & HD 24966          & 104 & A0    & --   & --    & --                           & --       & \\
105 & HD 30422          & 56  & A3    & --   & --    & --                           & --       & \\
106 & HD 31295          & 36  & A0    & --   & --    & --                           & --       & \\
107 & HD 35850          & 27  & F8    & 1.2  & 1.78  & 2.41 $\times$ 10$^{-12}$     & --       & 3 \\
108 & HD 38206          & 69  & A0    & --   & --    & --                           & --       & \\
109 & $\beta$ Pic       & 19  & A6    & --   & --    & --                           & 20       & 47 \\
110 & HD 42111          & 201 & A3    & --   & --    & --                           & --       & \\
111 & HD 54341          & 102 & A0    & --   & --    & --                           & --       & \\
112 & HD 71043          & 73  & A0    & --   & --    & --                           & --       & \\
113 & HD 71155          & 38  & A0    & --   & --    & --                           & --       & \\
114 & HD 78702          & 84  & A0    & --   & --    & --                           & --       & \\
115 & HD 136246         & 144 & A1    & --   & --    & --                           & --       & \\
116 & HD 159082         & 135 & B9    & --   & --    & --                           & --       & \\
117 & HD 164249         & 48  & F6    & --   & --    & --                           & --       & \\
118 & HD 166191         & 119 & F8    & --   & --    & --                           & --       & \\
119 & HD 181296         & 48  & A0    & --   & --    & --                           & --       & \\
120 & HD 182681         & 69  & B9    & --   & --    & --                           & --       & \\
121 & HD 218396         & 39  & A5    & --   & --    & --                           & --       & \\
122 & HD 220825         & 47  & A2    & --   & --    & --                           & --       & \\

    \hline
    \end{longtable}
    \tablefoot{References: 1 - \citet{Luhman+etal_2010}; 2 - \citet{Andrews+etal_2013}; 3 - \citet{Rigliaco+etal_2015}; 4 - \citet{Hartigan+Kenyon_2003}; 5 - \citet{Hartmann+etal_1998}; 6 - \citet{Hartigan+etal_1995}; 7 - \citet{Hernandez+etal_2004}; 8 - \citet{Mendigutia+etal_2011}; 9 - \citet{White+Ghez_2001}; 10 - \citet{Duchene+etal_1999}; 11 - \citet{White+etal_1999}; 12 - \citet{Hartigan+etal_1994}; 13 - \citet{Mathieu+etal_1997}; 14 - \citet{Czekala+etal_2016}; 15 - \citet{Espaillat+etal_2010}; 16 - \citet{Schneider+etal_2012}; 17 - \citet{Herczeg+etal_2014}; 18 - \citet{Vacca+etal_2011}; 19 - \citet{Garcia+etal_2006}; 20 - \citet{Fairlamb+etal_2015}; 21 - \citet{Natta+etal_2004}; 22 - \citet{Bast+etal_2011}; 23 - \citet{Salyk+etal_2013}; 24 - \citet{Andrews+etal_2009}; 25 - \citet{Ricci+etal_2010}; 26 - \citet{Reboussin+etal_2015}; 27 - \citet{Natta+etal_2006}; 28 - \citet{Isella+etal_2009}; 29 - \citet{Wilking+etal_2005}; 30 - \citet{Andrews+etal_2010}; 31 - \citet{Nattab+etal_2004}; 32 - \citet{Grosso+etal_2003}; 33 - \citet{Herbig+Bell_1988}; 34 - \citet{Mora+etal_2001}; 35 - \citet{Beskrovnaya+etal_1999}; 36 - \citet{Wade+etal_2007}; 37 - \citet{Donehew+etal_2011}; 38 - \citet{Meeus+etal_2012}; 39 - \citet{vanBoekel+etal_2005}; 40 - \citet{Manoj+etal_2006}; 41 - \citet{Czekala+etal_2015}; 42 - \citet{Hardy+etal_2015}; 43 - \citet{Cieza+etal_2012}; 44 - \citet{Montesinos+etal_2009}; 45 - \citet{Torres+etal_2006}; 46 - \citet{Pecaut+etal_2012}; 47 - \citet{Binks+etal_2014}; 48 - \citet{Kraus+etal_2014}; 49 - \citet{Bowler+etal_2014}; 50 - \citet{Pietu+etal_2003}; 51 - \citet{Lieman-Sifry+etal_2016}; 52 - \citet{Hernandez+etal_2005}; 53 - \citet{Williams+etal_2013}.
    \\ No reference means data comes from SIMBAD.  }
    \end{longtab}

%

%%%%%%%%%%%%%%%%%%%%%%%%%%%%
\section{Discussion}
\label{sec:discussion}

\subsection{Interpretation of the $S_{\rm CO}/F_{\rm cont}$ correlations}

Fig. \ref{fig:co_vs_dust_21} and \ref{fig:co_vs_dust_32} reveal an interesting correlation
that can be interpreted as follows:
different regimes of opacity correspond to different regions of the diagrams.
Depending on the dust/gas opacity, the emission is governed by different disk physical parameters.
%To better evaluate the positions of the disks in this diagram, it is important to take into account the different physical parameters which influence the emission of the disks.
Following the standard description of a dust disk \citep{Dutrey+etal_1996}, in the optically thin regime, the flux can be written as:
\begin{equation}
\label{eq:thin}
S_{\nu}(\tau_\nu \ll 1) = \frac{2kT}{D^2}\left(\frac{\nu}{c}\right)^2 \kappa M
,\end{equation}
where T is the (average) temperature, $\kappa$ the absorption coefficient, and M the mass
(each of these parameters being relative to the gas or dust component).

The continuum emission of dust grains is expected to be optically thin at the considered wavelengths even for
proto-planetary disks (typically, the optically thick area has a radius of  $R \simeq 10$ au at 1.3~mm for a disk such as
that surrounding DM Tau).
In contrast, we expect the $^{12}$CO gas emission to be mostly in the optically thick regime, where it should then write:

\begin{equation}
\label{eq:thick}
S_{\nu} (\tau_\nu \gg 1) \propto \frac{2\pi kT}{D^2}\left(\frac{\nu}{c}\right)^2 (R_{\rm out}^2 - R_{\rm in}^2) \times \cos(i)
,\end{equation}
where $R_{out}$ and $R_{in}$ are respectively the outer and inner radii of the disk.

To compare the optically thin emission to the optically thick one,
we can develop Equation \eqref{eq:thin} and write the mass as a function of the inner and outer disk radii.
Indeed, the mass can be derived from the surface density (see  Appendix \ref{annex:mass_diskfit} for details),
$\Sigma$(r)~=~$\Sigma_0 \left(\frac{R}{R_0}\right)^{-p}$:
\begin{equation}
M \propto \frac{2 \pi \Sigma_0 R_0^2}{2-p} \left(\frac{R_{\rm out}}{R_0} - \frac{R_{\rm in}}{R_0}\right)^{2-p} 
.\end{equation}

\begin{figure}
   \centering
   \includegraphics[width=0.45\textwidth,keepaspectratio]{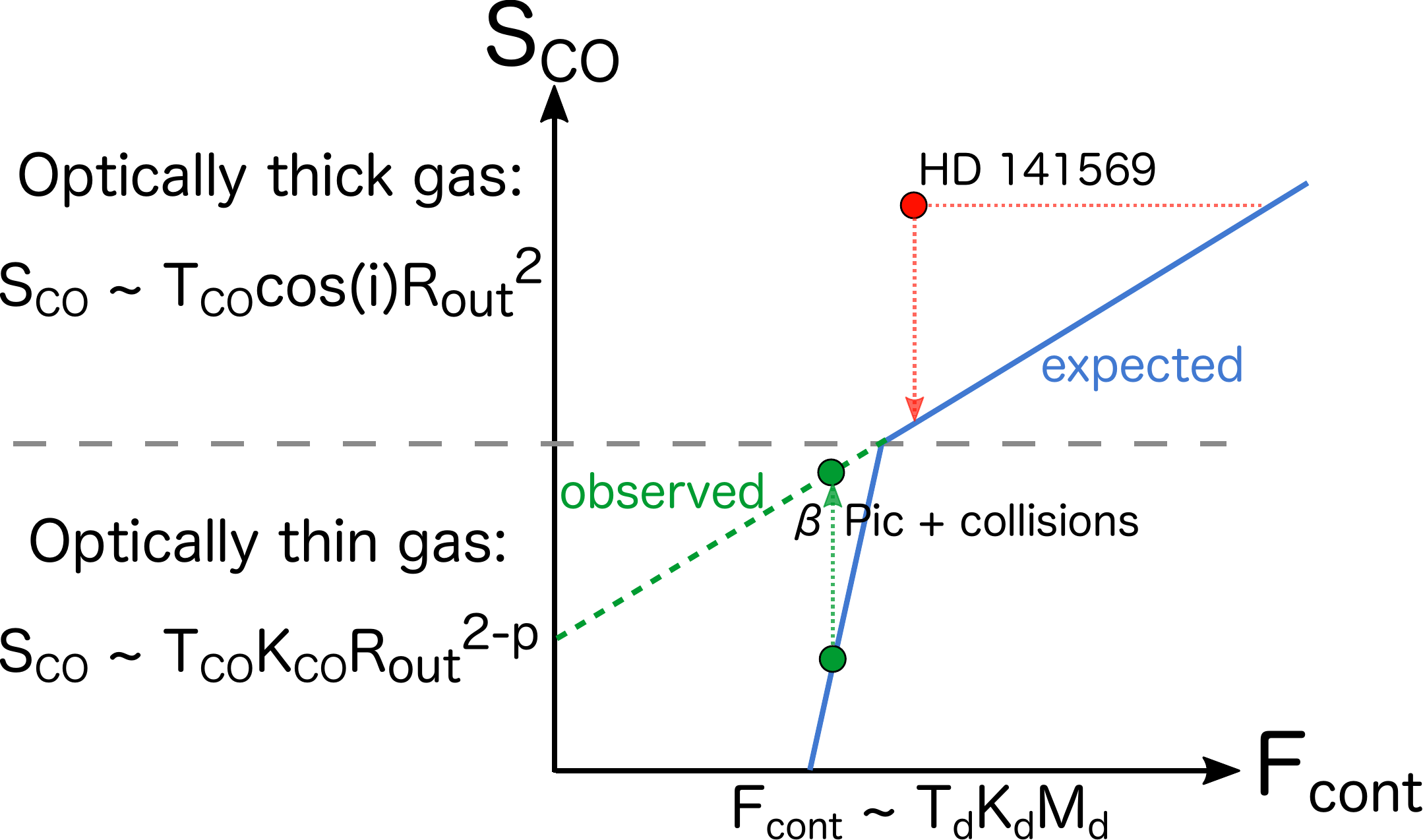}
   \caption{Diagram showing the physical parameters which dominate the emission of the CO
   gas and dust (see Section 4.2 for details). The higher part of the plot
   represents the optically thick regime of the gas, and the lower part represents the optically thin
   regime. In a hybrid disk such as HD~141569, the dust may evolve faster than the gas in
   a first step, moving the disk out of the correlation. In a second step, the natural
   gas dissipation moves downwards in position. On the contrary, debris disks are expected to
   lie in the optically thin region. For $\beta$~Pic, the position of the disk in the diagram could be higher than
   expected because of the gas enhancement produced by collisional events. For
   both sources, the arrows show the expected path in the diagram resulting from their
   possible evolutions. }
              \label{fig:theory_co_dust}%
\end{figure}

At first order, we thus see that the emission in the optically thick regime (upper part of the diagrams),
neglecting the inner radius, varies as R$_{out}^2$, while the optically thin emission varies as R$_{out}^{2-p}$.
The transition between the optically thick and thin regimes for the gas emission should then result in
a slope break, as sketched out in Fig.~\ref{fig:theory_co_dust}.
Others parameters, such as the inclination, the difference between the dust and the gas temperatures,
or the presence of a large inner hole may be responsible for the dispersion of the points
around the global trend in these diagrams, but should not affect the dependence on R$_{\rm out}$.

Interestingly, we do not observe a break in the correlations of Fig. \ref{fig:co_vs_dust_21} and \ref{fig:co_vs_dust_32}, all the observed sources
show the same slope suggesting optically thick CO emission all along.

\subsection{A possible diagram of evolution}

In Figs \ref{fig:co_vs_dust_21} and \ref{fig:co_vs_dust_32} , the systems span a large
range of properties (luminosity, mass and age etc.), from envelope-embedded young Class~II
sources to transition and young debris disks, and  a large fraction of the disk evolution
sequence is represented. Therefore, we investigate whether or not the correlation observed
between the CO emission and the thermal mm/sub-mm flux can be interpreted as a sequence
of evolution in the life of disks.

First of all, it is important to highlight that the detected hybrid objects are either isolated or in loose associations, and may not be representative of disk evolution in dense clusters.
Observations show that the fraction of disks seems lower in older clusters \citep{Haisch+etal_2001, Hernandez+etal_2007}, which indicates that the disk lifetime is approximately 5 Myr.
Survival of disks in general, however, might not be as short as the age determined in these studies.
Indeed, in clusters, the rich stellar environment may enhance the disk dispersal and lead to shorter lifetime estimates than for isolated stars \citep{Pfalzner+etal_2014, Vincke+etal_2015}.
\citet{Pfalzner+etal_2014} evaluate that at least 30\%, and possibly as much as 50\%, of the disks are living for more than 10~Myr.
The (isolated) hybrid disks we are discovering nowadays might thus be the first of a long series.
The fact that they are found around isolated A-stars contributes to the ease of their detection, since their disks tend to be more massive and luminous.

Nevertheless, as a proto-planetary disk evolves, its surface density becomes eroded under the effects of viscous evolution, photo-evaporation and planet formation.
As a consequence, its dust and gas emissions are expected to decrease until they reach the optically thin domain of debris disks
where, for a matter of sensitivity, only the continuum emission of (second generation) dust remains detectable in most cases.
The accretion rate is also often interpreted as an indicator of the evolution of the disks, since the accretion signatures are supposed to decrease when the gas dissipates.
\citet{Hartmann+etal_1998} have identified a decrease of the accretion rates with the stellar age, but the study was restricted to CTTS disks.
Several studies have also examined the evolution of the accretion rate in transitional disks compared to CTTS ones, since transitional disks are supposed to be in a more evolved stage than CTTS. It appears that the disk fraction showing accretion signatures is significantly lower in the transitional disk population, but when they are present, the accretion level is similar to the CTTS level, or less than an order of magnitude below \citep{Fang+etal_2013, Sicilia-Aguilar+etal_2010, Espaillat+etal_2012, Najita+etal_2007}. Fig. \ref{fig:ratio_macc} does
not show any correlations.

%If gas and dust evolve in the same way, no correlation between $S_{\rm CO}/F_{\rm cont}$ and the stellar type should be observed (as we found, see also Fig. \ref{fig:ratio_spt}, \ref{fig:ratio_mstar} and \ref{fig:ratio_lstar} ).
In Figs \ref{fig:co_vs_dust_21} and \ref{fig:co_vs_dust_32}, there is no significant difference between young (gas rich) disks orbiting Herbig Ae and T Tauri stars.
The linear shape of the CO/grains emission distribution in Figs \ref{fig:co_vs_dust_21} and \ref{fig:co_vs_dust_32},  and the small dispersion around the mean
trend suggest that the gas and dust components generally evolve concurrently, with comparable dissipation timescales, even if in our sample the distribution of
CTTS (including binaries, which are all along the main correlation) is more homogenous than that of Herbig Ae stars.

Although the number of faint disks simultaneously detected in CO line and continuum
remains too small for a robust analysis in the bottom-left region of the diagram,
all hybrid disk candidates (red symbols) with evolved dust and gas of (partially)
primordial origin clearly lie above the correlation line. This reveals either a deficit of
dust emission or an excess of CO emission. Similarly, we also observe an enhanced
flux ratio in Figs. \ref{fig:ratio_spt},  \ref{fig:ratio_mstar} and \ref{fig:ratio_lstar}  for the
hybrid disk population.

\subsection{Origin of hybrid disks}

% This high CO/dust flux ratio compared to most other objects of the diagram
% can be explained in several ways:
% \begin{enumerate}
% \item There is an excess of CO emission which can be due to a locally enhanced gas production,
% possibly associated with a recent collisional event in the disk.
% \item There is a deficit of mm dust emission. This  can be due for instance to
% a faster/more efficient grain growth followed by an inner drift occuring earlier in the
% Herbig Ae disk evolution.
% \item
% We cannot also exclude that

Herbig Ae stars may follow a different pathway, that is, on a specific trend lying above
that of low-mass stars or with a different slope in Figs \ref{fig:co_vs_dust_21} and
\ref{fig:co_vs_dust_32}, but the lack of Herbig stars in the range $F_{CO}=10-500$\,mJy
prevents us from drawing conclusions. However, in this region of the diagram, sensitivity is not
an issue and the apparent absence of Herbig Ae stars in the vicinity of the hybrid disks
group might be better explained by an effect of disk dissipation rather than by a
detection bias.
% \end{enumerate}

Two scenarios, corresponding to different evolution pathways in
Figs~\ref{fig:co_vs_dust_21} and \ref{fig:co_vs_dust_32}, may be invoked.

\subsubsection{Excess of CO by comet/planetesimal destruction}

% massive collisions of planetesimals can be invoked to release large amounts of CO.
In this first case, dramatic events such as analogs of the Late Heavy Bombardment in our
solar system \citep{Gomes+etal_2005} or a massive collision may be invoked to make faint
sources move towards the upper part of the diagram. Such a scenario is suspected for
$\beta$\,Pictoris, whose large CO intensity is thought to originate from a recent collision
between Mars-size objects or in an enhanced collision rate due to planetesimals trapped in
mean-motion resonances with an unseen planet \citep{Dent+etal_2014}.
%It might thus have temporarily reached the line
%where optically thick disks lie (see Fig. \ref{fig:theory_co_dust}).
%as it has been suggested for $\beta$ Pictoris by \citet{Dent+etal_2014}.
%One puzzling observation in our diagram is indeed
%\citet{Dent+etal_2014} have recently mapped the CO emission in this disk with ALMA,
%and showed that the CO emission is not uniform,
%but concentrated in one or two clumps and might result from a large collisional event or
%from an enhanced collision rate due to planetesimal trapped in mean-motion resonances with an unseen planet.
%The presence of the debris disk of $\beta$~Pic on the same line
%as the less evolved systems in the diagram (Fig~\ref{fig:co_vs_dust_32}) is indeed puzzling.
This suggests that its current location in our diagram, on the same line as optically
thick disks, is transient (see Fig. \ref{fig:theory_co_dust}), while the "steady-state"
regime of the icy debris collisional cascade may produce insufficient CO gas for most debris disks to be detectable
\citep[see recent \textit{ALMA} non-detections of CO around the nearby debris disks Fomalhaut and HD~107146,][]{Matra+etal_2015,Ricci+etal_2015}.

For HD~21997, a typical hybrid disk, \citet{Kospal+etal_2013} have evaluated the CO mass
from the C$^{18}$O emission. They constrain it to 4-8$\times$10$^{-2}$~M$_{\oplus}$, that is,
$\sim$4$\times$10$^{23}$~kg and $\sim$7$\times$10$^{20}$~kg for $^{12}$CO and C$^{18}$O (assuming an isotopic ratio of 560), respectively.
They estimated that such a CO mass cannot be purely produced by cometary collisions,
as at least 6,000 Hale-Bopp like comets would need to be totally destroyed each year.

To evaluate the relevance of a comet/planetesimal destruction scenario in explaining the
observed CO mass, we outline another approach. We use the following assumptions:
\begin{itemize}
\item $M_{disk} = 0.1 M_{\star} = 0.2 M_{\odot}$ for Herbig stars;
\item $M_{disk} \approx M_{H_2}$;
\item most of the O is in the form of H$_2$O; the H$_2$O content of the disk is
then given by cosmic oxygen abundance;
\item the CO content is derived using a $[\mathrm{CO/H}_2\mathrm{O}]$ molecular abundance
ratio in comets/planetesimals of 10\%. This value is found for CO-rich comets such as
Hale-Bopp and Lovejoy in the Solar System; the median being $\sim$4\% \citep[see
e.g,][]{Paganini+etal_2014,Mumma+Charnley_2011};
\item 10\% of this CO content ends up locked in comets/planetesimals.
\item $X18$, the $^{12}$CO/C$^{18}$O ratio in gas and in comets is similar; we use $X18 = 560$.
\end{itemize}
These assumptions give an estimate of the initial mass of $^{12}$CO and C$^{18}$O in the disk:
\begin{equation}
M_{CO}^{init} < M_{disk} \times m_{O/H} \times \left[\frac{\mathrm{CO}}{\mathrm{H}_2\mathrm{O}}\right] \times \frac{m_{CO}}{m_{O}}
,\end{equation}
where $m_{CO}$ is the CO molecular mass, $m_{O}$ the oxygen atomic mass, and
$m_{O/H}$ the fraction mass of oxygen compared to H. Taking the cosmic abundance
[O/H]~=~3.4$\times$10$^{-4}$ yields $m_{O/H} = 16 \times [O/H] \approx 0.005$. The
numerical application gives $M_{CO}^{init} <4\times10^{26}$~kg. The mass
ultimately locked in comets/planetesimals is thus $M_{CO}^{lock} < 4 \times 10^{25}$~kg and
$7 \times 10^{22}$~kg for $^{12}$CO and C$^{18}$O, respectively.

From the observed CO column density in HD~21997, neglecting the shielding from dust grains which have
a low column density and taking into account both the stellar UV flux and the ISM flux,
\citet{Kospal+etal_2013} found that the lifetime would be $<30,000$\,yr for $^{12}$CO and $<6,000$\,yr for C$^{18}$O.
The observed C$^{18}$O content would thus need to be replenished at a rate of $10^{17}$~kg.yr$^{-1}$.
The available reservoir of comets estimated above would then be exhausted in less than $6 \times 10^5$~yr.
Without a reservoir, the remaining $^{12}$CO would only last \textless30\,000 yr longer.

This should be compared to the required duration of the phenomenon: 3 Myr given the
detection rate (approximately 7 \%) and age of the oldest system (approximately 40 Myr). Thus, in spite of
the uncertainties in our assumptions (particularly in our conservative estimate
of the CO production rate), it is very unlikely that the scenario of
comets/planetesimals destruction could explain the whole CO emission observed in hybrid disks.

\subsubsection{Deficit of mm emission -- grain growth}

In the second scenario, the observed correlation suggests that disks share common
properties during most of their lifetime, differing in characteristic size and surface
density, but not in gas to dust ratio, and evolve along the correlation line because of
a global decrease in gas and dust content. Hybrid disks would appear when the dust
properties change significantly enough to affect the CO flux to mm flux ratio. A potential cause
is grain growth, as observed in the inner regions of disks
\citep{Guilloteau+etal_2011,Perez+etal_2012}, which lowers the dust opacity and thus the
mm flux without affecting the gas. A further reduction of the NIR to UV dust opacity can
allow the onset of photo-evaporation, which will then reduce the gas content faster than
the dust content \citep{Gorti+etal_2015}. In this scenario, the small number of hybrid
disks is in accordance with the short duration of the gas-rich, optically thin dust phase.

Unfortunately, the gas to dust ratio is difficult to estimate from the observations,
which only prove a high CO to continuum flux ratio in the hybrid disk phase. The more
detailed studies of HD~21997 \citep{Kospal+etal_2013} and HD~141569 (Di~Folco et al., in
prep.) indicate gas to dust mass ratios much larger than the canonical ISM value of 100
in the outer disk regions ($R>100$\,au) sampled by the observations. However, the paucity
of resolved hybrid disks precludes any firm conclusion on the gas to dust mass ratio, and
thus on the validity of the proposed scenario. We note however that a fast dust evolution
in these disks around A-type stars is also supported by the observational finding from
NIR excess that the (small) dust dispersal timescale was found to be shorter around
intermediate-mass stars than around low-mass stars
\citep[e.g,][]{Hillenbrand+etal_1993,Kennedy+Kenyon_2009,Ribas+etal_2015}.

Current searches have focussed on A stars: establishing the fraction of hybrid disks
around lower-mass stars would be an essential step in indentifying the cause of
their existence. This requires much more sensitive searches, only possible with ALMA
\citep[see][]{Lieman-Sifry+etal_2016}.
Spatially resolved studies such as those of HD~21997 and HD~141569 are also needed
to move from a global CO to dust flux ratio to a view of their respective distributions,
and to guide us towards the understanding of this elusive phase.

\begin{acknowledgements}
This research has made use of the SIMBAD database, operated at CDS, Strasbourg, France.
This publication is based on observations carried out with the IRAM 30-m telescope and the Atacama Pathfinder Experiment (APEX).
IRAM is supported by INSU/CNRS (France), MPG (Germany) and IGN (Spain).
APEX is a collaboration between the Max-Planck-Institut fur Radioastronomie, the European Southern Observatory, and the Onsala Space Observatory.
This work was supported by “Programme National de
Physique Stellaire” (PNPS) and “Programme National de Planétologie"
(PNP) from INSU/CNRS.
Finally, we warmly acknowledge the unknown referee for her/his detailed comments which 
helped to improve the paper.
\end{acknowledgements}

% %________________________________________________________________
% %________________________________________________________________
% %________________________________________________________________

% \section{Conclusion}

% %________________________________________________________________
% %________________________________________________________________
% %________________________________________________________________

\bibliographystyle{aa} % style aa.bst
  \bibliography{biblio_survey_co} % your references Yourfile.bib
% \begin{thebibliography}{}

%   % \bibitem[Baker(1966)]{baker} Baker, N. 1966,
%   %     in Stellar Evolution,
%   %     ed.\ R. F. Stein,\& A. G. W. Cameron
%   %     (Plenum, New York) 333

%    \bibitem[Balluch(1988)]{balluch} Balluch, M. 1988,
%       A\&A, 200, 58

%    \bibitem[Cox(1980)]{cox} Cox, J. P. 1980,
%       Theory of Stellar Pulsation
%       (Princeton University Press, Princeton) 165

%    \bibitem[Cox(1969)]{cox69} Cox, A. N.,\& Stewart, J. N. 1969,
%       Academia Nauk, Scientific Information 15, 1

%    \bibitem[Mizuno(1980)]{mizuno} Mizuno H. 1980,
%       Prog. Theor. Phys., 64, 544

%    \bibitem[Tscharnuter(1987)]{tscharnuter} Tscharnuter W. M. 1987,
%       A\&A, 188, 55

%    \bibitem[Terlevich(1992)]{terlevich} Terlevich, R. 1992, in ASP Conf. Ser. 31,
%       Relationships between Active Galactic Nuclei and Starburst Galaxies,
%       ed. A. V. Filippenko, 13

%    \bibitem[Yorke(1980a)]{yorke80a} Yorke, H. W. 1980a,
%       A\&A, 86, 286

%    \bibitem[Zheng(1997)]{zheng} Zheng, W., Davidsen, A. F., Tytler, D. \& Kriss, G. A.
%       1997, preprint
% \end{thebibliography}

%________________________________________________________________
%________________________________________________________________
%________________________________________________________________

\appendix

%________________________________________________________________
%________________________________________________________________
%________________________________________________________________

\section{Mass calculations}

%________________________________________________________________
%________________________________________________________________
%________________________________________________________________

\subsection{From integrated flux $S_{CO}$}
\label{annex:mass_scoville}

%________________________________________________________________
%________________________________________________________________
%________________________________________________________________

From \citet{Scoville+etal_1986}, the mean column density averaged over the beam writes:

    \begin{equation}
      N = \frac{k}{h\nu} \frac{3k}{8\pi^3B\mu^2} \frac{T_x + \frac{hB}{3k}}{(J+1).exp(-\frac{h\nu}{kT_x})} \frac{\tau}{1-e^{-\tau}} \int T_B dv
    ,\end{equation}

where B is the rotational constant of the observed molecule, $\mu$ its permanent dipole moment, J the rotational number of the lower state of the transition, T$_x$ is the excitation temperature, $\tau$ the optical depth and T$_B$ the brigntness temperature. This expression assumes the excitation temperature is high compared to the background temperature (see \citet{Scoville+etal_1986} for details).
The total H$_2$ mass is related to this column density thanks to the mean atomic weight of the gas $\mu_G$, the dihydrogen mass m$_{H_2}$, the abundance ratio of CO to $H_2$ X, the distance of the source D and the beam solid angle $\Omega_S$:

    \begin{equation}
      M_{H_2} = N \frac{\mu_G m_{H_2}}{X(CO/H_2)} D^2 \Omega_S
    .\end{equation}

Using the Planck's law,
    % \[
    %   T_b = \frac{S_{\nu}c^2}{2k\nu^2\Omega_S}
    % \]
we thus have

    \begin{equation}
      M_{H_2} = \frac{k}{h\nu} \frac{3cD^2\mu_G m_{H_2}}{16\nu^2\pi^3 B\mu^2 X} \frac{T_x + \frac{hB}{3k}}{(J+1).exp(-\frac{h\nu}{kT_x})} \frac{\tau}{1-e^{-\tau}}.10^{-23}S_{CO}
    ,\end{equation}

where all terms are in S.I. units, except the integrated flux S$_{CO}$ which should be expressed in Jy.km.s$^{-1}$.
We can sum up by

    \begin{equation}
    \label{eq:mass_formula}
      M_{H_2} (M_{\oplus}) = K_1 \times 10^{-11} \frac{D^2(pc)}{X}\frac{T_x+{\frac{hB}{3k}}}{exp(-K_2/T_x)}\frac{\tau}{1-e^{-\tau}} S_{CO}
    .\end{equation}

For the CO molecule, B~=~1.93~cm$^{-1}$ , that is, 5.79~$\times$~10$^{10}$~Hz, $\mu$~=~0.1098~Debye , that is, $\mu^2$~=~1.2~$\times$~10$^{-51}$~S.I., and $\mu_G$~=~1.36, which implies for the J~=~2$\rightarrow$1 transition: $K_1$~=~4.83, $K_2$~=~11.1 and for the J~=~3$\rightarrow$2 transition: $K_1$~=~0.953 and $K_2$~=~16.6.
We have taken the usual value of 10$^{-4}$ for X, the abundance of CO to H$_2$.
In this paper we have set the values of the excitation temperature and opacity to T$_{ex}$~=~40~K and $\tau$~=~1.

%________________________________________________________________
%________________________________________________________________
%________________________________________________________________
\subsection{From DiskFit modeling}
\label{annex:mass_diskfit}
%________________________________________________________________
%________________________________________________________________
%________________________________________________________________

Error bars on $\Sigma_0$ determined by the code DiskFit can be used to determine the $H_2$ mass as follows:

   \[
    \Sigma_m(r) = \Sigma_0 \left(\frac{r}{r_0}\right)^{-p} (cm^{-2}),
    \]
    \[
    N_{CO} = \int^{r_{out}}_{r_{int}} \int_0^{2\pi} \Sigma_m(r)r drd\theta,
    \]
    % \[
    % R = r/r_0
    % \]
    % \[
    % N_{CO} = \int_{R_{int}}^{R_{out}} \int_0^{2\pi} \Sigma_m(r) (Rr_0)(r_0dR)d\theta
    % \]
    % \[
    % N_{CO} = 2\pi r_0^2 \int_{\frac{r_{int}}{r_0}}^{\frac{r_{out}}{r_0}} \Sigma_0R^{-p}R dR
    % \]
    % \[
    % N_{CO} = 2\pi \Sigma_0 r_0^2 \int_{\frac{r_{int}}{r_0}}^{\frac{r_{out}}{r_0}} R^{1-p} dR
    % \]
    % \[
    % N_{CO} = 2\pi \Sigma_0 r_0^2 \left[\frac{R^{2-p}}{2-p}\right]_{\frac{r_{int}}{r_0}}^{\frac{r_{out}}{r_0}}
    % \]
    \[
    N_{CO} = \frac{2\pi \Sigma_0 r_0^2}{2-p} \left(\frac{r_{out}}{r_0}^{2-p} - \frac{r_{out}}{r_0}^{2-p} \right),
    \]
    \[
    M_{CO} = N_{CO}m_{CO},
    \]

where $\Sigma_m$ is the molecular surface density, $\Sigma_0$ its value at radius $r_0$, $r_{out}$ the outer radius of the disk, $N_{CO}$ the number of CO molecules in the disk, $m_{CO}$ the molecular mass of the CO, and $M_{CO}$ the CO mass in the disk (kg).
To retrieve the $H_2$ mass: $X = N_{CO}/N_{H_2} = 1 \times 10^{-4}$, then $N_{H_2} = N_{CO}/X$ and $M_{H_2}=N_{H_2}m_{H_2}$ where $N_{H_2}$ is the number of molecule of $H_2$, $m_{H_2}$ its molecular mass and $M_{H_2}$ the total mass of $H_2$.

%________________________________________________________________
%________________________________________________________________
%________________________________________________________________

\subsection{Comparison of the two methods}
\label{annex:comparison_mass}

%________________________________________________________________
%________________________________________________________________
%________________________________________________________________

\begin{figure}
   \centering
   \includegraphics[width=0.45\textwidth,keepaspectratio]{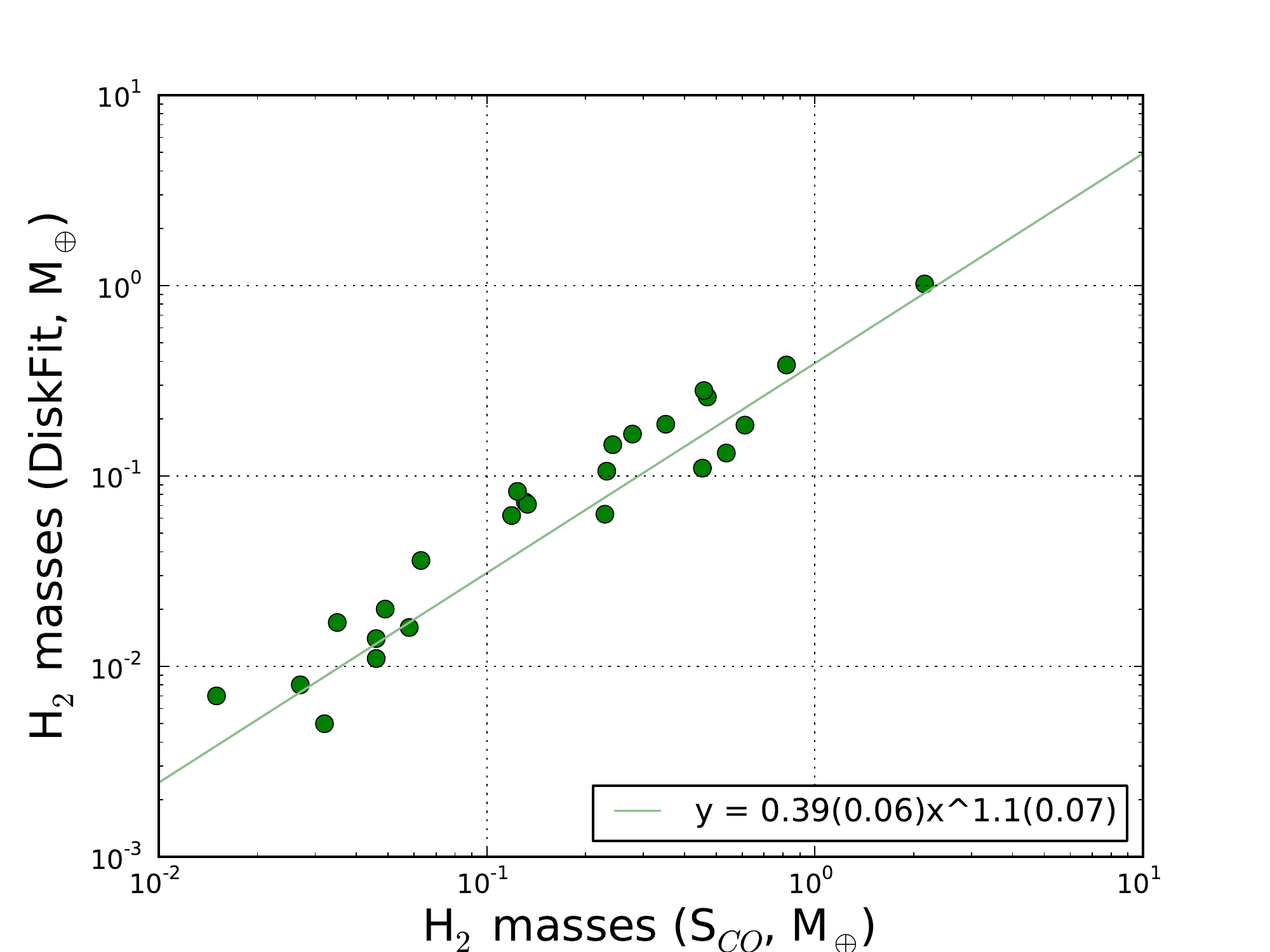}
   \caption{Masses of H$_2$ determined for the debris disks observed with \textit{APEX/IRAM} with the DiskFit method on Y axis and with formula \eqref{eq:mass_formula} on X axis. }
              \label{fig:mass_methods}%
\end{figure}

The Fig. \ref{fig:mass_methods} presents the masses determined by the two methods.
The methods give consistent results within a factor three. % (check something like 1$\sigma$ instead of 3$\sigma$ in the calculations).
The points showing strong deviations are HD~2772, HD~159082 (below the line), and HD~23642 (above the line).
% Reason unknown for the moment, to check.

%%%%%%%%%%%%%%%%%%%%%%%%%%%%%%%%%%%%%%%%%%%%%%%%%%%%%%%%%%%%%%%%%%%%%%%%%%%%%%%%%%%%%%%%%%%%%%%%%%%%%%%%%%%%%%%%%%%%%%%%%%
%%%%%%%%%%%%%%%%%%%%%%%%%%%%%%%%%%%%%%%%%%%%%%%%%%%%%%%%%%%%%%%%%%%%%%%%%%%%%%%%%%%%%%%%%%%%%%%%%%%%%%%%%%%%%%%%%%%%%%%%%%

%________________________________________________________________
%________________________________________________________________
%________________________________________________________________

\section{Interpolation of CO and dust emission}
\label{appendix:interpol}
%________________________________________________________________
%________________________________________________________________
%________________________________________________________________

\begin{figure*}
  \centering
  \begin{subfigure}{0.49\textwidth}
  \centering
  \includegraphics[width=0.99\textwidth,keepaspectratio]{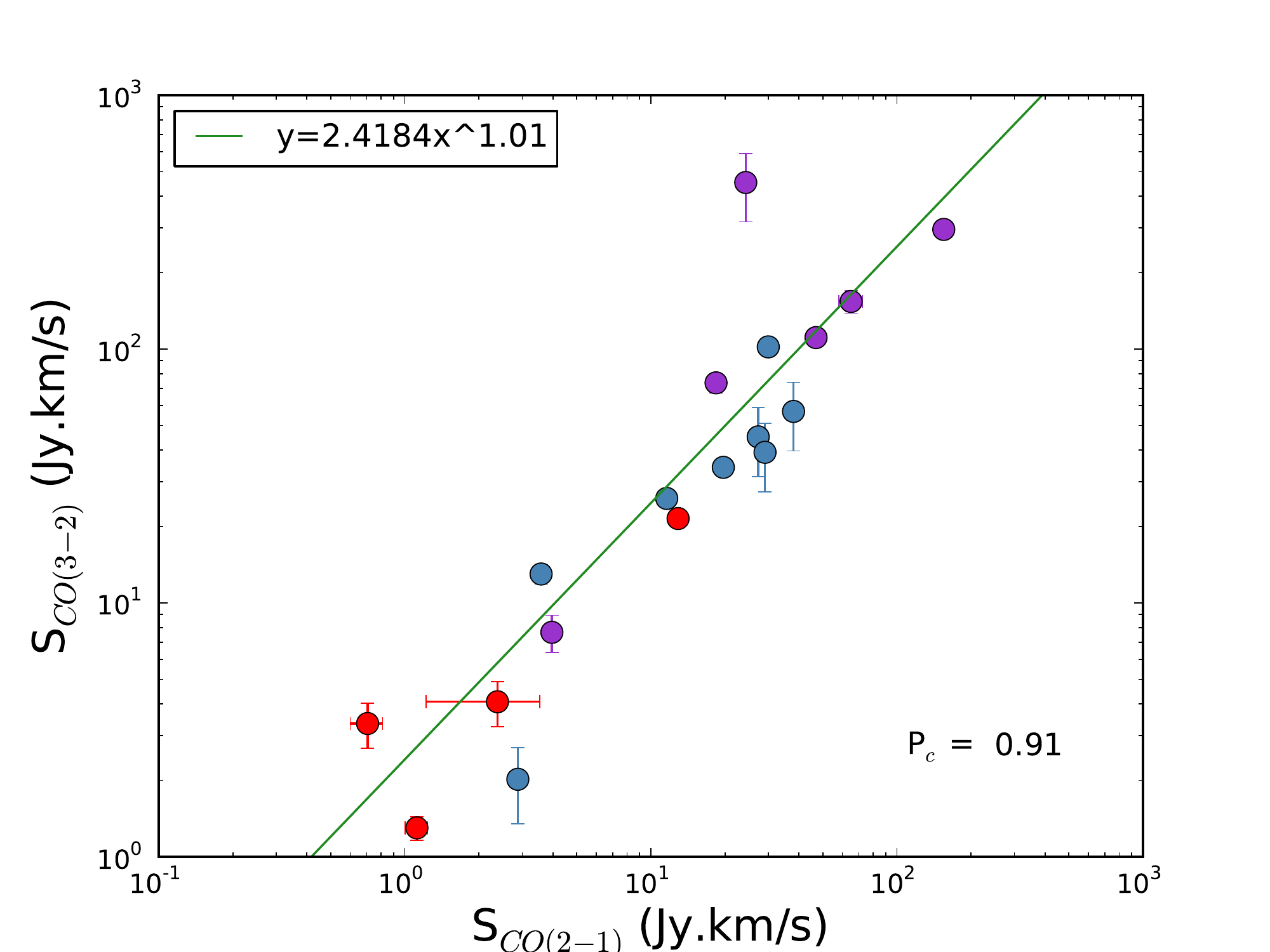}
  \caption{}
             \label{fig:co21_vs_co32}%
\end{subfigure}
\begin{subfigure}{0.49\textwidth}
\centering
\includegraphics[width=0.99\textwidth,keepaspectratio]{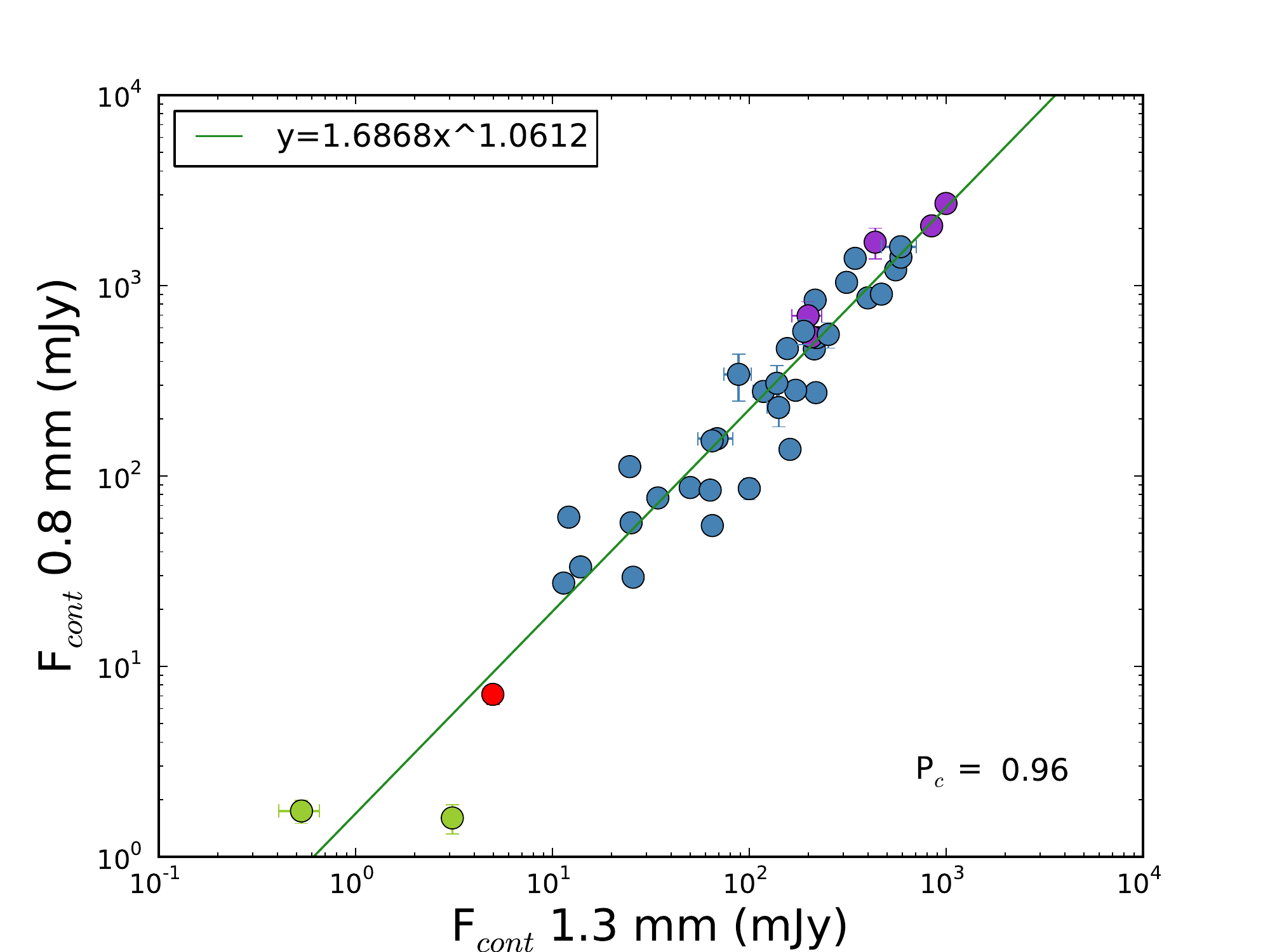}
  \caption{}
             \label{fig:d21_vs_d32}%
\end{subfigure}
\caption{(a) Emission of the \cotd{} plotted against the \codu{} emission. The Pearson
coefficient of correlation of the data, P$_c$, is indicated in the lower right-hand corner.\\
(b) Emission of the continuum at 0.8 mm plotted against the 1.3 mm continuum emission.
The Pearson coefficient of correlation of the data, P$_c$, is indicated in the lower
right-hand corner. Fluxes are rescaled at 100 pc. Same symbols as for
Fig.\ref{fig:co_vs_dust_nointerpol}}
\end{figure*}

Fig. \ref{fig:co21_vs_co32} and \ref{fig:d21_vs_d32} show the correlation between the
observations of the continuum and gas emission at the two wavelengths from the literature.
A linear regression of the data gives the two following equations, that we have used to
fill in the missing values (see section \ref{subsec:compilation}). When no error is given in
the literature, we have assumed a 10\% error. % - \textbf{AD ::: 10 \% or 20 \% ???}.

\begin{equation}
\label{eq:corr_co}
S_{CO(3-2)} = 1.7^{\pm0.4} \times \left(S_{CO(2-1)}\right)^{1.05^{\pm0.05}}
,\end{equation}

\begin{equation}
\label{eq:corr_dust}
F_{cont}^{0.8mm} = 1.3^{\pm0.5} \times \left(F_{cont}^{1.3mm}\right)^{1.09^{\pm0.06}}
.\end{equation}

\begin{figure*}
  \centering
  \begin{subfigure}{0.49\textwidth}
  \centering
  \includegraphics[width=0.99\textwidth,keepaspectratio]{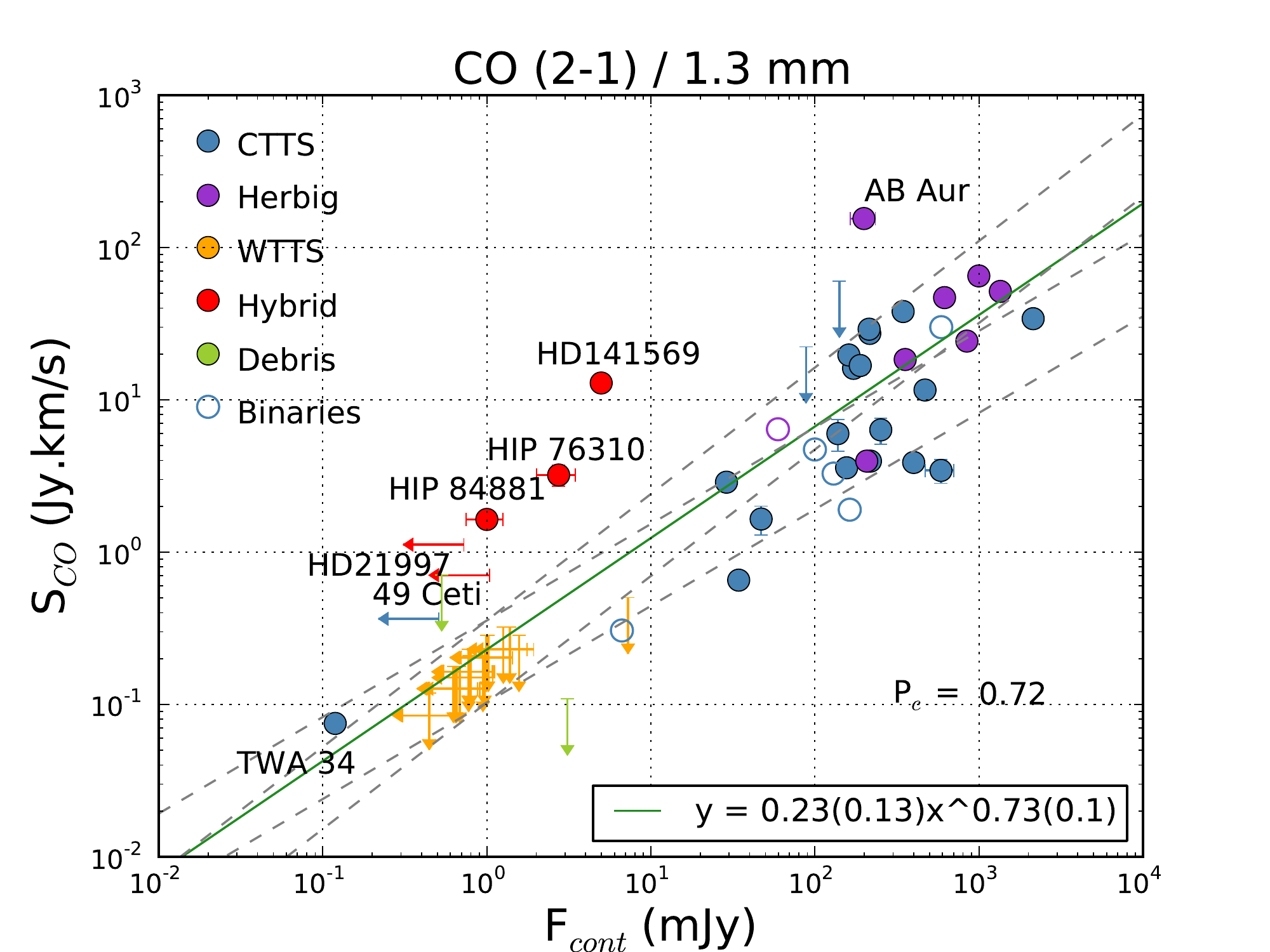}
  \caption{}
  \end{subfigure}
  \begin{subfigure}{0.49\textwidth}
  \includegraphics[width=0.99\textwidth,keepaspectratio]{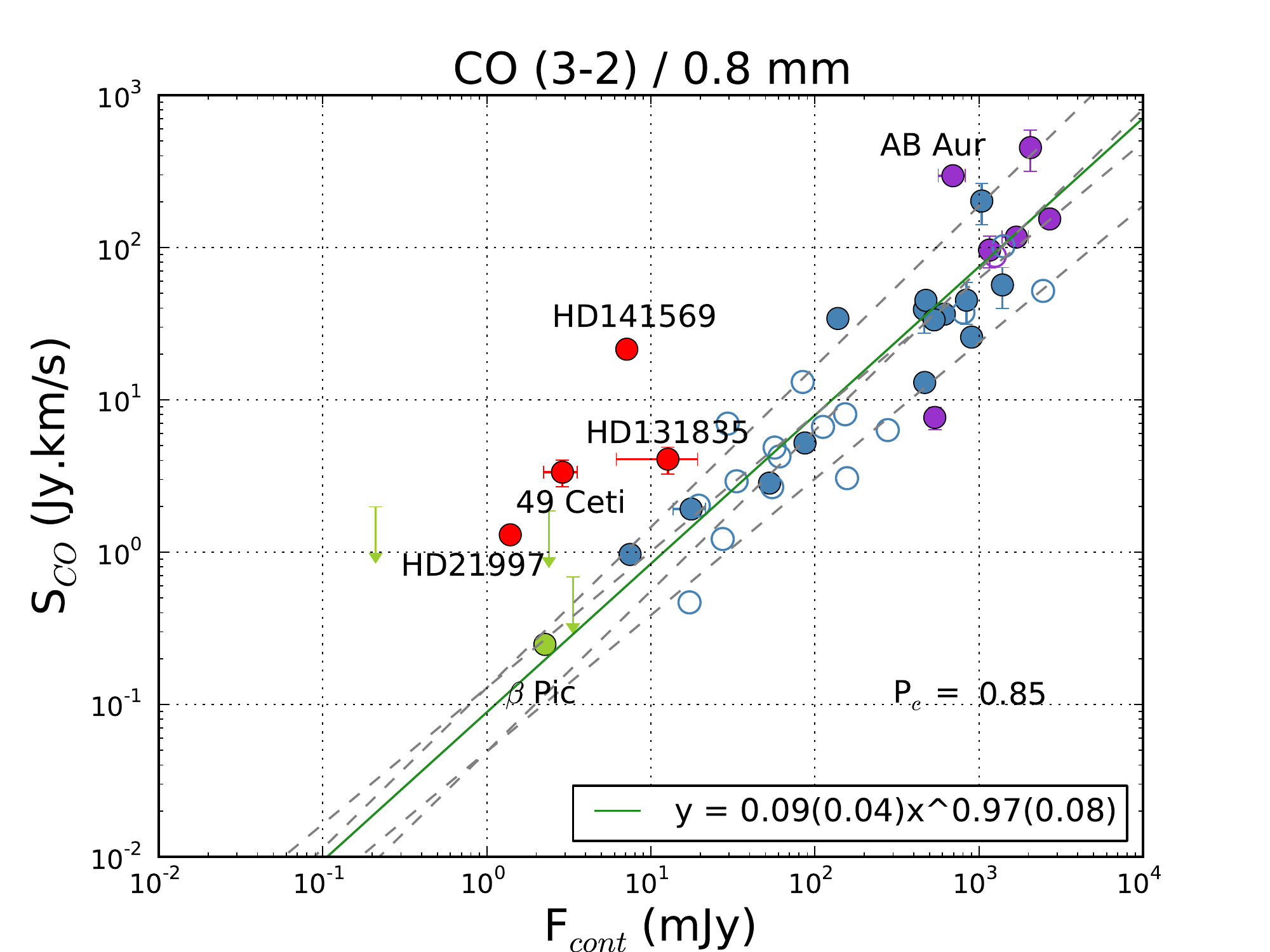}
  \caption{}
  \end{subfigure}
  \caption{(a) Emission of \codu{} emission plotted against the 1.3~mm continuum emission. The Pearson coefficient of correlation of the data, P$_c$, is indicated in the lower
right-hand corner. No extrapolated data are used here, only observations are reported and used to do the linear regression (green full line, equation indicated in the lower
right-hand corner). Right: same as upper left, for the \cotd{} emission against the 0.8~mm continuum emission. Fluxes are rescaled at 100 pc.}
             \label{fig:co_vs_dust_nointerpol}%
\end{figure*}

Fig. \ref{fig:co_vs_dust_nointerpol} shows the correlation between the CO and dust
emission, without interpolating the missing values. Only detections of CTTS and Herbig
are used to calculate the linear regression. When the error on a measurement was unknown,
it was fixed to 20\% of the value. A 10\% uncertainty was added to those given in Table
\ref{table:1} for the linear regression.
% The same figures with the interpolation are shown for the purpose of comparison (Fig. \ref{fig:co_vs_dust_interpol}).

\begin{table}
   \caption{Ratio of fluxes and derived H$_2$ masses for sources of the debris disks surveys where CO is detected.}
   \label{table:flux_mass_detections}
   \centering
   \begin{tabular}{lcc}     % 7 columns
   \hline\hline
     Source & CO(3-2)/CO(2-1) & M$_{H_2}^{32}$/M$_{H_2}^{21}$ \\
   \hline

     49~Ceti & 4.75 & 1.08 \\
     HD~141569 & 1.68 & 0.38 \\
     HD~21997 & 1.16 & 0.26 \\
     HD~131835 & 1.71 & 0.78 \\
   \hline
   \end{tabular}
   \end{table}

\section{Correlations with stellar parameters}
\label{appendix:correl-star}

\begin{figure*}
   \centering
     \includegraphics[width=0.48\textwidth,keepaspectratio]{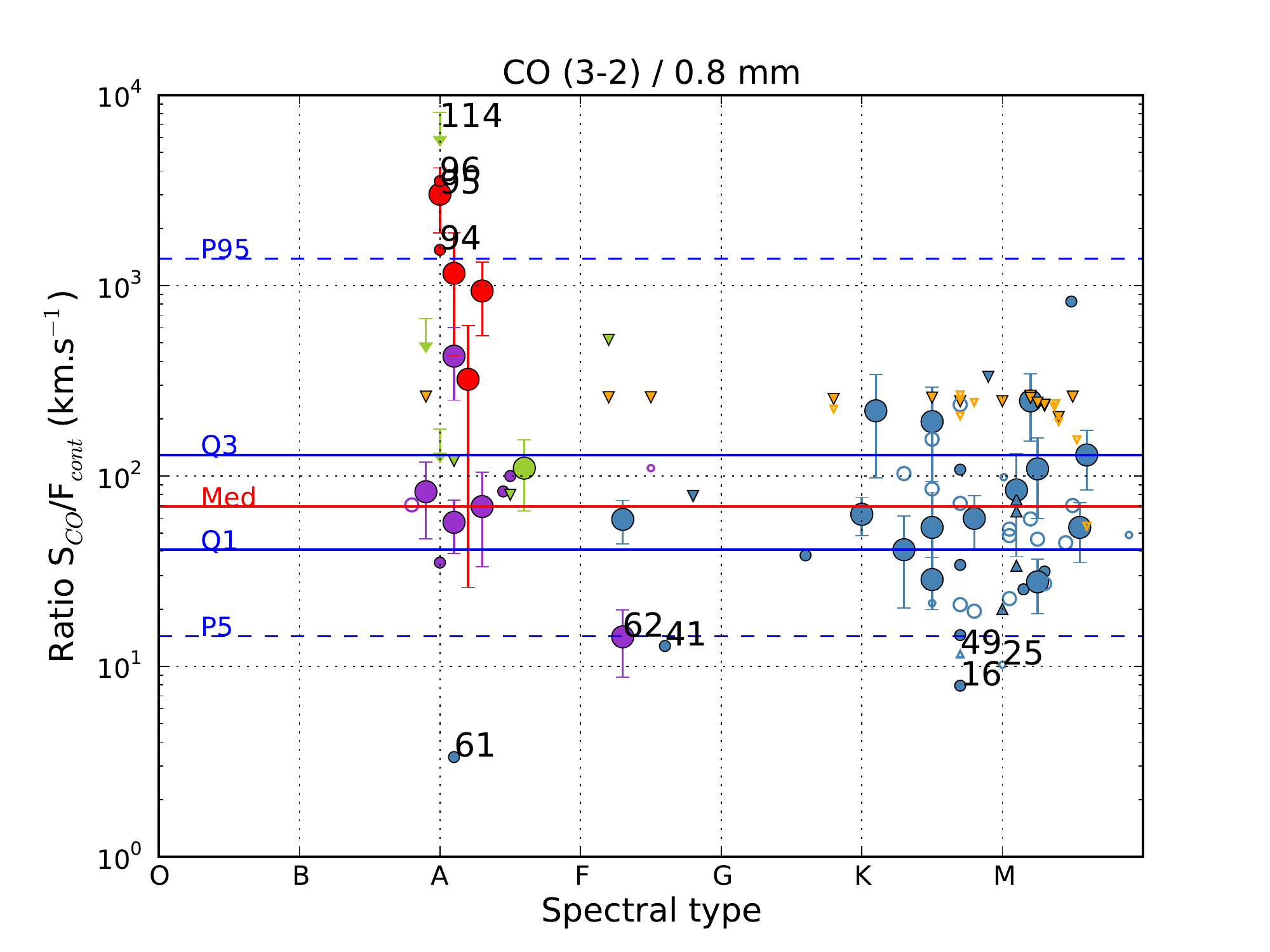}
   \caption{Ratio S$_{\rm CO}$/F$_{\rm cont}$ plotted against the spectral types of the stars, at 0.8 mm. The color code is the same as in Fig. \ref{fig:co_vs_dust_21} and \ref{fig:co_vs_dust_32}: blue for CTTS, violet for Herbig, orange for WTTS, red for hybrid and green for debris disks. Some statistics is represented by the lines: the red full line shows the median value of the ratio for the distribution of points, the blue full lines represent the first and third quartiles (50\% of the points are between these two lines) and the blue dashed lines shows the 5th and 95th percentiles (90\% of the points between the lines).}
              \label{fig:ratio_spt32}%
\end{figure*}

\begin{figure*}
   \centering
     \includegraphics[width=0.48\textwidth,keepaspectratio]{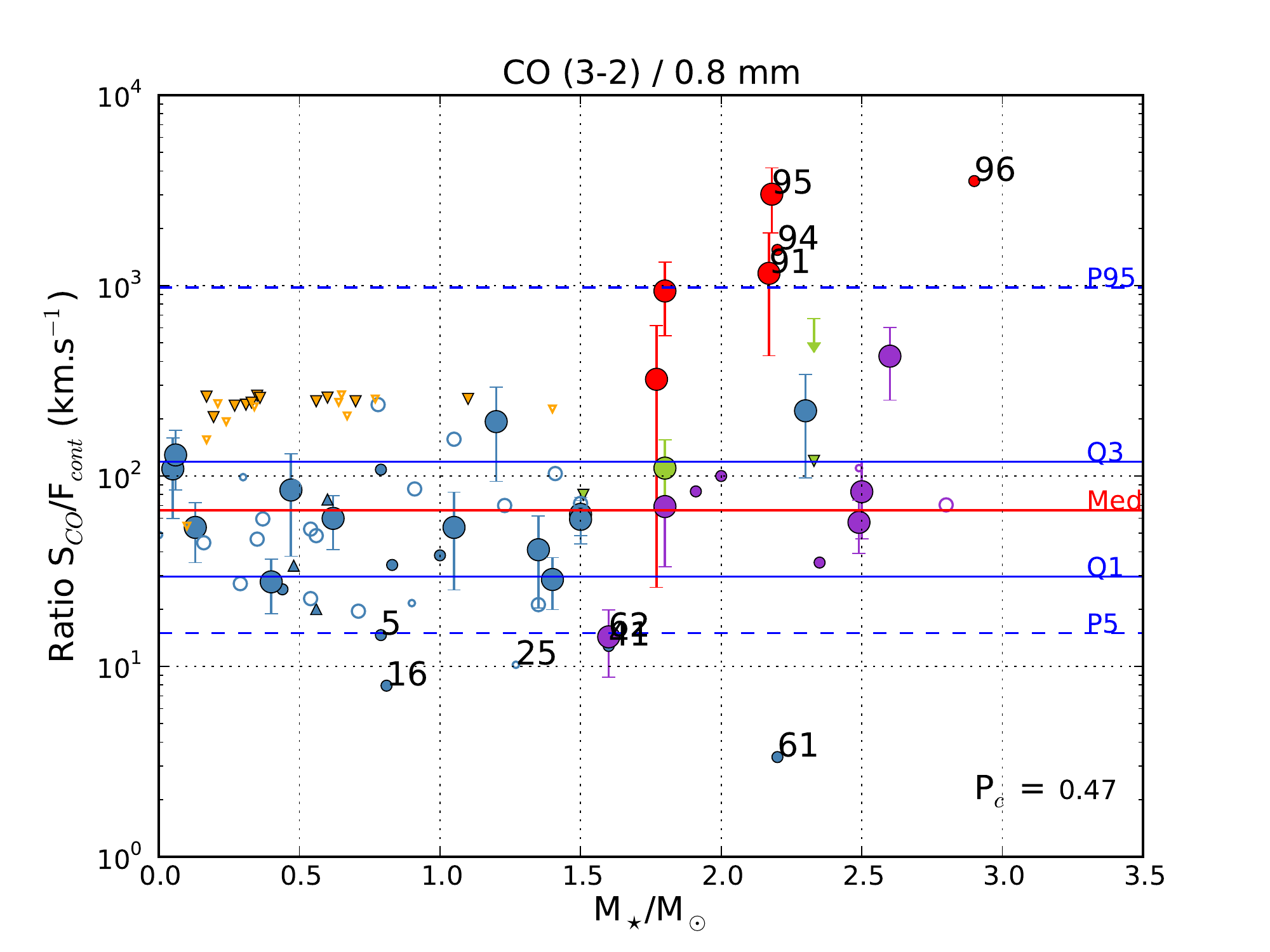}
      \includegraphics[width=0.48\textwidth,keepaspectratio]{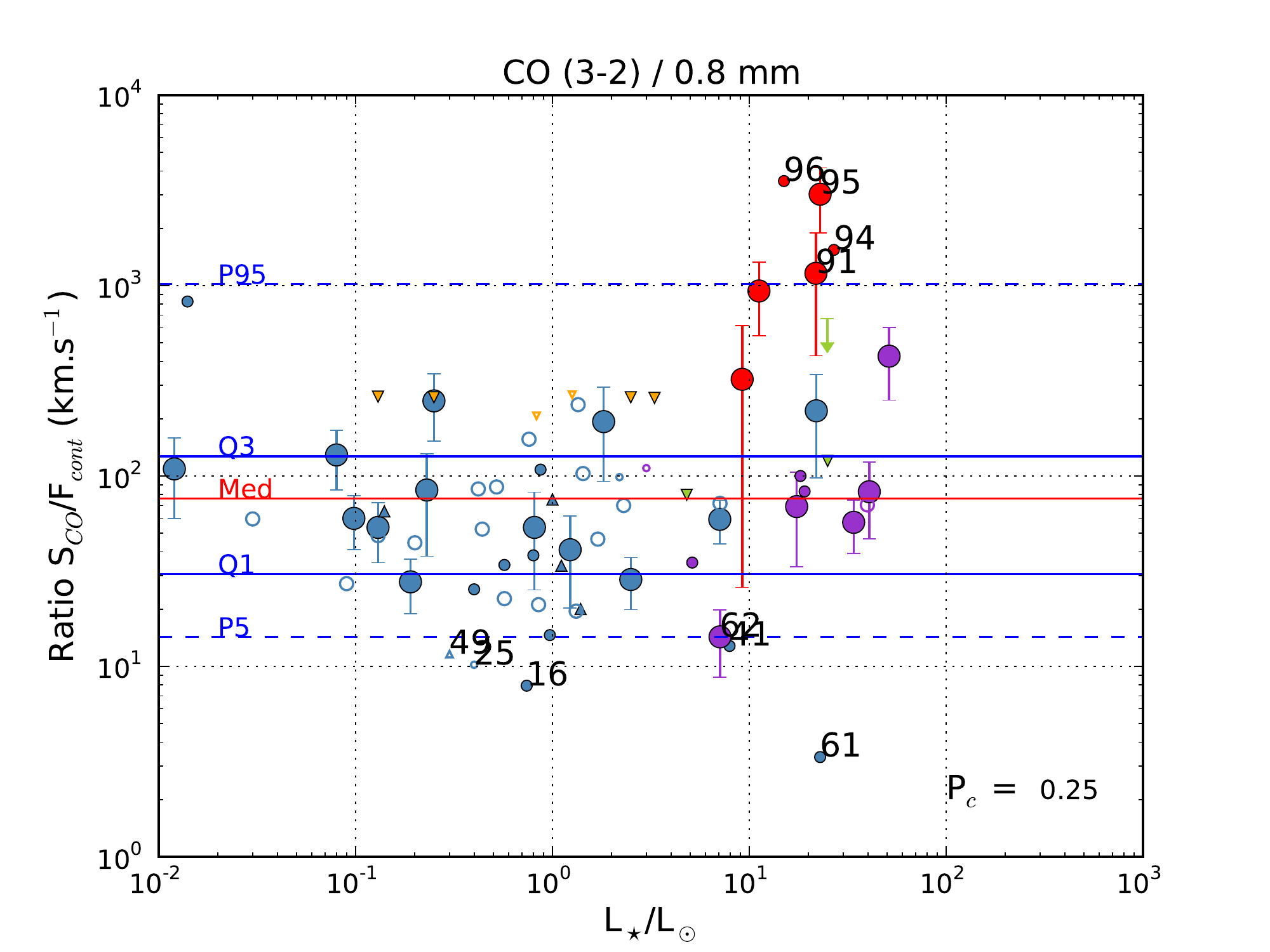}
    \caption{Left: Ratio S$_{\rm CO}$/F$_{\rm cont}$ plotted against the stellar mass, at 0.8 mm. The color code and lines are the same as in Fig. \ref{fig:ratio_mstar}. The Pearson coefficient of correlation of the data, P$_c$, is indicated in the lower
right-hand corner. Right: Ratio S$_{\rm CO}$/F$_{\rm cont}$ plotted against the stellar luminosity, at 0.8 mm. The color code and lines are the same as in Fig. \ref{fig:ratio_lstar}. The Pearson coefficient of correlation of the data, P$_c$, is indicated in the lower
right-hand corner.}
              \label{fig:ratio_mstar32}%
\end{figure*}

%\begin{figure*}
%   \centering
%      \includegraphics[width=0.48\textwidth,keepaspectratio]{../figures/new_figures/figures_ratio_lstar_idnumber_co21.pdf}
%     \includegraphics[width=0.48\textwidth,keepaspectratio]{../figures/new_figures/figures_ratio_lstar_idnumber_co32.pdf}
%    \caption{Ratio S$_{\rm CO}$/F$_{\rm cont}$ plotted against the stellar luminosity, at 0.8 mm. The color code and lines are the same as in Fig. \ref{fig:ratio_lstar}. The Pearson coefficient of correlation of the data, P$_c$, is indicated in the right lower part.}
%              \label{fig:ratio_lstar}%
%\end{figure*}

\begin{figure*}
   \centering
     \includegraphics[width=0.48\textwidth,keepaspectratio]{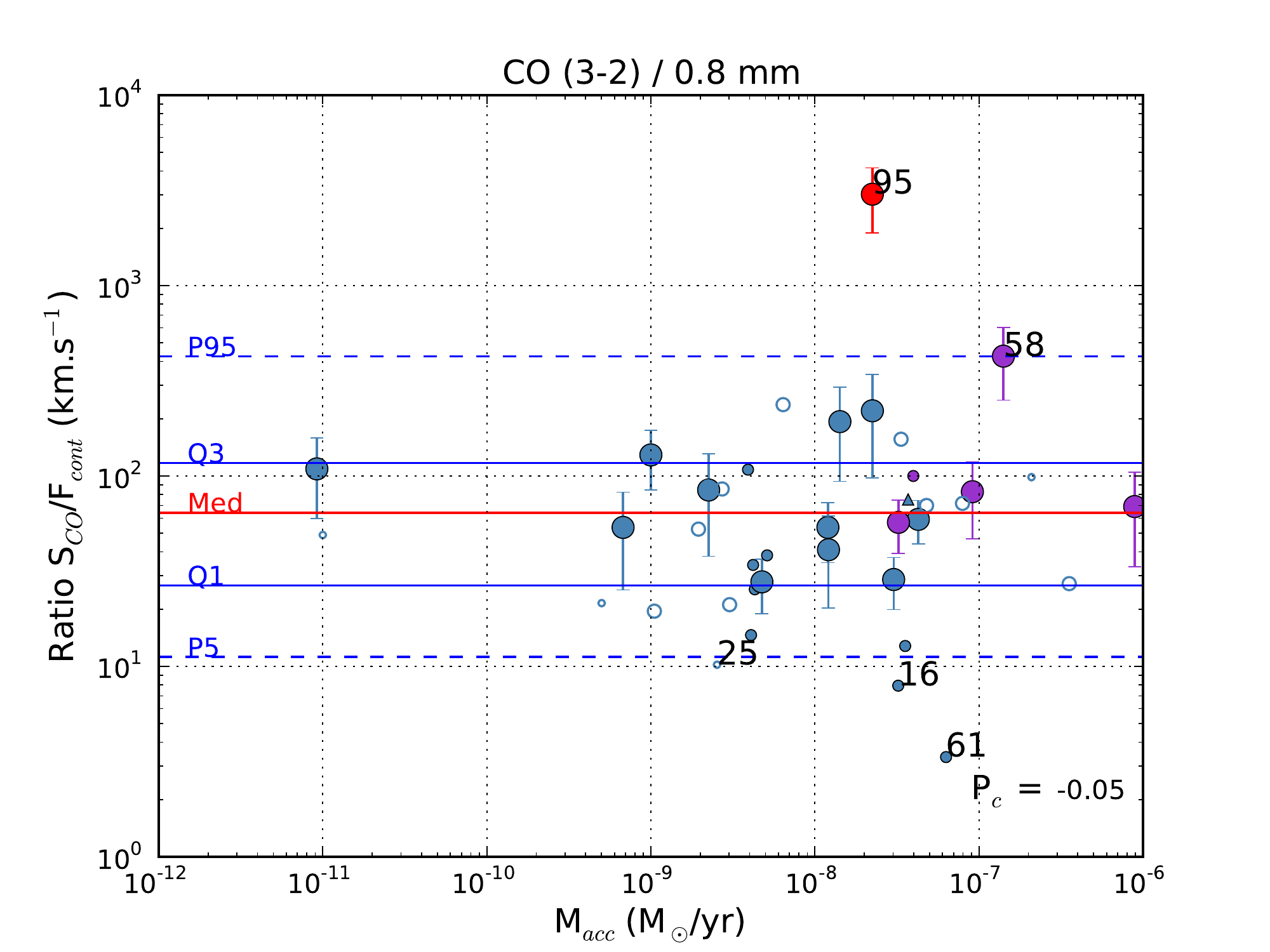}
          \includegraphics[width=0.48\textwidth,keepaspectratio]{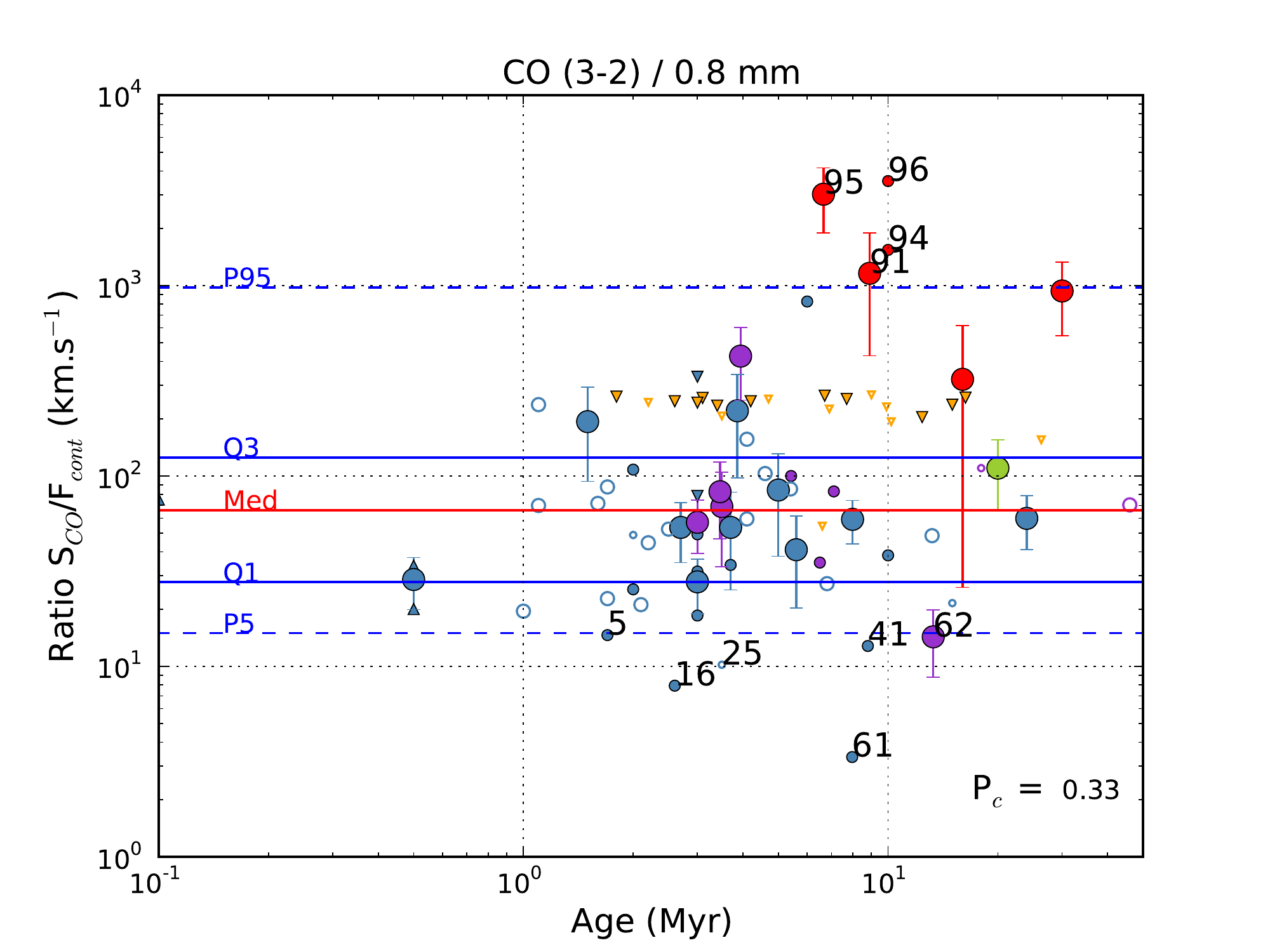}
    \caption{Left: Ratio S$_{\rm CO}$/F$_{\rm cont}$ plotted against the accretion rate, at 0.8 mm. The color code and lines are the same as in Fig. \ref{fig:ratio_macc}. The Pearson coefficient of correlation of the data, P$_c$, is indicated in the lower
right-hand corner. Right: Ratio S$_{C\rm O}$/F$_{\rm cont}$ plotted against the age of the system, at 0.8 mm. The color code and lines are the same as in Fig. \ref{fig:ratio_age}. The Pearson coefficient of correlation of the data, P$_c$, is indicated in the lower
right-hand corner.}
              \label{fig:ratio_macc32}%
\end{figure*}

%\begin{figure*}
%   \centering
%      \includegraphics[width=0.48\textwidth,keepaspectratio]{../figures/new_figures/figures_ratio_age_idnumber_co21.pdf}
%     \includegraphics[width=0.48\textwidth,keepaspectratio]{../figures/new_figures/figures_ratio_age_idnumber_co32.pdf}
%    \caption{Ratio S$_{C\rm O}$/F$_{\rm cont}$ plotted against the age of the system, at 0.8 mm. The color code and lines are the same as in Fig. \ref{fig:ratio_age}. The Pearson coefficient of correlation of the data, P$_c$, is indicated in the right lower part.}
%              \label{fig:ratio_age}%
%\end{figure*}
%
%
%

\end{document}